\documentclass[useAMS,usenatbib]{mn2e}

\usepackage{lscape}
\usepackage{aas_macros}
\usepackage{times}

\newbox\grsign \setbox\grsign=\hbox{$>$} \newdimen\grdimen \grdimen=\ht\grsign
\newbox\labox \newbox\gabox \newbox\simpropbox \newbox\wtildebox 
\setbox\gabox=\hbox{\raise.5ex\hbox{$>$}\llap
    {\lower.5ex\hbox{$\sim$}}}\ht1=\grdimen\dp1=0pt
\setbox\labox=\hbox{\raise.5ex\hbox{$<$}\llap
    {\lower.5ex\hbox{$\sim$}}}\ht2=\grdimen\dp2=0pt
\def\ga{\mathrel{\copy\gabox}}
\def\la{\mathrel{\copy\labox}}

\newcommand{\be}{\mbox{\begin{equation}}}
\newcommand{\ee}{\mbox{\end{equation}}}

\newcommand{\Cref}{\mbox{$m_{\rm ref}$}}

\newcommand{\msun}{\mbox{M$_\odot$}}

\title{On the fraction of star formation occurring in bound stellar clusters}   
\author{J.~M.~Diederik Kruijssen\\	
Max-Planck Institut f\"{u}r Astrophysik, Karl-Schwarzschild-Stra\ss e 1, 85748 Garching, Germany; kruijssen@mpa-garching.mpg.de}

\begin{document}

\date{Accepted 2012 August 10.  Received 2012 August 7; in original form 2012 June 15.}

\pagerange{\pageref{firstpage}--\pageref{lastpage}} \pubyear{2012}
\label{firstpage}

\maketitle

\begin{abstract}
We present a theoretical framework in which bound stellar clusters arise naturally at the high-density end of the hierarchy of  the interstellar medium (ISM). Due to short free-fall times, these high-density regions achieve high local star formation efficiencies, enabling them to form bound clusters. Star-forming regions of lower density remain substructured and gas-rich, ending up unbound when the residual gas is expelled. Additionally, the tidal perturbation of star-forming regions by nearby, dense giant molecular clouds imposes a minimum density contrast required for the collapse to a bound cluster. The fraction of all star formation that occurs in bound stellar clusters (the cluster formation efficiency or CFE) follows by integration of these local clustering and survival properties over the full density spectrum of the ISM, and hence is set by galaxy-scale physics. We derive the CFE as a function of observable galaxy properties, and find that it increases with the gas surface density, from $\Gamma\sim1$\% in low-density galaxies to a peak value of $\Gamma\sim70$\% at densities of $\Sigma_{\rm g}\sim10^{3}~\msun~{\rm pc}^{-2}$. This explains the observation that the CFE increases with the star formation rate density in nearby dwarf, spiral, and starburst galaxies. Indeed, comparing our model results with observed galaxies yields excellent agreement. The model is applied further by calculating the spatial variation of the CFE within single galaxies. We also consider the variation of the CFE with cosmic time and show that it increases with redshift, peaking in high-redshift, gas-rich disc galaxies. It is estimated that {up to} $30$--$35$\% of all stars in the Universe once formed in bound stellar clusters. We discuss how our theory can be verified with {\it Gaia} and ALMA, and provide possible implementations for theoretical work and for simulations of galaxy formation and evolution.
\end{abstract}


\setcounter{footnote}{0}
\section{Introduction} \label{sec:intro}
The formation of stars from a turbulent interstellar medium (ISM) is caused by the fragmentation of hierarchically collapsing giant molecular clouds \citep[e.g.][]{larson81,elmegreen04,maclow04,mckee07}. While this by necessity implies that stars generally do not form alone, it has been known for a long time that the star formation process does not exclusively produce stars in bound stellar clusters \citep{elmegreen83,lada87}. Instead, some fraction of stars is born in unbound stellar associations over a broad range of physical scales \citep[e.g.][]{blaauw64,clarke00,megeath04,portegieszwart10,gieles11,bastian12}, and in rare cases individual stars may even form in relative isolation \citep{parker07,krumholz09,bressert12}. Surprisingly though, star clusters are often still considered to be a fundamental unit of star formation \citep[e.g.][]{pflamm07,pfalzner09,assmann11}. Whereas galaxy formation studies augmented the scenario of monolithic collapse \citep[see e.g.~the seminal paper by][]{eggen62} with the current picture of hierarchical galaxy formation \citep[e.g.][]{white78,searle78,white91} several decades ago, the concept of star formation occurring in quantized, gravitationally bound systems has remained remarkably popular. 

It was pointed out by \citet{lada03} that in the Milky Way, embedded stellar groups constitute the vast majority of star formation and that these groups are about 10--20 times more numerous than gas-rid star clusters, suggesting that only a small fraction is capable of surviving the gas-embedded phase. With their review, the concept of {\it infant mortality} became firmly established -- in this scenario, most (if not all) stars form in clusters, but the population of embedded clusters is then decimated due to the expulsion of residual gas by feedback \citep[e.g.][]{tutukov78,hills80,lada84,adams00,geyer01,boily03,boily03b,bastian06b,goodwin06,baumgardt07,parmentier08}.

However, a recent study by \citet[also see \citealt{parker12b}]{bressert10} of the spatial distribution of young stellar objects (YSOs) in the solar neighbourhood has shown that star formation follows a continuous spectrum of number densities, indicating that there is no separate or critical density scale for star cluster formation imprinted in the star formation process. Their results imply that surface density thresholds used for the identification of stellar clusters \citep[e.g.][]{lada03,jorgensen08,gutermuth09} are arbitrary, and do not correspond to a physical scale. It is thus impossible to conclude that the majority of stars form in clusters by dividing the density spectrum of star formation at a certain density. In other work, it has been shown that YSOs are strongly correlated with the hierarchically structured ISM \citep{testi00,allen07,gutermuth11,bressert12b}, and that the impact of gas expulsion on bound stellar structure in a dynamical environment resulting from such a hierarchy is modest, if present at all \citep{kruijssen12,girichidis12,cottaar12}. Rather than the scenario of star cluster formation in which infant mortality (i.e.~gas expulsion) plays an important role, these results favour a framework in which the observed scarcity of bound, gas-rid clusters is a simple result of the star formation process: only some fraction of star formation reaches the densities required to attain high star formation efficiencies and result in bound stellar clusters, while the rest forms in a more dispersed fashion throughout the natal cloud \citep[e.g.][]{elmegreen01}.

The fraction of star formation occurring in bound stellar clusters is often quantified as the {\it cluster formation efficiency} \citep[CFE or $\Gamma$, see e.g.][]{bastian08,goddard10,adamo11,silvavilla11}. The CFE is a crucial quantity in many respects. Not only can it enable a better understanding of the star formation process itself \citep[e.g.][]{elmegreen02}, but it is also a key ingredient for work that aims to trace the (star) formation histories of galaxies using their star cluster populations \citep[e.g.][]{larsen01,bastian05,smith07,konstantopoulos09,fedotov11}, or studies of star cluster disruption \citep[e.g.][]{gieles05}. On galactic scales, the CFE may be used to infer whether the most massive stars inject their feedback energy within the same bubbles and hence how efficiently feedback couples to the ISM \citep[cf.][]{strickland99,krause12}. Additionally, the CFE is an important parameter in numerical simulations on galactic or cosmological scales that aim to model the assembly history of star cluster systems \citep[e.g.][]{prieto08,kruijssen11,kruijssen12c}. Thusfar, such simulations have had to assume that the CFE is constant throughout cosmic time.

Observational studies of extragalactic cluster populations at ages older than the embedded phase suggest that the CFE increases with the star formation rate surface density of the host galaxy \citep{larsen00b,goddard10,adamo11,silvavilla11}, but at present there is no theoretical understanding of the variation of the CFE with the galactic environment. It is notoriously hard to model the formation of stellar clusters in numerical simulations due to the large dynamic range that needs to be covered to resolve the necessary physics. Current state-of-the art simulations can model systems up to masses of a few $10^3~\msun$ \citep[e.g.][]{tilley07,bonnell08,krumholz12}, but this is still insufficient by several orders of magnitude to address the formation of cluster populations on the desired galactic scales. 

The aim of this paper is to derive and apply a theory of cluster formation that is based on simple analytical considerations. In part, it will follow the recent approach by \citet{elmegreen08}, who showed that the CFE can be related to the density spectrum of the ISM, but did so by defining a critical density (or pressure) for cluster formation. While not suitable to do quantitative predictions due to an arbitrarily defined density threshold, the work is a big step forward from the idealized picture in which centrally concentrated gas is expelled from spherically symmetric, bound clusters. In the present paper, the avenue suggested by \citet{elmegreen08} is explored further and expanded to provide a self-consistent framework for star cluster formation in galaxies.

The paper is organized as follows. In \S\ref{sec:clform} the required physical mechanisms are discussed that should be included in a theory of cluster formation. The model is derived and presented in \S\ref{sec:model}--\ref{sec:cfe}. In \S\ref{sec:param} the parameter space is explored and it is addressed how the CFE varies among different galaxy types. The calculated CFEs are compared to observed cluster populations in \S\ref{sec:obs}, giving excellent agreement. In \S\ref{sec:spatial} the spatial variation of the CFE within single galaxies is addressed. In \S\ref{sec:hubble} the variation of the CFE with cosmic time is discussed, accounting for the evolution of galaxy properties in a cosmological context. A discussion of possible caveats is presented in \S\ref{sec:disc}, together with an outlook to the observational verification of the model, and potential applications in theoretical and numerical work. The conclusions of this paper are given in \S\ref{sec:concl}. We provide Fortran and IDL routines for calculating the CFE with the model of this paper at http://www.mpa-garching.mpg.de/cfe (see Appendix~\ref{sec:appsupp}).

\section{Formation of bound stellar structure in star-forming regions} \label{sec:clform}
This section discusses the physical mechanisms that should be included in a theory of the CFE, and presents the derivation of a model in which these mechanisms are accounted for. In the remainder of the paper, the model is applied, verified and discussed.

\subsection{Relevant physical mechanisms} \label{sec:phys}
The fraction of stellar structure that eventually ends up being bound after star formation has ceased is the result of several physical mechanisms that act on a range of spatial scales. These mechanisms postulate a set of requirements for a theory designed to describe and explain the CFE. Observations, simulations and theory show that star formation is likely a universal process \citep[e.g.][]{elmegreen97,kennicutt98b,elmegreen02,krumholz05,mckee07}, in the sense that the physics of how a collapsing cloud converts its gas to stars seem to hold everywhere in the Universe. This implies that the blueprint for the initial stellar population is to a large degree already present in the characteristics of the ISM at the onset of star formation \citep[see e.g.][]{gutermuth11,longmore12,bressert12c}. It is therefore essential that a theory of the CFE is based on the structure and properties of the ISM that govern the process of star and cluster formation {(see point (i) below)}.

In a hierarchical ISM, the traditional picture of monolithic star cluster formation \citep[e.g.][]{baumgardt07} breaks down, as is illustrated by star-forming regions such as M16, M17, Taurus, Ophiuchus or even in the region around the Orion Nebula Cluster, which is the local archetype of clustered star formation. These regions may contain local concentrations of stars (which are generally unbound after subtracting the gravitational potential of the gas, see e.g.~\citealt{dezeeuw99,scally05,huff06}), but on a global scale they are constituted by a stellar population that is hierarchical in nature \citep[][]{efremov98,testi00,clarke00,bressert12b}. Per consequence, observational and theoretical studies do not find any evidence for discrete modes of star formation in the density spectrum of stellar structure in such regions \citep{bressert10,gutermuth11,gieles11,kruijssen12}. Instead, dispersed and clustered stellar structure represent opposite extremes of a continuous density spectrum of star formation. Although it is tempting to define a critical density that separates both extremes, it should be noted that this is not possible, especially for a hierarchically structured distribution -- at intermediate densities, only a certain part of the structure is bound \citep{goodwin97,dezeeuw99,kruijssen12} and no hard separation exists. This calls for a continuous description of cluster formation, in which the (partial) boundedness of stellar structure is evaluated locally {(see point (ii) below)}. This will enable the formulation of the CFE across the stellar density spectrum, which in turn allows the CFE to be connected to the density spectrum of the ISM if one adopts a local prescription for star formation \citep[e.g.][]{elmegreen02,krumholz05,padoan11}.

It is irrelevant to the problem at hand whether unbound stars result from gas expulsion or from a possibly intrinsically unbound state of the (stellar) substructure. The hierarchical nature of star-forming regions implies that there can exist spatially concentrated groups of stars that each individually may or may not be unbound \citep[see][]{kruijssen12,girichidis12}, while other, more distributed sequences of stellar groups could have ended up merging and becoming a star cluster had the gas between them not been expelled \citep[depending on the virial state of the cloud, see e.g.][]{dobbs11}. This does not mean that gas expulsion is not important, but it does indicate that the classical picture of infant mortality \citep{lada03} due to gas expulsion from centrally concentrated embedded clusters does not hold in a picture of hierarchical star formation. In such a dynamic environment, a singular observational definition of a cluster in terms of current or future potential groups is not possible, and therefore the CFE {requires specifying an age $t$ at which it is evaluated (see point (iii) below)}.

While the relative amounts of clustered and distributed star formation may already be largely set by the local properties of the ISM, it has also been shown that environmental effects can play an important role during the aftermath of star and cluster formation. This is to be expected -- given a universal star formation process, the characteristics of the ISM would only contain the complete blueprint for the state of the new-born stellar structure if star formation were an instantaneous mechanism. But because the star formation process takes time, environmental effects come into play as the conversion of gas into stars proceeds. The truncation of star formation by feedback and the potential dispersal of the stellar structure may appear to be mostly an internal process, but theoretical and numerical work have shown that the effectiveness of gas expulsion is largely determined by how the feedback couples to the ISM \citep[e.g.][]{elmegreen97,pelupessy12}. This implies that the time-scale on which the inflow of gas onto a star forming region can be halted depends on a combination of the porosity and pressure of the ambient ISM {(see point (iv) below)}.

Another environmental effect is entirely external in nature. In \citet{kruijssen11,kruijssen12} we identified the {\it cruel cradle effect}, which refers to the tidal disruption of star-forming regions or young stellar clusters by encounters with giant molecular clouds (GMCs) or other substructure in the gas of the dense, natal environment \citep[also see][]{elmegreen10b}. If the density contrast between the star-forming region and surrounding GMCs is low enough, tidal shocks\footnote{The similarity in nomenclature with {\it hydrodynamic} shocks is unfortunate, especially in the context of this paper. However, tidal shocks are completely unrelated to fluid discontinuities, and refer to the transient injection of energy into a gravitational system by the passage of a massive object such as for instance a GMC \citep[see e.g.][]{spitzer87,kundic95,gnedin99c,gieles06,kruijssen11}. Throughout the paper, both types of shocks are distinguished using the adjectives `tidal' and `supersonic'.} are capable of destroying the new-born stellar structure before the primordial gas has been cleared. The cumulative effect of this mechanism increases as star formation proceeds. The cruel cradle effect thus needs to be included in a theory of the CFE {(see point (v) below)}.

It is clear that the fraction of star formation producing bound stellar clusters is the result of a combination of factors, some of which are already imprinted in the ISM before the onset of star formation, while others are environmental effects that act on the stellar structure throughout. The requirements for a theory of the cluster formation efficiency are summarized as follows.
\begin{enumerate}
\item
The theory must account for the hierarchical structure of the ISM, and do so in a continuous manner, i.e.~without invoking any arbitrary thresholds for the formation of stars or stellar clusters.
\item
The dispersed and unbound fractions of star formation should follow from a local criterion that can be applied to the entire hierarchy of the ISM.
\item
The theory should enable predictions that are resolved in time, to account for the continuous nature of cluster formation.
\item
The truncation of star formation by feedback has to be accounted for, as well as its dependence on the ambient ISM (unless star formation is so efficient that the gas is depleted before then).
\item
The disruption of substructure by tidal shocks acting within the star formation time-scale should also be included.
\end{enumerate}
A model that satisfies the above requirements enables the prediction of the CFE as a function of the galactic environment. This can be between individual galaxies as well as locally resolved, both in space and time.

\subsection{Outline of the model} \label{sec:model}
The cluster formation efficiency $\Gamma$ can be formulated as the product of two fractions:
\begin{equation}
\label{eq:cfe}
\frac{\Gamma}{100\%}=f_{\rm bound}f_{\rm cce} .
\end{equation}
The first fraction, $f_{\rm bound}$, accounts for the naturally bound part of star formation and includes the effects of gas expulsion by feedback, thus incorporating the second, third and fourth points of the set of requirements postulated in the previous section. The second fraction, $f_{\rm cce}$, covers the fifth of these points and indicates the fraction of potentially clustered star formation that is left after applying the tidal disruption due to the cruel cradle effect.

Expressions for $f_{\rm bound}$ and $f_{\rm cce}$ are derived in detail below, but the theory can be summarized as follows. We translate the density spectrum of the hierarchical ISM into a local clustering of star formation. This can then be integrated to determine which part of star formation occurs in bound stellar clusters. The model consists of the following steps.
\begin{enumerate}
\item
The starting point is the overdensity probability distribution function (PDF) of the ISM. This PDF reflects the distribution of density contrasts with respect to the mean density. (\S\ref{sec:ismpdf})
\item
By assuming that star formation occurs in a gas disc that obeys hydrostatic equilibrium, the mid-plane density can be derived, allowing the overdensity PDF to be written as a PDF of the absolute density. (\S\ref{sec:disk})
\item
Given a certain (absolute) density, the local free-fall time is known. By adopting a specific star formation rate per free-fall time,\footnote{This is the fraction of the gas that is converted into stars per free-fall time.} and determining when the feedback pressure is strong enough to prevent the further inflow of cold gas and hence halt star formation, it is possible to determine the final star formation efficiency at each density. (\S\ref{sec:sfe})
\item
For a local star formation efficiency, it is possible to formulate a local fraction of stars that remains bound upon instantaneous gas expulsion. To establish this relation, we use a numerical simulation of star formation in a turbulent ISM. This then provides the naturally bound fraction of star formation at each density. (\S\ref{sec:fbound})
\item
It follows from the classical theory of tidal shocks in stellar clusters that there is a certain overdensity above which regions can survive the tidal perturbations by their environment. The theory is adjusted slightly to be applicable to gas-rich star-forming regions. (\S\ref{sec:cce})
\item
The final expression for the CFE is obtained by integration of the naturally bound fraction of star formation over the density range of the PDF where it survives the cruel cradle effect, and dividing it by the integral of the star formation efficiency over the entire density range of the PDF. (\S\ref{sec:cfe})
\end{enumerate}
The reader who is not so much interested in the details of the derivation below is referred to \S\ref{sec:cfe}, where the theory is summarized graphically in Figure~\ref{fig:xpdf}.

\subsection{The density spectrum of the interstellar medium} \label{sec:ismpdf}
The hierarchical structure of the ISM is driven by a combination of turbulence and local gravitational contraction. It has been known for well over a decade now that the PDF of the mass density in isothermal, supersonically turbulent clouds and larger structures of cold gas is well-described by a log-normal function \citep{vazquez94,padoan97,padoan02,federrath08,federrath10,padoan11,hill12}. When formulated as the PDF of the {\it over}density with respect to the mean density in the turbulent region $x=\rho_{\rm g}/\rho_{\rm ISM}$, it can be written independently of the physical scale as:
\begin{equation}
\label{eq:pdf}
\frac{{\rm d} p}{{\rm d} x}=\frac{1}{\sqrt{2\pi\sigma_\rho^2}x}\exp{\left[-\frac{(\ln{x}-\overline{\ln{x}})^2}{2\sigma_\rho^2}\right]} ,
\end{equation}
where the logarithmic mean $\overline{\ln{x}}$ is related to the standard deviation $\sigma_\rho$ as
\begin{equation}
\label{eq:meanx}
\overline{\ln{x}}=-\frac{\sigma_\rho^2}{2} ,
\end{equation}
and
\begin{equation}
\label{eq:sigma}
\sigma_\rho^2=\ln{\left(1+b^2{\cal M}_{\rm 3D}^2\right)}=\ln{\left(1+3b^2{\cal M}^2\right)},
\end{equation}
with $b\approx0.5$ \citep[e.g.][]{padoan02}, which is consistent with a mix of solenoidal and compressive turbulence forcing \citep{federrath10}, and $\cal{M}_{\rm 3D}$ and $\cal{M}$ the three-dimensional and linear Mach numbers, respectively. Equations~(\ref{eq:pdf})--(\ref{eq:sigma}) hold at scales smaller than the scale at which the turbulence is driven, which in the case of disc-like gas distributions has an upper limit equal to the disc scale height \citep[e.g.][]{maclow04}. The smallest scale on which these equations hold is set by the ambipolar diffusion length, i.e.~a few hundreths of a parsec, or the length-scale of protostellar cores \citep{ossenkopf02}. It was recently shown that the form of the overdensity PDF is very similar in the presence of strong magnetic fields, requiring only a correction factor in the second term of equation~(\ref{eq:sigma}) of $\beta_0/(\beta_0+1)$, where $\beta_0\equiv P_{\rm th}/P_{\rm mag}$ is the ratio of the thermal pressure to the magnetic pressure \citep{padoan11,molina12}. For simplicity, magnetic fields are omitted in the present model for the CFE, although it should be noted that the adopted value of $b=0.5$ is roughly consistent with recent results of magnetohydrodynamical simulations with weak magnetic field strengths \citep[in which the Alfv\'{e}nic velocity is comparable to or smaller than the sound speed, see][]{molina12}.

\subsection{Gas discs in hydrostatic equilibrium} \label{sec:disk}
The above set of equations shows that the density PDF solely depends on the Mach number of the gas ${\cal M}$ and the mid-plane density of the galaxy disc $\rho_{\rm ISM}$. Following \citet{krumholz05} we assume that the gas disc of the galaxy is in hydrostatic equilibrium,\footnote{{Perhaps surprisingly, this assumption is not too inaccurate for (e.g. merger-induced) starburst galaxies, in which the rapid cooling of star-forming gas implies that a substantial fraction of the star-forming regions follows a disc-like morphology (see \citealt{hopkins09b} and \S\ref{sec:assum}).}} which allows both quantities to be expressed in terms of the mean surface density $\Sigma_{\rm g}$, angular velocity $\Omega$, and the \citet{toomre64} stability parameter $Q$. This implies that the overdensity PDF becomes a PDF of the absolute density, which is set by those three variables. In this formulation, the mid-plane density is given by
\begin{eqnarray}
\label{eq:rhoism}
\nonumber \rho_{\rm ISM}&=&\frac{\phi_P\Omega^2}{\pi G Q^2} \\
&=&2.8\times10^{-21}Q^{-2}\Omega_0^2~{\rm g}~{\rm cm}^{-3} ,
\end{eqnarray}
where $\phi_P=3$ is a constant to account for the gravity of the stars and $\Omega_0\equiv\Omega/{\rm Myr}^{-1}$ is a convenient scaling of the angular velocity. By deriving an expression for the velocity dispersion in GMCs and comparing it to the typical sound speed in the cold ISM, the Mach number of the star-forming regions in a galaxy disc is approximated as
\begin{equation}
\label{eq:mach}
{\cal M}=2.82\phi_{\overline{P}}^{1/8}Q\Omega_0^{-1}\Sigma_{\rm g,2} ,
\end{equation}
with $\Sigma_{\rm g,2}\equiv\Sigma_{\rm g}/10^2~\msun~{\rm pc}^{-2}$. In this equation, $\phi_{\overline{P}}$ is the ratio of the mean GMC pressure to the mid-plane disc pressure and can be expressed as
\begin{equation}
\label{eq:phi}
\phi_{\overline{P}}=10-8f_{\rm GMC}\approx10-8\left(1+0.025\Sigma_{\rm g,2}^{-2}\right)^{-1},
\end{equation}
where $f_{\rm GMC}$ is the mass fraction of the ISM in GMCs. The \citet{toomre64} $Q$ parameter is defined as
\begin{equation}
\label{eq:q}
Q\equiv\frac{\kappa\sigma_{\rm g}}{\pi G\Sigma_{\rm g}}\approx\frac{\sqrt{2}\Omega\sigma_{\rm g}}{\pi G\Sigma_{\rm g}},
\end{equation}
in which $\kappa$ is the epicyclic frequency, $\sigma_{\rm g}$ the one-dimensional velocity dispersion of the gas, and $\Omega$ the angular velocity within the galaxy. The second equality assumes a flat galaxy rotation curve. Gas discs with $Q<1$ are considered unstable to gravitational collapse, whereas $Q>1$ indicates stability by kinetic support. As a fiducial value $Q=1.5$ is assumed, but it is known to vary between 0.5 and~6 \citep[e.g.][]{kennicutt89,martin01}. Finally, the angular velocity can be related to the surface density as
\begin{equation}
\label{eq:omega}
\Omega_0=0.058\Sigma_{\rm g,2}^{0.49},
\end{equation}
with an intrinsic scatter of about 0.5~dex in the range $\Sigma_{\rm g,2}=10^{-2}$--$10^3$. Note that this is a fit to observed, nearby galaxies in their entirety, implying that $\Omega$ should still be treated as an independent variable when the spatially resolved CFE within a model galaxy is calculated, or when high-redshift galaxies are considered. Moreover, if the rotation curve is not flat, $\Omega$ should be replaced by
\begin{equation}
\label{eq:kappa}
\Omega\rightarrow\frac{\kappa}{\sqrt{2}}\equiv\frac{v_{\rm c}}{R_{\rm gc}}\sqrt{1+\frac{{\rm d}\ln{v_{\rm c}}}{{\rm d}\ln{R_{\rm gc}}}} ,
\end{equation}
in all model equations. In this definition, $v_{\rm c}$ indicates the circular velocity, and $R_{\rm gc}$ represents the galactocentric radius. The quantity $\kappa/\sqrt{2}$ increases from $\Omega$ for flat rotation curves to $\sqrt{2}\Omega$ for solid body rotation. Neglecting this variation thus introduces an error of up to a factor $\sqrt{2}$ in the angular velocity, which translates to a smaller uncertainty on the CFE (see \S\ref{sec:param} for the variation of $\Gamma$ with $\Omega$).

For details regarding the derivation of equations~(\ref{eq:mach}), (\ref{eq:phi}), and~(\ref{eq:omega}) the reader is referred to \citet{krumholz05}. In combination with equations~(\ref{eq:pdf})--(\ref{eq:sigma}), this set of equations defines the absolute density PDF as a function of the gas surface density $\Sigma_{\rm g}$, the Toomre $Q$ parameter, and the angular velocity $\Omega$.

\subsection{The local star formation efficiency} \label{sec:sfe}
Given the density PDF of the turbulent gas, we can compute the fraction of the mass that ends up in stars (the star formation efficiency, SFE or $\epsilon$) as a function of density and thereby establish how much each density contributes to the total amount of star formation in a galactic region. 

\subsubsection{The specific star formation rate per free-fall time}
Assuming that star formation proceeds on a free-fall time $t_{\rm ff}$ \citep[e.g.][]{elmegreen00}, the star formation efficiency can be expressed in terms of the specific star formation rate per free-fall time ${\rm sSFR}_{\rm ff}$, i.e.~the mass fraction of a GMC that is converted into stars each free-fall time. For this, it is needed to assume a star formation law, which has been the topic of extensive debate in the recent literature \citep[e.g.][]{elmegreen02,krumholz05,krumholz07,elmegreen07,padoan11}. A star formation law specifies ${\rm sSFR}_{\rm ff}$, which interestingly exhibits little variation over a large dynamic range. Observational results on galactic scales suggest that ${\rm sSFR}_{\rm ff}\sim0.01$ \citep{kennicutt98b,elmegreen02}, whereas in the Cores to Disks (c2d) {\it Spitzer} Legacy project, a value ${\rm sSFR}_{\rm ff}\sim0.04$ is found within nearby star-forming regions \citep{evans09}. Considering the difference in physical scales, this relative consistency is remarkable. In this paper, the CFE is derived using two different star formation laws. The first is based on the empirically motivated assumption that ${\rm sSFR}_{\rm ff}$ is approximately constant \citep{elmegreen02}:
\begin{equation}
\label{eq:ssfrff_e02}
{\rm sSFR}_{\rm ff}^{\rm E02}=0.012 .
\end{equation}
The second star formation law that is used in this work is the prescription from \citet{krumholz05}. They estimate sSFR$_{\rm ff}$ from the part of the overdensity range that can be interpreted as protostellar cores:
\begin{equation}
\label{eq:ssfrff_km05}
{\rm sSFR}_{\rm ff}^{\rm KM05}=0.13\left[1+{\rm erf}\left(\frac{\sigma_\rho^2-\ln{0.68\alpha_{\rm vir}^2{\cal M}^4}}{2^{3/2}\sigma_\rho}\right)\right] ,
\end{equation}
where the Mach number is estimated from $\Sigma_{\rm g}$, $Q$, and $\Omega$ as before, and $\alpha_{\rm vir}$ is the virial parameter of a typical GMC \citep[see][]{bertoldi92}. The virial parameter is given by
\begin{equation}
\label{eq:alpha}
\alpha_{\rm vir}=\frac{5\sigma_{\rm int}^2R}{GM} ,
\end{equation}
in which $M$ is the cloud mass, $R$ is the cloud radius, and $\sigma_{\rm int}$ is the internal velocity dispersion, which is assumed to equal the turbulent velocity dispersion because cold GMCs are highly supersonic. GMCs with $\alpha_{\rm vir}\sim1$ are in virial equilibrium, while those with $\alpha_{\rm vir}>1$ are not self-gravitating\footnote{{Note that this does not imply zero star formation, since part of the GMC will be gravitationally unstable.}} and those for which $\alpha<1$ are contracting or supported by magnetic pressure. A fiducial value of $\alpha_{\rm vir}=1.3$ \citep{mckee03} is adopted here, but it should be noted that values in the range $\alpha_{\rm vir}=0.2$--$10$ are found in observations \citep{solomon87,heyer09} and numerical work \citep{tasker09,dobbs11,hopkins12}. For typical values of $\alpha_{\rm vir}=1.3$ and ${\cal M}=100$ (this is appropriate for $\Sigma_{\rm g,2}\sim1$), equation~(\ref{eq:ssfrff_km05}) gives ${\rm sSFR}_{\rm ff}^{\rm KM05}=0.014$, very similar to ${\rm sSFR}_{\rm ff}^{\rm E02}=0.012$. In our fiducial model, the empirical star formation law of equation~(\ref{eq:ssfrff_e02}) will be used; the influence of the adopted star formation law {(and, per consequence, of the GMC virial parameter)} on the CFE is discussed in more detail in \S\ref{sec:param}.

\subsubsection{The end point of the star formation process}
If the star formation process freely continues until it is halted by feedback,\footnote{The Galactic Center provides an exception to such a scenario due to its high density and angular velocity: the Arches cluster is thought to have stopped forming stars due to a cloud-cloud collision, which would explain why it is no longer associated with a GMC at its present age of $\sim2$--$4$~Myr \citep{wang06}. For regions where the time-scale for cloud-cloud collisions is much shorter than the feedback time-scale, i.e.~$t_{\rm coll}\ll t_{\rm fb}$, the feedback time-scale in equation~(\ref{eq:sfe}) can be replaced by the cloud-cloud collision time-scale $t_{\rm coll}=2Q\Sigma_{\rm GMC}/\phi_P\Omega\Sigma_{\rm g}$ \citep[see e.g.][]{silk97}, with $\Sigma_{\rm GMC}$ the typical surface density of GMCs. However, on a global scale the formation of stellar clusters in galaxy discs is generally unaffected, because the cloud-cloud collision time-scale increases approximately linearly with galactocentric radius, and greatly exceeds $t_{\rm fb}$ outside the central 100~pc of the Milky Way.} one can formulate a feedback time-scale $t_{\rm fb}$ to characterise the duration of star formation. The SFE then becomes a simple function of sSFR$_{\rm ff}$, $t_{\rm ff}$, and $t_{\rm fb}$:
\begin{equation}
\label{eq:sfe}
\epsilon=\frac{{\rm sSFR}_{\rm ff}}{t_{\rm ff}}t_{\rm fb} .
\end{equation}
In this equation, the free-fall time $t_{\rm ff}$ is defined for each density $\rho_{\rm g}$ as
\begin{equation}
\label{eq:tff}
t_{\rm ff}=\sqrt{\frac{3\pi}{32G\rho_{\rm g}}} .
\end{equation}

The feedback time, i.e.~the time it takes to halt star formation, is taken to depend on when pressure equilibrium is attained between feedback and the surrounding ISM. There is some debate in the literature as to which feedback mechanisms halts star formation on the spatial scale of entire GMCs. There are indications that supernova feedback is the most important agent for truncating the star formation process on these scales \citep{larson74,mckee77,korpi99,joung06,dobbs11b}. However, it has recently also been shown that supernova feedback could be ineffective in high-density galaxies due to efficient cooling and instead radiative feedback would be the dominant agent \citep[e.g.][]{thompson05,fall10,murray10,dale12}. Throughout this paper, we will assume that supernovae truncate the star formation process on the largest spatial scales, but in Appendix~\ref{sec:apprad} we also consider the (additional) effect of radiative feedback. The relevance of supernovae is supported to some degree by numerical simulations of embedded clusters -- those clusters that span the largest spatial scales and have the weakest coupling between feedback and the ISM, reflecting a high degree of inhomogeneity, have dynamical histories that are most strongly impacted by supernovae rather than winds from massive stars \citep{pelupessy12}. Crucially though, the description of feedback in our model serves the purpose of truncating star formation on a time-scale that is broadly consistent with observations, rather than reflecting the details of the feedback process. Additionally, Appendix~\ref{sec:apprad} shows that varying the feedback prescription has only a modest influence on the predicted CFE. 

We write the feedback time as the sum of the time since the onset of star formation until the first supernova $t_{\rm sn}$ and the subsequent time until pressure equilibrium between feedback and the surrounding ISM $t_{\rm eq}$:
\begin{equation}
\label{eq:tfb}
t_{\rm fb}=t_{\rm sn}+t_{\rm eq} .
\end{equation}
At times $t<t_{\rm fb}$, star formation still is still ongoing because the ambient ISM pressure $P_{\rm ISM}$ is higher than the outward pressure provided by feedback $P_{\rm fb}$, and gas is continuously fed into the star-forming region. The ambient pressure that needs to be overcome to halt star formation in GMCs is dominated by turbulent motion and is therefore given by
\begin{equation}
\label{eq:pamb}
P_{\rm ISM}=\rho_{\rm ISM}\sigma_{\rm g}^2=\frac{\pi^2G^2Q^2\Sigma_{\rm g}^2\rho_{\rm ISM}}{2\Omega^2} ,
\end{equation}
where equation~(\ref{eq:q}) was used in the second equality. The condition for pressure equilibrium depends on how the feedback couples to the ISM -- since the structure of the ISM is hierarchical, its porosity causes the pressure support by feedback to be inefficient \citep[e.g.][]{silk97}. Taking this into account, equilibrium can then be formulated as
\begin{equation}
\label{eq:pressure}
P_{\rm fb}=\frac{E_{\rm fb}}{V}=\phi_{\rm fb}\rho_{\rm s} t_{\rm eq}=\phi_{\rm fb}\epsilon\rho_{\rm g} t_{\rm eq}=\frac{\pi^2G^2Q^2\rho_{\rm ISM}\Sigma_{\rm g}^2}{2\Omega^2} ,
\end{equation}
where $E_{\rm fb}$ is the feedback energy providing the outward pressure, $V$ is the volume of the region, $\rho_{\rm s}$ is the stellar density, and $\phi_{\rm fb}$ is a constant that represents the rate at which feedback injects energy into the ISM per unit stellar mass. It thus includes a certain efficiency factor smaller than unity and includes the porosity of the ISM as well as additional energy loss due to cooling. While it is known that the feedback--ISM coupling is crucial in shaping the properties of galaxies in numerical simulations \citep[e.g.][]{efstathiou00,springel03,abadi03,robertson05,dubois08,dallavecchia08}, its characteristics are still uncertain \citep{silk97,maclow99,navarro00,dib06}, and therefore the value of $\phi_{\rm fb}$ has not been determined conclusively. In Appendix~\ref{sec:appfb}, $\phi_{\rm fb}$ is chosen by considering order-of-magnitude estimates from the literature, and it is shown that the influence of this choice on the CFE is minor. In the remainder of the paper, we adopt $\phi_{\rm fb}\approx 0.16~{\rm cm}^2~{\rm s}^{-3}=3.2\times10^{32}~{\rm erg}~{\rm s}^{-1}~\msun^{-1}$. From equation~(\ref{eq:pressure}), the equilibrium time $t_{\rm eq}$ follows as
\begin{equation}
\label{eq:teq}
t_{\rm eq}=\frac{\pi^2G^2Q^2\Sigma_{\rm g}^2}{2\phi_{\rm fb}\epsilon x\Omega^2} ,
\end{equation}
where $x=\rho_{\rm g}/\rho_{\rm ISM}$ as in equation~(\ref{eq:pdf}).

Equations~(\ref{eq:sfe}), (\ref{eq:tfb}), and (\ref{eq:teq}) can be solved for the SFE $\epsilon$ and the feedback time $t_{\rm fb}$. The solution for $t_{\rm fb}$ reads
\begin{equation}
\label{eq:tfbfull}
t_{\rm fb}=\frac{t_{\rm sn}}{2}\left(1+\sqrt{1+\frac{2\pi^2G^2t_{\rm ff}Q^2\Sigma_{\rm g}^2}{\phi_{\rm fb}{\rm sSFR}_{\rm ff}t_{\rm sn}^2\Omega^2x}}\right) ,
\end{equation}
and the SFE is given by
\begin{equation}
\label{eq:sfefb}
\epsilon_{\rm fb}= \frac{{\rm sSFR}_{\rm ff}}{t_{\rm ff}}\frac{t_{\rm sn}}{2}\left(1+\sqrt{1+\frac{2\pi^2G^2t_{\rm ff}Q^2\Sigma_{\rm g}^2}{\phi_{\rm fb}{\rm sSFR}_{\rm ff}t_{\rm sn}^2\Omega^2x}}\right) ,
\end{equation}
where the subscript `fb' indicates that this is the SFE after star formation is halted by feedback. For a solar neighbourhood-like parameter set, an overdensity of $x=10^1$, and $t_{\rm sn}=3$~Myr, the feedback time-scale is typically $t_{\rm fb}\sim5$~Myr (with $\epsilon_{\rm fb}\sim0.01$), and it quickly approaches $t_{\rm fb}=t_{\rm sn}$ for $x\geq 10^2$.

In practice, star formation need not be halted by feedback as the region may either be observed when the star formation process is still ongoing, or the density may have been so high that all of the available gas was consumed before $t_{\rm fb}$. If star formation is still ongoing, the SFE at time $t$ follows from a slight modification of equation~(\ref{eq:sfe}) as
\begin{equation}
\label{eq:sfeinc}
\epsilon_{\rm inc}=\frac{{\rm sSFR}_{\rm ff}}{t_{\rm ff}}t ,
\end{equation}
where the subscript `inc' indicates that the conversion of gas to stars is still incomplete. Of course, $\epsilon_{\rm inc}$ depends on the moment of evaluating the CFE, and a standard value of $t=10$~Myr is assumed here. However, by varying $t$ one can obtain the time-evolution of the CFE, which is shown in \S\ref{sec:tview}. If star formation proceeds at such a high density that it is optimally efficient before the onset of feedback, the SFE equals the core star formation efficiency $\epsilon_{\rm core}$:
\begin{equation}
\label{eq:sfemax}
\epsilon_{\rm max}=\epsilon_{\rm core} ,
\end{equation}
where the subscript `max' indicates that this is the maximum SFE. The value of $\epsilon_{\rm core}$ follows from the mass fraction of protostellar cores that ends up in the actual stars after accounting for the mass loss in outflows. According to \citet{matzner00}, $\epsilon_{\rm core}=0.25$--$0.75$, and a constant value of $\epsilon_{\rm core}=0.5$ is therefore adopted in this paper. Finally, this implies that the actual SFE is given by
\begin{equation}
\label{eq:sfefinal}
\epsilon=\min{\{\epsilon_{\rm max},\epsilon_{\rm fb},\epsilon_{\rm inc}\}} .
\end{equation}

The above treatment of the feedback-induced truncation of star formation assumes that the specific star formation rate per free-fall time sSFR$_{\rm ff}$ is constant until star formation is halted. This might not actually hold once the first supernovae have started to inject energy into the ISM, as the heating might either prevent or stimulate some regions from undergoing collapse \citep{elmegreen77,dale05,wunsch08}. Due to the ambiguous effects of feedback, it is not clear {\it a priori} whether a gradual truncation of star formation would be appropriate. A simple correction is explored in Appendix~\ref{sec:appfb}, which shows that the change of the resulting CFE is very marginal either way. This justifies the use of a constant sSFR$_{\rm ff}$.

Another word of caution is in order regarding the possible variation of the onset and strength of feedback, covered in the constants $t_{\rm sn}$ and $\phi_{\rm fb}$. These may depend weakly on metallicity (see the discussion in Appendix~\ref{sec:appfb}), but they are affected much more by the stochasticity of star formation at the high-mass end of the mass range. The sampling of the masses of the few most massive stars determines whether a continuous (and in that sense statistical) approach as in equation~(\ref{eq:sfefb}) is suitable. It should therefore be kept in mind that the treatment of feedback is idealised in the sense that it may hold for an ensemble of star-forming GMCs, but not necessarily for individual examples.

\subsection{The naturally bound part of star formation} \label{sec:fbound}
It has long been known that the CFE depends on the SFE on the local scale of young stellar clusters. The less efficient star formation is locally, the smaller the part of the stellar structure that remains bound after the gas has been expelled \citep{tutukov78,hills80,adams00,geyer01,boily03,boily03b,goodwin06,baumgardt07,parmentier08,smith11}. However, the details of the relation between the local CFE $\gamma$ and the SFE $\epsilon$ have so far always been determined for a static background potential\footnote{Except for a normalization of the gas potential that decreases with time when the gas is expelled.} and/or some degree of equilibrium between the gas and stars. It is an important property of star-forming environments in nature that the gas and stars can evolve independently of each other. The decoupled kinematics of both components \citep[see e.g.][]{offner09} and the accretion of the gas in the direct vicinity of the stars imply that stellar structure can evolve to a gas-poor state while remaining embedded in an evacuated cocoon of gas \citep{kruijssen12,girichidis12}. As such, the assumption of dynamical equilibrium between gas and stars does not hold. For the purpose of this paper it is therefore not possible to follow the `classical' equilibrium-type results to obtain $\gamma(\epsilon)$. Instead, numerical simulations of turbulent fragmentation can be used to establish what the fraction of stars is that is gravitationally bound on some scale, for a given gas-to-stellar mass ratio or SFE.

In Appendix~\ref{sec:appbound}, the analysis of the \citet{bonnell08} simulations by \citet{maschberger10} and \citet{kruijssen12} is used to address the local CFE as a function of the local SFE in a hierarchical star-forming region. Since the simulation does not include feedback, the effect of gas expulsion is accounted for by ignoring the gravitational potential of the gas when calculating the boundedness of the stellar structure. This reflects the most violent form of feedback, as the gas is removed instantaneously. Obviously this is an idealized approach. A more realistic treatment of feedback \citep[see e.g.][]{wang10,krumholz12} may work in two ways. Either it could accelerate the evolution of stellar subclusters to a gas-poor state, in which case the approach is still valid, or it could slow down star formation to such a degree that the stars and gas retain some degree of dynamical equilibrium. It is shown in Appendix~\ref{sec:appbound} that the extreme case of adopting the results from $N$-body simulations that assume complete dynamical equilibrium between gas and stars does not substantially change the resulting CFE. Another implication of the absence of feedback is that star formation is quite rapid in the \citet{bonnell08} simulation, with ${\rm sSFR}_{\rm ff}\sim0.3$. As a result, the collapse into stars proceeds so rapidly that turbulence may not have reached statistical equilibrium yet, which {could drive the density PDF away from the characteristic log-normal shape if it is generated by the turbulence. However, it has been shown that turbulence is not required to produce a log-normal density PDF \citep{tassis10} and indeed we find that the gas in the central 5~pc of the simulation is roughly consistent with the log-normal of equation~(\ref{eq:pdf}) for ${\cal M}=3$--$10$.}

The analysis in Appendix~\ref{sec:appbound} shows that in the boundedness of stellar structure in the \citet{bonnell08} simulation can be described with 30\% accuracy by the simple relation $\gamma(\epsilon)=\epsilon$. This does not account for protostellar outflows, i.e.~the maximum SFE in the analysis is 100\%, while in the model of this section, a maximum SFE equal to $\epsilon_{\rm core}=0.5$ is assumed. The question thus rises whether the local CFE could still be optimal at a maximum SFE that is smaller than unity due to outflows. Recent observational and numerical results on protostellar outflows show that they should be capable of driving the turbulence on scales of $\sim 1$~pc, but not on the scales of entire GMCs \citep[e.g.][]{arce10,wang12,hansen12,buckle12}. As such, their influence may not reach very far, contrary to earlier theoretical expectations \citep[e.g.][]{matzner00}. The typical protostellar outflow velocities of $\sim 0.5$~km~s$^{-1}$ in the numerical work of \citet{nakamura07} are consistent with this picture, and for protoclusters with radii of 1~pc they imply crossing times of 2~Myr. Since this is longer than the typical free-fall time of such protoclusters \citep[cf.][]{evans09}, the stellar structure can respond adiabatically to any mass loss in protostellar outflows. The mere existence of bound, young stellar clusters without a massive halo of escaping stars \citep[e.g.][]{rochau10,cottaar12} evidences that in at least some cases protostellar outflows alone are not sufficient to unbind stellar clusters. The local CFE should thus not be strongly affected by the outflows. This leads to the following adopted relation for the local CFE:
\begin{equation}
\label{eq:cfelocal}
\gamma=\epsilon/\epsilon_{\rm core} ,
\end{equation}
which increases linearly from 0 to 1 as the SFE increases from 0 to $\epsilon_{\rm core}$.

The naturally bound part of star formation $f_{\rm bound}$ from equation~(\ref{eq:cfe}) can now be obtained by integration of the overdensity PDF of the ISM:
\begin{equation}
\label{eq:fbound}
f_{\rm bound}=\frac{\int_{-\infty}^\infty \gamma(x)\epsilon(x) x({\rm d} p/{\rm d} x){\rm d} x}{\int_{-\infty}^\infty \epsilon(x) x({\rm d} p/{\rm d} x){\rm d} x} ,
\end{equation}
where the numerator represents the part of star formation that results in bound structure, and the denominator denotes the total amount of star formation. The quantities that set $f_{\rm bound}$ are the gas surface density $\Sigma_{\rm g}$, the Toomre $Q$ parameter, and the angular velocity $\Omega$.\footnote{As mentioned previously, if the galaxy rotation curve is not flat, $\Omega$ should be replaced by equation~(\ref{eq:kappa}) in all equations.}

\subsection{The cruel cradle effect} \label{sec:cce}
Star-forming regions are characterized by their enhanced densities relative to the mean ambient density in a galaxy disc. Over the course of a star formation time-scale, it is possible that these regions encounter other density peaks or GMCs with which they interact gravitationally. Such interactions result in tidal shocks, during which the perturbation injects energy into the star-forming region. This may prevent collapse on a global scale, but because the susceptibility of an object to tidal shocks is set by its density, local condensations into protostellar cores are not inhibited. Tidal shocks thus decrease the CFE, but not the SFE.

\subsubsection{The Spitzer theory of tidal shocks}
The disruption time-scale of a hierarchically structured star-forming region due to tidal shocks $t_{\rm cce}$ can be derived from the energy considerations that also hold for stellar clusters \citep[see e.g.][]{spitzer87,kundic95,gnedin99c,gieles06,prieto08,kruijssen11}. The only modifications lie in the adopted proportionality constants, which account for the physical and structural differences between star clusters and hierarchical star-forming regions. The time-scale is written as
\begin{equation}
\label{eq:tsh}
t_{\rm cce}=\left|\frac{E}{f}\left(\frac{{\rm d} E}{{\rm d} t}\right)^{-1}\right| ,
\end{equation}
where $f$ is a constant that accounts for the fraction of the energy gain that is used to unbind the star-forming region, $E$ is the internal energy per unit mass of the region, and the time derivative indicates the energy gain per unit mass due to tidal perturbations. It is assumed that the relevant regions experience more than one encounter before $t=10$~Myr, which is verified in \S\ref{sec:adiab} below.

In order to calculate the internal energy of a star-forming region, it is convenient to assume a certain radial density profile. While such spherical symmetry in principle violates the philosophy of this work to treat the ISM as a hierarchy, it is a reasonable approach since stars and stellar clusters are born in local density peaks. This assumption is particularly appropriate in the context of tidal perturbations: the details of the density profile of the perturbed region translate into second-order changes in the proportionality constants \citep[e.g.][]{spitzer87}. For mathematical simplicity, the regions are thus considered to follow a \citet{plummer11} potential:
\begin{equation}
\label{eq:plummer}
\Phi=-\frac{GM}{R_0}\left(1+\frac{R^2}{R_0^2}\right)^{-1/2} ,
\end{equation}
where $\Phi$ is the potential energy, $M$ the mass of the region, $R_0$ the characteristic (Plummer) radius, and $R$ the radius. For the above potential, the internal energy per unit mass is given by
\begin{equation}
\label{eq:e}
E=-\frac{3\pi}{64}\frac{GM}{R_0} .
\end{equation}

The rate of energy change per unit mass ${\rm d} E/{\rm d} t$ of the star-forming region can be derived according to the formalism for the perturbation of stellar clusters of \citet{spitzer58,spitzer87}, but applied to the gas-rich, star-forming region as a whole. As laid out by \citet{binney87}, ${\rm d} E/{\rm d} t$ is given by the phase-space integration of the product of the encounter rate and the energy injected by each encounter. Generalizing their result as in \citet{gieles06} and assuming a Maxwellian velocity distribution, one obtains
\begin{eqnarray}
\label{eq:dedt}
\nonumber\frac{{\rm d} E}{{\rm d} t}&=&\frac{4\pi^{3/2}gG^2\phi_{\rm sh}}{3\sigma_{\rm g}}\Sigma_{\rm GMC}\rho_{\rm ISM}\overline{R^2}\phi_{\rm ad}\\ 
&=&\frac{4\sqrt{2\pi}gG\phi_{\rm sh}\Omega}{3\Sigma_{\rm g}Q}\Sigma_{\rm GMC}\rho_{\rm ISM}\overline{R^2}\phi_{\rm ad} ,
\end{eqnarray}
where $g\approx 1.5$ is a correction factor to account for the extended nature of the perturbing clouds (see below), $\phi_{\rm sh}\approx 2.8$ accounts for the higher-order energy gain\footnote{Calculated for \citet{plummer11} parameters using equation~(23) of \citet{kruijssen11}.} \citep{kundic95,gnedin97,prieto08,kruijssen11}, $\Sigma_{\rm GMC}$ is the surface density of GMCs, and equation~(\ref{eq:q}) was used in the second equality. The factor $\phi_{\rm ad}$ is the \citet{spitzer87} adiabatic correction, which accounts for the dampening of the energy injection by the adiabatic expansion of the region. The expression for $\phi_{\rm ad}$ is derived below in \S\ref{sec:adiab}.  For the Milky Way and galaxies in the Local Group, observations show $\Sigma_{\rm GMC}^{\rm LG}\sim100~\msun~{\rm pc}^{-2}$ \citep{solomon87,bolatto08,heyer09}, but pressure equilibrium implies that for galaxies with $\Sigma_{\rm g}>100~\msun~{\rm pc}^{-2}$ the GMC surface density is set by the disc surface density $\Sigma_{\rm GMC}\sim\Sigma_{\rm g}$ \citep[also see][]{hopkins12}. We therefore adopt
\begin{equation}
\label{eq:surfgmc}
\Sigma_{\rm GMC}=\max{\{\Sigma_{\rm GMC}^{\rm LG},\Sigma_{\rm g}\}} .
\end{equation}

Contrary to the distant encounters to which star clusters are often subjected, in the case of star-forming regions it is crucial to account for the spatially extended nature of the perturber. There are several corrections for tidal shocking by extended mass distributions available in the literature \citep{aguilar85,binney87,gnedin99c,gieles06}. Here, the approach of \citet{gieles06} is adopted for mathematical simplicity and consistency -- both in terms of the general derivation and the use of Plummer profiles. Their analysis shows that if the perturber is about an order of magnitude more massive than the perturbed object, their extended nature can be accounted for by adopting $g\approx1.5$ in equation~(\ref{eq:dedt}). Such a mass ratio is plausible because the GMC mass spectrum covers about two orders of magnitude, distributed according to a power law with an index of about $-1.7$ \citep[]{solomon87,elmegreen96,williams97}. This indicates that most of the tidal disruption will typically come from those structures that are more massive than the region itself, but likely not by more than an order of magnitude.

In $N$-body simulations of the disruption of star clusters by tidal shocks, a fraction $f=0.25$ of the injected energy is used to unbind the cluster \citep{gieles06}. This low efficiency is due to the excess energy that is carried away by escaping stars. However, in a dissipative medium such as the ISM the energy is not removed by a small number of high-velocity escapers, but instead it can be radiated away by the dissipation of turbulent energy.\footnote{This assumes that the kinematics of the star-forming region are dominated by the gas, i.e.~the SFE is low. As will be shown in Figure~\ref{fig:xpdf} below, at the ages of interest this is indeed the case for those overdensities where the cruel cradle effect is important.} The influence of turbulent energy decay on the value of $f$ is addressed in Appendix~\ref{sec:appcce} for a large set of Monte-Carlo experiments, which are used to model the stochasticity of the strength and time-separation of tidal perturbations. It is found that for those regions where $t_{\rm cce}\leq t$, the fraction of the tidally injected energy that is used to unbind a region is about $f\sim0.7$, independently of the environmental conditions. This relatively large fraction arises because regions are only disrupted if they happen to encounter a rapid sequence of tidal perturbations, thereby not allowing the energy to be dissipated. Hence, a large fraction of the energy can be used to overcome the gravitational potential, and those regions that do not survive until time $t$ are thus characterized by efficient disruption. It is verified in Appendix~\ref{sec:appcce} that this holds for the complete parameter range that is relevant for the presented theory (see \S\ref{sec:cfe} and Table~\ref{tab:vars} below).

\subsubsection{The adiabatic correction} \label{sec:adiab}
In the harmonic approximation \citep[cf.][]{spitzer87}, the adiabatic correction of equation~(\ref{eq:dedt}) is given by $\phi_{\rm ad}=\exp{(-2\phi_t)}$, in which $\phi_t\equiv t_{\rm enc}/t_{\rm diss}$, with $t_{\rm enc}$ the typical time interval between encounters and $t_{\rm diss}$ the time-scale for the dissipation of turbulent energy. This assumes that the duration of the tidal shock is of the same order as the time interval between subsequent perturbations, as should be expected due to the continuous nature of the ISM. The dissipation time-scale is approximately given by \citep{mckee07}
\begin{equation}
\label{eq:tdiss}
t_{\rm diss}=\frac{R}{\sigma_{\rm int}} ,
\end{equation}
which in combination with equations~(\ref{eq:alpha}) and~(\ref{eq:tff}) becomes
\begin{equation}
\label{eq:tdiss2}
t_{\rm diss}=\sqrt{\frac{40}{\pi^2\alpha_{\rm vir}}}t_{\rm ff}\approx 1.5\left(\frac{\alpha_{\rm vir}}{1.3}\right)^{-1/2}t_{\rm ff} ,
\end{equation}
implying that $t_{\rm diss}$ falls in the range $0.5t_{\rm ff}$--$3.8t_{\rm ff}$ for the values of $\alpha_{\rm vir}$ found in observations and numerical work. The encounter time-scale is defined from simple kinetic theory as \citep{binney87}:
\begin{equation}
\label{eq:tenc}
t_{\rm enc}=\left(4\sqrt{\pi}n_{\rm GMC}\sigma_{\rm g}b_{\rm max}^2\right)^{-1} ,
\end{equation}
where $n_{\rm GMC}$ is the number density of GMCs and $b_{\rm max}\equiv(48/\pi n_{\rm GMC})^{1/3}$ is the mean cloud separation. The GMC number density can be specified by writing $n_{\rm GMC}=\rho_{\rm ISM}/M_{\rm GMC}$, with $M_{\rm GMC}$ the typical GMC mass. If one assumes that the characteristic GMC mass is set by the Jeans mass, it can be written as \citep{krumholz05}:
\begin{equation}
\label{eq:mgmc}
M_{\rm GMC}=\frac{\pi^4 G^2\Sigma_{\rm g}^3Q^4}{4\Omega^4} .
\end{equation}
Substitution of equations~(\ref{eq:rhoism}), (\ref{eq:q}) and~(\ref{eq:mgmc}) into equation~(\ref{eq:tenc}) then yields
\begin{equation}
\label{eq:tenc2}
t_{\rm enc}=\frac{\sqrt{2}\pi^{1/3}\phi_P^{1/6}}{768^{2/3}\sqrt{G\rho_{\rm ISM}}}\approx 0.025\phi_P^{1/6}(G\rho_{\rm ISM})^{-1/2} .
\end{equation}
Since at the solar galactocentric radius $\Omega_0\approx 0.026$~Myr$^{-1}$ (assuming a circular velocity of 220~km~s$^{-1}$), equations~(\ref{eq:rhoism}) and~(\ref{eq:tenc2}) imply that $t_{\rm enc}\sim2.5$~Myr in the solar neighbourhood, which is a relatively low-density environment. The number of encounters increases with angular frequency, so at the radii where the bulk of the Galactic star formation occurs \citep[between galactocentric radii of~2 and~7~kpc, see e.g.][]{mckee97} always $t_{\rm enc}\ll t$. This justifies the earlier assumption that a star-forming region typically suffers multiple encounters before $t=10$~Myr. Finally, the combination of the final expressions for $t_{\rm diss}$ and $t_{\rm enc}$ now provides
\begin{equation}
\label{eq:phit}
\phi_t\equiv\frac{t_{\rm enc}}{t_{\rm diss}}\approx 3.1\left(\frac{\alpha_{\rm vir}}{1.3}\right)^{1/2}\left(\frac{x}{10^4}\right)^{1/2} .
\end{equation}
This expression for $\phi_t$ as a function of $\alpha_{\rm vir}$ and $x$ now allows the specification of $\phi_{\rm ad}$, which becomes
\begin{equation}
\label{eq:phiad}
\phi_{\rm ad}={\rm e}^{-2\phi_t}\approx\exp{\left[-6.2\left(\frac{\alpha_{\rm vir}}{1.3}\right)^{1/2}\left(\frac{x}{10^4}\right)^{1/2}\right]} .
\end{equation}
For gaseous regions, this \citet{spitzer87} formulation of the adiabatic correction is preferred over the \citet{weinberg94b} correction, of which the power law tail at large $\phi_t$ is caused by stellar oscillations that do not necessarily apply to a turbulent medium. Since the presented theory approximates gas-rich, star-forming aggregates locally with \citet{plummer11} potentials, which are parabolic in the centre, the harmonic approach is more appropriate.

\subsubsection{A critical overdensity for surviving the cruel cradle effect}
Substitution of equations~(\ref{eq:e}) and (\ref{eq:dedt}) into equation~(\ref{eq:tsh}) now gives the disruption time-scale of star-forming regions due to the cruel cradle effect:
\begin{eqnarray}
\label{eq:tcce}
\nonumber t_{\rm cce}&=&\frac{9\sqrt{\pi}}{2^{17/2}fg\phi_{\rm sh}\phi_{\rm ad}}\frac{Q\Sigma_{\rm g}}{\Omega\Sigma_{\rm GMC}}\left(\frac{R_{\rm h}}{R_0}\right)\left(\frac{R_{\rm h}^2}{\overline{R^2}}\right)x\\
&\approx&\frac{\sqrt{\pi}}{124 fg\phi_{\rm sh}\phi_{\rm ad}}\frac{Q\Sigma_{\rm g}}{\Omega\Sigma_{\rm GMC}}x ,
\end{eqnarray}
where $R_{\rm h}$ is the half-mass radius of the star-forming region, and the second equality uses $R_{\rm h}=1.3R_0$ and $R_{\rm h}^2/\overline{R^2}\approx 0.25$ for a Plummer profile \citep{gieles06}. By writing $t_{\rm cce}=t$, it is possible to derive a threshold overdensity $x_{\rm cce}$ above which star-forming regions will survive the cruel cradle effect at least until time $t$, which is given by:
\begin{equation}
\label{eq:xcce}
x_{\rm cce}=\frac{124fg\phi_{\rm sh}\Omega\Sigma_{\rm GMC}t}{\sqrt{\pi}Q\Sigma_{\rm g}}\phi_{\rm ad}(x_{\rm cce}) ,
\end{equation}
where $\phi_{\rm ad}$ is a monotonically decreasing function of $x_{\rm cce}$, which indicates that equation~(\ref{eq:xcce}) has a unique solution. Because the relation is implicit, $x_{\rm cce}$ has to be obtained by solving equation~(\ref{eq:xcce}) numerically. Regions with overdensities $x<x_{\rm cce}$ cannot form bound stellar clusters because they are disrupted by tidal perturbations due to the primordial environment before time $t$.\footnote{{Note that this does not introduce a certain threshold density that can be used to define bound clusters observationally. Firstly because $x_{\rm cce}$ varies in space and time, but also because most regions that survive the cruel cradle effect (i.e. $x>x_{\rm cce}$) are often intrinsically unbound, i.e. $f_{\rm bound}<f_{\rm cce}'$ (see e.g. Fig.~\ref{fig:xpdf} and Table~\ref{tab:galaxies}).}} A typical value for solar neighbourhood-type parameters is $x_{\rm cce}\sim10^{2}$ (see the discussion of Figure~\ref{fig:xpdf} in \S\ref{sec:cfe} below). Also note that the proportionality $x_{\rm cce}\propto\Omega Q^{-1}\Sigma_{\rm g}^{-1}$ implies $x_{\rm cce}\propto{\cal M}^{-1}$. Because for $\Sigma_{\rm g}>\Sigma_{\rm GMC}^{\rm LG}$ we have $\Sigma_{\rm GMC}\sim\Sigma_{\rm g}$, high-surface density galaxies follow $x_{\rm cce}\propto\Omega Q^{-1}\propto\rho_{\rm ISM}^{1/2}$.

The influence of the cruel cradle effect on the CFE can now be quantified in two ways. Firstly, the fraction of the naturally bound part of star formation that survives the cruel cradle effect, i.e.~$f_{\rm cce}$ from equation~(\ref{eq:cfe}), is obtained by integration of the overdensity PDF of the ISM:
\begin{equation}
\label{eq:fcce}
f_{\rm cce}=\frac{\int_{x_{\rm cce}}^\infty \gamma(x)\epsilon(x) x({\rm d} p/{\rm d} x){\rm d} x}{\int_{-\infty}^\infty \gamma(x)\epsilon(x) x({\rm d} p/{\rm d} x){\rm d} x} ,
\end{equation}
where the numerator represents the bound structure that survives the cruel cradle effect, and the denominator denotes the part of star formation that results in bound structure. This quantity does not reflect the absolute contribution of the cruel cradle effect as equation~(\ref{eq:fbound}) does for the naturally bound part of star formation, because the parts of the density range that are dispersed by both mechanisms overlap. As a result, $f_{\rm cce}$ only gives the {\it additional} decrease of the CFE due to tidal perturbations. To directly compare the individual contributions to dispersed star formation of naturally unbound star formation or the cruel cradle effect, it is thus useful to define the fraction of {\it all} star formation that survives tidal perturbations by the dense primordial environment:
\begin{equation}
\label{eq:fcce2}
f_{\rm cce}'=\frac{\int_{x_{\rm cce}}^\infty \epsilon(x) x({\rm d} p/{\rm d} x){\rm d} x}{\int_{-\infty}^\infty \epsilon(x) x({\rm d} p/{\rm d} x){\rm d} x} ,
\end{equation}
in which the denominator now represents all star formation. This definition allows one to determine the CFE if all star formation were intrinsically bound, giving an absolute measure for the disruptive potential of the cruel cradle effect. As for $f_{\rm bound}$ in equation~(\ref{eq:fbound}), the quantities that set $f_{\rm cce}$ and $f_{\rm cce}'$ are the gas surface density $\Sigma_{\rm g}$, the Toomre $Q$ parameter, and the angular velocity $\Omega$. 

\subsection{The cluster formation efficiency} \label{sec:cfe}
\begin{figure}
\center\resizebox{\hsize}{!}{\includegraphics{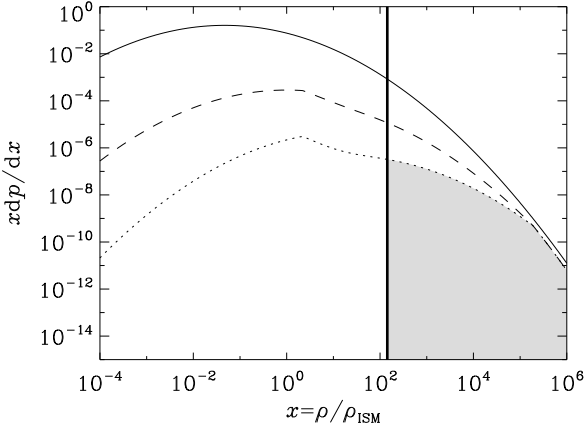}}\\
\caption[]{\label{fig:xpdf}
      The determination of the cluster formation efficiency (CFE). The solid curve shows the overdensity probability distribution function of the interstellar medium. Multiplication with the star formation efficiency gives the part that is converted into stars, which is represented by the dashed curve. The subset of stars that are formed in naturally bound stellar clusters is shown as a dotted line, of which the part that has an overdensity larger than a certain environmentally dependent, critical value (thick vertical line) survives tidal disruption by the dense star-forming giant molecular clouds. The CFE is given by the ratio of the grey-shaded region to the integral of the dashed curve. The figure shows the distributions for a parameter set that is characteristic for the Milky Way (see text).
                 }
\end{figure}
The final expression for the CFE is obtained by combination of equations~(\ref{eq:fbound}) and~(\ref{eq:fcce}), which provide the naturally bound fraction of star formation and the fraction thereof that survives the cruel cradle effect:
\begin{equation}
\label{eq:cfefinal}
\frac{\Gamma}{100\%}=f_{\rm bound}f_{\rm cce}=\frac{\int_{x_{\rm cce}}^\infty \gamma(x)\epsilon(x) x({\rm d} p/{\rm d} x){\rm d} x}{\int_{-\infty}^\infty \epsilon(x) x({\rm d} p/{\rm d} x){\rm d} x} .
\end{equation}
This integral is to be evaluated numerically, and is illustrated in Figure~\ref{fig:xpdf}. It shows the overdensity PDF of the ISM $x{\rm d} p/{\rm d} x$ before and after multiplication by $\epsilon(x)$ and an additional factor $\gamma(x)$, respectively, as well as the environmentally dependent threshold overdensity that ensures survival after the cruel cradle effect. The parameter set is chosen to reflect the properties of the solar neighbourhood, with $\Sigma_{\rm g}=12~\msun~{\rm pc}^{-2}$, $Q=1.5$, and $\Omega=0.026~{\rm Myr}^{-1}$ being substituted in equation~(\ref{eq:mach}) to obtain the Mach number and set the overdensity PDF. While higher Mach numbers decrease the value of the peak overdensity somewhat, they also lead to a much broader PDF. The curves that include the factors $\epsilon$ and $\gamma$ nicely illustrate the three regimes of star formation given by equation~(\ref{eq:sfefinal}). In this example, star formation is still ongoing at age $t$ for overdensities $x\la 10^{0}$, whereas for $x\ga 10^5$ it has been so efficient that an optimal SFE of $\epsilon_{\rm core}$ is reached. At intermediate overdensities, star formation is halted by feedback. Due to the cruel cradle effect, bound stellar clusters are only formed at high densities, i.e.~either in the regime where star formation is halted by feedback or has been optimally efficient. The CFE equals the ratio of the grey-shaded region in Figure~\ref{fig:xpdf} to the integral of the dashed curve, which represents the product $\epsilon(x)x{\rm d} p/{\rm d} x$ and reflects the total amount of star formation. The figure illustrates that the cruel cradle effect and the naturally bound part of star formation both enable a fairly similar part of the overdensity range to remain bound. As such, the cruel cradle effect accelerates mainly the dispersal of unbound star formation, whereas its effect on the bound part of star formation is smaller.

The presented theory of the cluster formation efficiency satisfies the five criteria listed in \S\ref{sec:phys}.
\begin{enumerate}
\item
The fraction of star formation that is locally bound is formulated in a scale-free and continuous manner, and is based on the hierarchical nature of the ISM and the star formation process.
\item
The global bound fraction of star formation is obtained from a local criterion that is integrated over the entire overdensity spectrum of the ISM.
\item
The formation of bound stellar clusters is considered as a continuous process, which allows the naturally bound fraction of star formation, the cruel cradle effect, and hence the cluster formation efficiency to be calculated at any time $t$.
\item
By addressing the CFE at a specific time $t$, the theory accounts for three possible outcomes of the star formation process: it can occur on such a short time-scale that the gas is consumed before the onset of feedback, it can be truncated by large-scale supernova feedback, or it can still be ongoing at the moment of observation. This order reflects a sequence of decreasing density.
\item
The effect of the tidal disruption of star-forming regions by the primordial environment is included and quantified at the time of observation $t$.
\end{enumerate}

\begin{table}
 \centering
  \caption{List of variables and their typical values.}\label{tab:vars}
  \begin{tabular}{@{}c c c c | c c@{}}
  \hline
  Variable & Minimum & Typical & Maximum & Quiescent & Starburst \\
  (1) & (2) & (3) & (4) & (5) & (6)\\
 \hline
$\Sigma_{\rm g,0}$ & $10^0$ & $10^1$ & $10^5$ & $10^{0.7}$ & $10^3$ \\
$Q$ & $0.5$ & $1.5$ & $6$ & $3$ & $0.5$ \\
$\Omega_0$ & $10^{-2.5}$ & $10^{-1.7}$ & $10^{0.5}$ & $10^{-1.9}$ & $10^{-0.7}$\\
\hline
\end{tabular}\\
$\Sigma_{\rm g,0}$ and $\Omega_0$ are in units of $\msun~{\rm pc}^{-2}$ and ${\rm Myr}^{-1}$, respectively.
\end{table}

\begin{table}
 \centering
  \caption{List of parameters.}\label{tab:param}
  \begin{tabular}{@{}l c@{}}
  \hline
  Parameter & Typical value \\
  (1) & (2) \\
 \hline
$\phi_P$ & 3 \\
$\alpha_{\rm vir}$ & 1.3 \\
$t_{\rm sn}$ & 3~Myr \\
$t$ & 10~Myr \\
$\phi_{\rm fb}$ & $0.16~{\rm cm}^2~{\rm s}^{-3}$ \\
$\epsilon_{\rm core}$ & 0.5 \\
$f$ & 0.7 \\
$g$ & 1.5 \\
$\phi_{\rm sh}$ & 2.8 \\
$\Sigma_{\rm GMC}^{\rm LG}$ & $100~\msun~{\rm pc}^{-2}$\\
\hline
\end{tabular}
\end{table}

\begin{figure*}
\center\resizebox{\hsize}{!}{\includegraphics{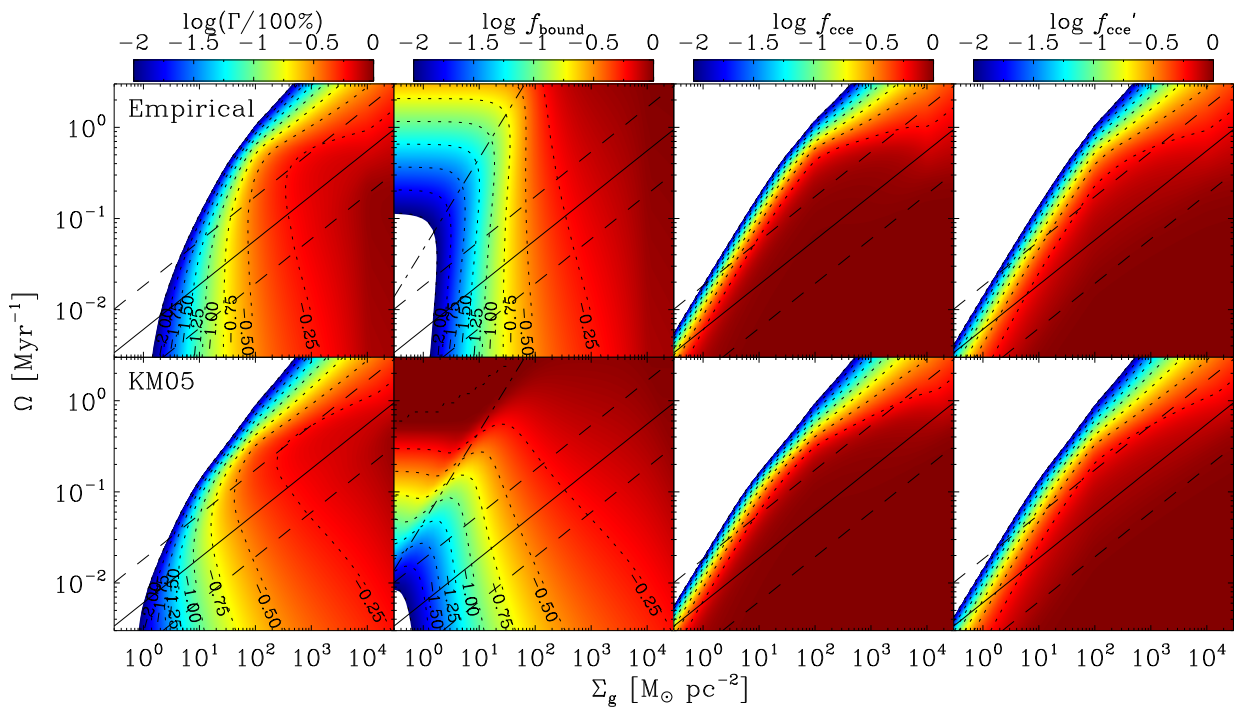}}
\caption[]{\label{fig:cfesflaw}
      Influence of the adopted star formation law on the cluster formation efficiency. From left to right, the panels show the cluster formation efficiency $\Gamma$, naturally bound fraction of star formation $f_{\rm bound}$, fraction of {\it bound} star formation surviving the cruel cradle effect $f_{\rm cce}$, and the fraction of {\it all} star formation surviving the cruel cradle effect $f_{\rm cce}'$ as a function of surface density and angular velocity. The displayed range is indicated in the legends at the top of each column, and the dotted contours denote fixed values as given by their labels. The solid lines mark the relation between $\Sigma_{\rm g}$ and $\Omega$ from equation~(\ref{eq:omega}) along which most nearby galaxies are located, with dashed lines representing a scatter of a factor of three. The dash-dotted lines in the second column separate the subsonic (top left) and supersonic regimes. {\it Top panels}: using the empirically motivated star formation law of equation~(\ref{eq:ssfrff_e02}) and \citet{elmegreen02}. {\it Bottom panels}: using the theoretical star formation law of equation~(\ref{eq:ssfrff_km05}) and \citet{krumholz05}. All other parameters are as in Table~\ref{tab:param}, with $Q=1.5$.
                 }
\end{figure*}
The model depends on two or three main variables: (1) the gas surface density $\Sigma_{\rm g}$, (2) the Toomre $Q$ parameter, (3) the angular velocity $\Omega$ or epicyclic frequency $\kappa$,\footnote{If the CFE is determined globally for a low-redshift galaxy, $\Omega$ and $\Sigma_{\rm g}$ are roughly related according to equation~(\ref{eq:omega}).} which are summarized in Table~\ref{tab:vars} together with their ranges. The `typical' set of variable values reminisces the conditions in the solar neighbourhood and is broadly characteristic of disc galaxies. The `quiescent' and `starburst' variable sets\footnote{{Throughout the paper, `starburst' is used to refer to galaxies with short ($t_{\rm depl}\la300$~Myr) gas depletion time-scales. The `starburst' parameter set of Table~\ref{tab:vars} is chosen to be representative of high-density starburst galaxies, but does not rely on any criterion for gas depletion.}} are chosen to reflect the typical extremes of star-forming galaxies, whereas the absolute extremes of the values themselves are listed in the `minimum' and `maximum' columns (see e.g.~\citealt{kennicutt98b,bigiel08} for discussions of the observed parameter range). The small set of other relevant parameters for which constant, fiducial values can be adopted are listed in Table~\ref{tab:param}. These tables will be referred back to in the remainder of the paper when discussing examples of cluster-forming galaxies.

\section{The variation of the cluster formation efficiency} \label{sec:param}
Before making specific predictions for the CFE in different galaxies and comparing them to observed values, it is relevant to explore the parameter space and reach an understanding of CFE variations across a range of galactic environments. In this section, we discuss the variation of the CFE as a function of gas surface density and angular velocity, while each time varying one of the other model components. In order of appearance, these are the star formation law, the \citet{toomre64} $Q$ parameter, the time-evolution of the CFE, and some miscellaneous parameters that have a weak influence on the CFE. The main results of the analysis in this section are summarized in \S\ref{sec:summparam}.

\subsection{General behavior and the influence of the star formation law} \label{sec:cfesflaw}
We show the CFE as a function of the gas surface density $\Sigma_{\rm g}$ and the angular velocity $\Omega$ in Figure~\ref{fig:cfesflaw} for the two different star formation laws \citep{elmegreen02,krumholz05} that are adopted in this paper. Because the behavior of the CFE is comparable in both cases, we first focus on common features between both star formation laws, before pointing out the more subtle differences. The panels in Figure~\ref{fig:cfesflaw} do not only show the variation of the CFE, but also of the naturally bound fraction of star formation $f_{\rm bound}$, the fraction of {\it bound} star formation that survives the cruel cradle effect $f_{\rm cce}$, and the fraction of {\it all} star formation that survives the cruel cradle effect $f_{\rm cce}'$. We reiterate that the CFE is given by the product of $f_{\rm bound}$ and $f_{\rm cce}$, which of course also applies to the panels in Figure~\ref{fig:cfesflaw}.

The naturally bound fraction of star formation exhibits a pronounced increase with the gas surface density. This can be understood in terms of the Mach number $\cal{M}$ as defined in equation~(\ref{eq:mach}). We mentioned in the discussion of Figure~\ref{fig:xpdf} that the Mach number sets the shape of the overdensity PDF of the ISM. Because ${\cal M}\propto\Sigma_{\rm g}$, the width of the overdensity PDF increases with the surface density, which enables a larger fraction of star formation to reside in the high-density range that leads to high local formation efficiencies of stars and clusters. This effect is large enough to overcome the decrease of the peak overdensity with increasing ${\cal M}$. The value of $f_{\rm bound}$ is remarkably insensitive to the angular velocity for most of the parameter space, because in an equilibrium disc the decrease of the Mach number with $\Omega$ and the corresponding narrowing of the overdensity PDF is compensated by an increase of the average mid-plane density. This implies that while the overdensities may be smaller, the absolute density remains similar, and $f_{\rm bound}$ only weakly changes with $\Omega$. The only exception occurs in the top left of the parameter space, at low surface densities and high angular velocities. In this region, $f_{\rm bound}$ is insensitive to the surface density, but instead becomes a function of the angular frequency. The reason is that above the line $\Omega_0=4.6\Sigma_{\rm g,2}$ the gas becomes subsonic and the overdensity PDF freezes. This removes the dependence on the Mach number, and implies that $f_{\rm bound}$ only depends on the angular frequency through the mid-plane density of equation~(\ref{eq:rhoism}).

The fraction of naturally bound (or total) star formation that survives the cruel cradle effect $f_{\rm cce}$ ($f_{\rm cce}'$) almost exclusively depends on the Mach number. This occurs because the critical overdensity that is required to survive external perturbations scales as $x_{\rm cce}\propto\sigma_{\rm g}\propto{\cal M}^{-1}$ due to the effect of gravitational focusing, while the PDF of the overdensity $x$ also is a function of the Mach number only. Both dependences result in larger survival fractions for higher Mach numbers, and as a result the contours of constant $f_{\rm cce}$ in Figure~\ref{fig:cfesflaw} mostly follow lines of constant Mach number, and mark a rather sudden transition between strong tidal disruption and complete survival. Only at high surface densities a `knee' appears in the contours, of which the location corresponds to the point where $\Sigma_{\rm g}=\Sigma_{\rm GMC}^{\rm LG}=100~\msun~{\rm pc}^{-2}$. Towards higher densities, the tidal perturbations grow because $\Sigma_{\rm GMC}=\Sigma_{\rm g}$, and hence the dependence of $x_{\rm cce}$ on the surface density vanishes, implying $x_{\rm cce}\propto\Omega Q^{-1}\propto\rho_{\rm ISM}^{1/2}$. The upturn of the contours at high surface densities and angular velocities marks the point where the critical overdensity for survival $x_{\rm cce}$ coincides with the overdensity at which the free-fall time is so short that the SFE is optimal and all available gas is turned into stars ($\epsilon=\epsilon_{\rm core}$). In this regime, the tidal perturbations start to unbind structure that is entirely gravitationally bound. Finally, the difference between $f_{\rm cce}$ and $f_{\rm cce}'$ is mostly a horizontal stretch in which a smaller fraction of all star formation survives the cruel cradle effect than of the naturally bound part -- as anticipated in the discussion of Figure~\ref{fig:xpdf}.

The total cluster formation efficiency $\Gamma$ is given by the product of the second ($f_{\rm bound}$) and third ($f_{\rm cce}$) panels in each row of Figure~\ref{fig:cfesflaw}. The cruel cradle effect provides a clear threshold for cluster formation at the low-density end of the parameter space, but when focusing on the region where most observed galaxies lie (solid and dashed lines), the CFE is mostly set by the naturally bound fraction of star formation. Only at very low and very high gas surface densities (cf. Table~\ref{tab:vars}), the cruel cradle effect notably suppresses the CFE. The combination of all effects leads to a CFE that increases with the gas surface density, from a few per cent in quiescent, low-density galaxies to a maximum of about 60\% in high-density environments. It decreases again for the most extreme densities ($\Sigma_{\rm g}\ga10^{3}~\msun~{\rm pc}^{-2}$) due to the cruel cradle effect.

The differences between the two star formation laws are minor. The \citet{krumholz05} prescription leads to CFEs that can be up to 0.2~dex higher than for the empirical star formation law, but this falls within the error margins \citep{kennicutt98b}. The main other difference is a stronger dependence on the angular velocity for the \citet{krumholz05} star formation prescription than for the empirically motivated one. This is caused by an explicit dependence of their specific star formation rate per free-fall time on the Mach number, which causes the rate of star formation to increase with decreasing Mach numbers. That aside, the overall variation of the CFE for both star formation laws is very similar. In the following, we therefore assume equation~(\ref{eq:ssfrff_e02}) unless it is clarifying to also show the results for the \citet{krumholz05} prescription of star formation.

\subsection{The influence of the Toomre $Q$ parameter} \label{sec:cfeQ}
\begin{figure*}
\center\resizebox{\hsize}{!}{\includegraphics{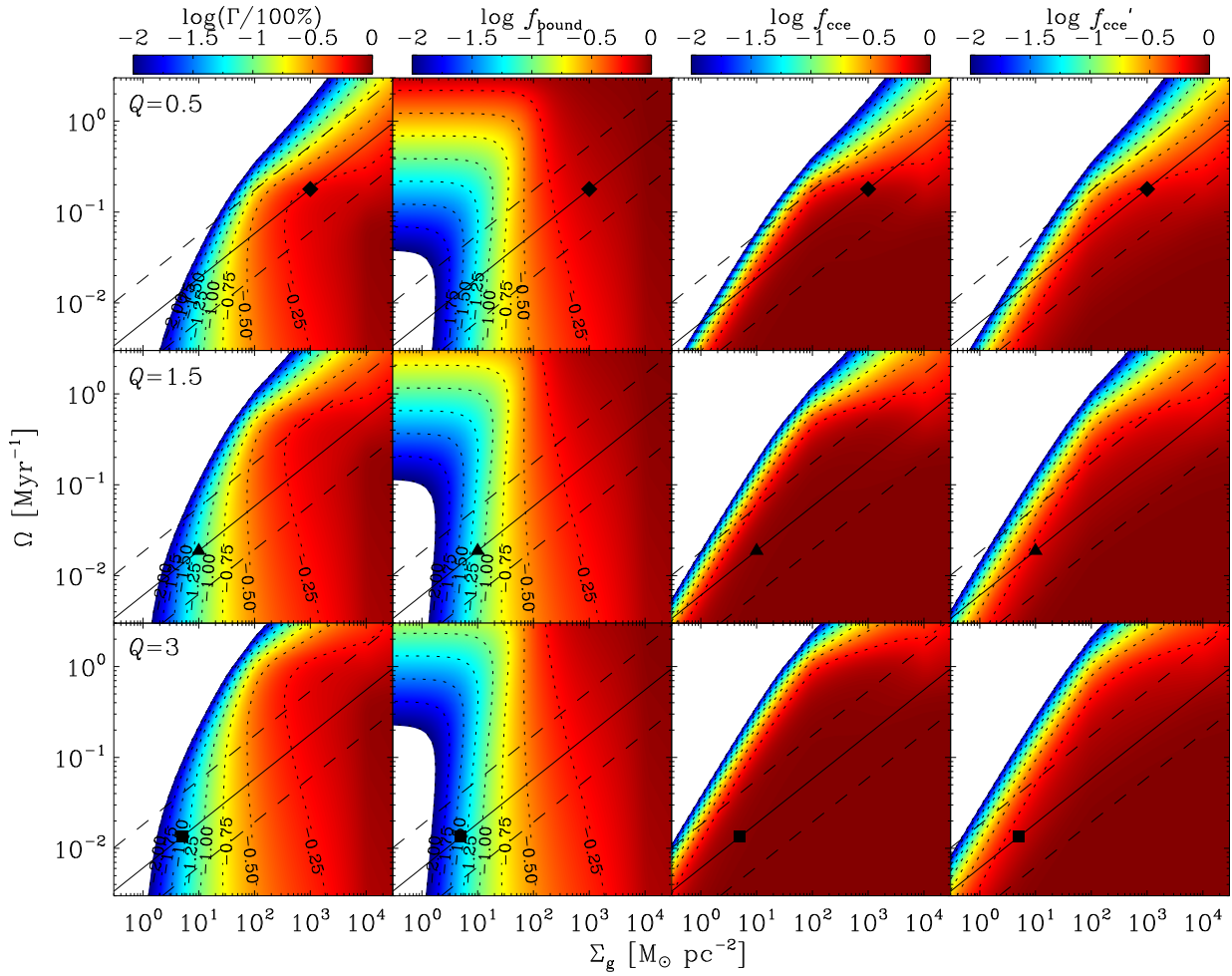}}
\caption[]{\label{fig:cfeQ}
      Influence of the \citet{toomre64} $Q$ parameter on the cluster formation efficiency. Panels and parameters are the same as in Figure~\ref{fig:cfesflaw}, this time adopting the empirical star formation law of equation~(\ref{eq:ssfrff_e02}) and \citet{elmegreen02}. From top to bottom, $Q$ varies as indicated in the top left corner of each row. The diamonds, triangles, and squares indicate the `starburst', `typical', and `quiescent' variable sets from Table~\ref{tab:vars}, respectively.
                 }
\end{figure*}
The variation of the CFE as a function of $\Sigma_{\rm g}$ and $\Omega$ is again shown in Figure~\ref{fig:cfeQ}, this time addressing the influence of the \citet{toomre64} $Q$ parameter, which indicates the stability of the gas disc. When $\Sigma_{\rm g}$ and $\Omega$ are specified, this parameter reflects the velocity dispersion of the gas as $Q\propto\sigma_{\rm g}$. As for Figure~\ref{fig:cfesflaw}, the physics of the CFE are best understood by separately considering the naturally bound fraction of star formation and the fraction thereof that survives the cruel cradle effect. Again, the Mach number plays a central role.

The Mach number scales as ${\cal M}\propto\sigma_{\rm g}\propto Q$ at fixed $\Sigma_{\rm g}$ and $\Omega$, which leads to a slight decrease of $f_{\rm bound}$ with increasing Toomre $Q$. This is easily understood in physical terms: at fixed $\Sigma_{\rm g}$ and $\Omega$, a more stable gas disc implies a higher velocity dispersion, which lowers the mid-plane density as $\rho_{\rm ISM}\propto\sigma_{\rm g}^{-2}$ and causes the fraction of star formation occurring in high-density peaks to become smaller with increasing $Q$. The effect is largely offset by the broadening of the overdensity PDF due to the growth of the Mach number, but this is insufficient to overcome the global density decrease entirely.

Interestingly, the fraction of naturally bound star formation that survives the cruel cradle effect $f_{\rm cce}$ shows the opposite behavior. As mentioned above, the overdensity PDF broadens as $Q$ increases, while the critical overdensity for survival $x_{\rm cce}\propto Q^{-1}\propto\sigma_{\rm g}^{-1}$ due to gravitational focusing. Both effects imply that the cruel cradle effect is more disruptive in unstable (low-$Q$) discs. When considering unstable, $Q=0.5$ galaxies along the typical $\Sigma_{\rm g}$--$\Omega$ relation (solid line in Figure~\ref{fig:cfeQ}), the cruel cradle effect plays an important ($>0.3$~dex) role in setting the CFE for $\Sigma_{\rm g}<10~\msun~{\rm pc}^{-2}$. It is also non-negligible for $\Sigma_{\rm g}>10^3~\msun~{\rm pc}^{-2}$, where it limits the CFE by disrupting high-density stellar structure that otherwise would have remained bound due to the high local SFE.

Combining the variation of $f_{\rm bound}$ and $f_{\rm cce}$ with the surface density, angular velocity, and $Q$ parameter, we see that the stability of the gas disc is instrumental mainly in setting the fraction of bound star formation that survives the cruel cradle effect. In particular, unstable galaxies with high angular velocities (upper dashed line in Figure~\ref{fig:cfeQ}) have reduced CFEs due to tidal perturbations by the star-forming environment. The naturally bound part of star formation is much less affected by the disc stability and follows the same trends as discussed in \S\ref{sec:cfesflaw}. Figure~\ref{fig:cfeQ} also indicates the `quiescent', `typical', and `starburst' galaxies from Table~\ref{tab:vars}, with CFEs of $\Gamma=\{4,7,59\}$\%, respectively. As by the above discussion should be expected, the figure clearly shows that the CFE for the high-density set of variables is mainly affected by the cruel cradle effect (i.e. $f_{\rm cce}'<f_{\rm bound}$), whereas the `quiescent' and `typical' model galaxies reside in the part of the parameter space where a low naturally bound fraction of star formation determines the CFE (i.e. $f_{\rm bound}<f_{\rm cce}'$).

\subsection{The time-evolution of the cluster formation efficiency} \label{sec:tview}
\begin{figure*}
\center\resizebox{\hsize}{!}{\includegraphics{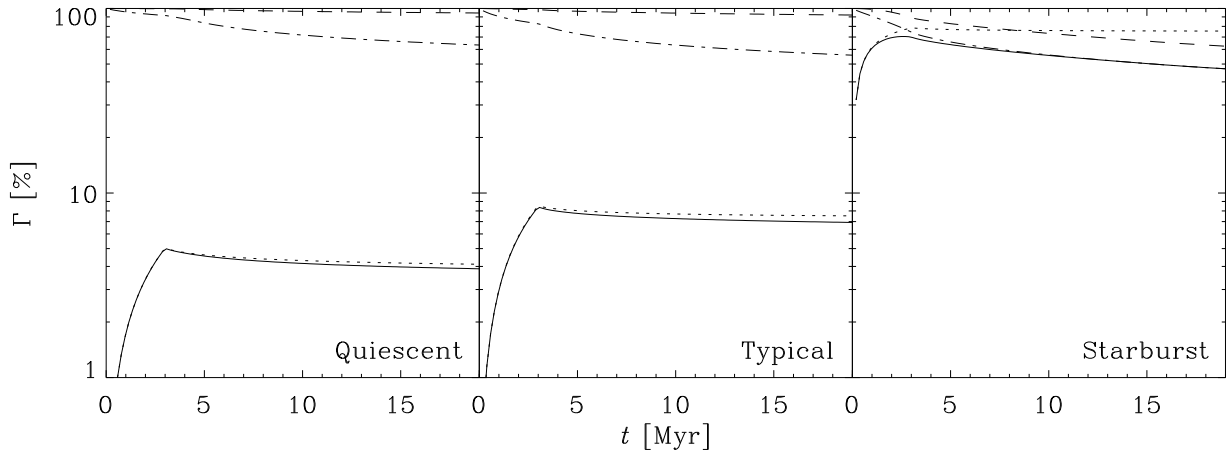}}
\caption[]{\label{fig:tview}
      Time-evolution of the cluster formation efficiency. The panels indicate the `quiescent', `typical', and `starburst' parameter sets from Table~\ref{tab:vars}, which are also marked with symbols in Figure~\ref{fig:cfeQ}. The solid, dotted, dashed, and dash-dotted lines indicate the cluster formation efficiency, the naturally bound fraction of star formation, the fraction of bound star surviving the cruel cradle effect, and the fraction of all star formation surviving the cruel cradle effect, respectively.
                 }
\end{figure*}
By varying the time $t$ at which the CFE is determined, it is possible to address the time-evolution of the CFE across the ensemble of all star-forming regions in a galaxy, for the hypothetical situation in which they are all coeval. This is a useful thought experiment for two reasons. Firstly, it can provide indications of how star and cluster formation proceeds. Secondly, it should be verified how sensitive the obtained CFE is to the time at which it is determined. Evidently, a high sensitivity could obstruct a reliable comparison to observations.

Figure~\ref{fig:tview} shows the time-evolution of the CFE for the `quiescent', `typical', and `starburst' parameter sets from Table~\ref{tab:vars} and Figure~\ref{fig:cfeQ}. We see that the CFE increases with time until the onset of feedback at $t_{\rm sn}=3$~Myr, which is a natural result of the ongoing collapse of gas into stars and bound structure while no gas is yet being expelled.\footnote{When including radiative feedback, the CFE during this phase can be up to 0.5~dex lower. However, this occurs only in a small part of the parameter space, as is shown in Appendix~\ref{sec:apprad}.} When the first supernovae explode, star formation is halted on a global scale in regions with high SFE and density, while in low-SFE regions star formation can still continue because the feedback pressure is smaller than the ISM pressure. Dense stellar clusters thus collapse and complete the star formation process first, while the remaining star formation is only possible in those regions with increasingly low SFEs. Most of the residual, ongoing star formation is unbound, and further star formation therefore contributes very little to the CFE. This causes $f_{\rm bound}$ to slowly decrease for $t>t_{\rm sn}$, but otherwise it hardly exhibits any variation. 

The slight decline of the CFE is aided further by the critical overdensity above which structure can survive the cruel cradle effect, which increases with time as is shown explicitly in equation~(\ref{eq:xcce}). This has an obvious explanation: over a longer timespan, a star-forming region is subjected to a larger number of perturbations. The resulting decrease of $f_{\rm cce}$ provides the largest time-evolution for ages $t>t_{\rm sn}$, and most prominently so in high-density environments (see the third panel of Figure~\ref{fig:tview}).

Combining the above considerations, we conclude that the CFE is approximately time-independent -- as long as $t$ is chosen such that the densest regions that dominate the clustered part of star formation have been able to collapse and complete the star formation process (i.e.~$t>t_{\rm sn}$).\footnote{This is a natural choice since the CFE is established observationally by considering unembedded clusters, implying either a high SFE or some influence of feedback.} This also explains why the CFE is insensitive to the adopted feedback energy input $\phi_{\rm fb}$ (see Appendix~\ref{sec:appfb}). In the high-density regions where bound stellar clusters are born, either star formation is optimally efficient ($\epsilon=\epsilon_{\rm core}$) or the SFE is so high that feedback is sufficiently strong to halt star formation on a short time-scale. The details of how star formation is stopped elsewhere do not notably shift the balance between bound and unbound stellar structure. The only important time-evolution may occur in high-density ($\Sigma_{\rm g}\ga10^3~\msun~{\rm pc}^{-2}$) galaxies, where the cruel cradle effect can lead to a decrease of the CFE with time.

\subsection{The influence of miscellaneous parameters}
\begin{figure*}
\center\resizebox{\hsize}{!}{\includegraphics{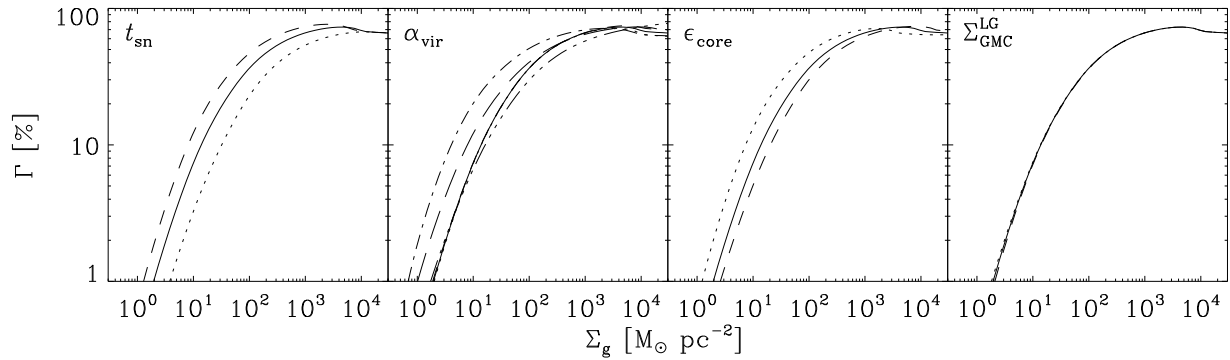}}
\caption[]{\label{fig:cfeline}
      Influence of miscellaneous parameters on the cluster formation efficiency (CFE), adopting the parameters from Table~\ref{tab:param} and taking $Q=1.5$. The CFE is shown as a function of surface density $\Sigma_{\rm g}$ only, obtained by writing the angular velocity as a function of $\Sigma_{\rm g}$ (see equation~(\ref{eq:omega}) and the solid lines of Figures~\ref{fig:cfesflaw} and~\ref{fig:cfeQ}). From left to right, the panels show the effects of the time of the first supernova $t_{\rm sn}$, the GMC virial parameter $\alpha_{\rm vir}$, the core SFE $\epsilon_{\rm core}$, and the giant molecular cloud (GMC) surface density $\Sigma_{\rm GMC}$. The dotted, solid, and dashed lines indicate values of $t_{\rm sn}=\{1.5,3,5\}~{\rm Myr}$, $\alpha_{\rm vir}=\{0.5,1.3,3\}$, $\epsilon_{\rm core}=\{0.25,0.5,0.75\}$, and $\Sigma_{\rm GMC}=\{40,100,250\}~\msun~{\rm pc}^{-2}$. In the second panel, the dash-dotted, long-dashed, and dash-triple-dotted lines indicate $\alpha_{\rm vir}=\{0.5,1.3,3\}$ for the \citet{krumholz05} star formation law, which depends on the adopted value of $\alpha_{\rm vir}$.
                 }
\end{figure*}
The influence of some key parameters from Table~\ref{tab:param} on the CFE is addressed in Figure~\ref{fig:cfeline}. The first panel shows the impact of varying the characteristic time-scale $t_{\rm sn}$ on which star formation is halted by feedback. The CFE increases with surface density due to the growth of the naturally bound fraction of star formation, until it reaches a peak and decreases at the highest densities due to the cruel cradle effect. The choice of $t_{\rm sn}=1.5$--$5$~Myr most strongly affects the CFE at the lowest gas surface densities, where the variation can amount up to 0.6~dex. However, this spread decreases towards higher densities. Over the entire density range, shorter supernova time-scales yield lower CFEs. This is easily understood in terms of the presented model. A short $t_{\rm sn}$ implies a lower SFE, which in turn implies that a smaller fraction of star formation will be bound after gas dispersal.

The second panel of Figure~\ref{fig:cfeline} shows the impact of the GMC virial parameter $\alpha_{\rm vir}$ on the CFE. For the empirically motivated star formation law, this influence is negligible because the virial parameter only enters as a second-order effect in the cruel cradle effect,\footnote{Specifically, it sets the adiabatic correction of equation~(\ref{eq:phiad}).} which in itself does not dominate the CFE over most of the density range. However, when adopting the \citet{krumholz05} star formation law, the range of virial parameters $\alpha_{\rm vir}=0.5$--$3$ produces a variation that is comparable in magnitude to that brought about by setting $t_{\rm sn}=1.5$--$5$~Myr in the previous panel. This is caused by the specific star formation rate per free-fall time of equation~(\ref{eq:ssfrff_km05}), which decreases with $\alpha_{\rm vir}$ and thus leads to a lower SFE and CFE at higher virial parameters.

The third panel of Figure~\ref{fig:cfeline} addresses the variation of the CFE with the maximum SFE $\epsilon_{\rm core}$. This also behaves in a predictable way -- higher maximum SFEs increase the star formation rate in the high-density tail of the star formation spectrum, but at lower densities they also decrease the local bound fraction $\gamma(\epsilon)=\epsilon/\epsilon_{\rm core}$ by requiring a larger SFE to attain the same $\gamma$. The latter effect ends up dominating the influence of $\epsilon_{\rm core}$ on the CFE, but its magnitude is not as large as for the parameters in the previous panels because both effects compete. It is important to realize that the influence of this parameter is purely driven by the prior postulate that a region reaching $\epsilon=\epsilon_{\rm core}$ is bound in its entirety. As such, there is no real physics present in this panel.

The fourth and last panel of Figure~\ref{fig:cfeline} shows the influence of the GMC surface density $\Sigma_{\rm GMC}^{\rm LG}$ on the CFE. This quantity only affects the disruptive power of tidal perturbations, and only at surface densities $\Sigma_{\rm g}<\Sigma_{\rm GMC}^{\rm LG}$ because at higher densities $\Sigma_{\rm GMC}=\Sigma_{\rm g}$. Therefore, $\Sigma_{\rm GMC}^{\rm LG}$ affects the CFE only the low-density regime, and only if the cruel cradle effect plays a non-negligible role. For the parameter set shown here (with $Q=1.5$) this does not occur, but the small variations that are visible are indicative of the dependence in unstable discs (i.e.~$Q=0.5$). At low ($\Sigma_{\rm g}<10~\msun~{\rm pc}^{-2}$) surface densities, the CFE decreases with increasing $\Sigma_{\rm GMC}$ due to the correspondingly stronger tidal perturbations. For stable galaxies this effect is entirely negligible.

In summary, we see that the variation induced by changing the most important, `fixed' model parameters is modest, and in all cases retains the characteristic increase of the CFE with the gas surface density.

\subsection{Summary of main model dependences} \label{sec:summparam}
The main conclusions of the parameter study in this section are as follows.
\begin{enumerate}
\item
The CFE increases with the gas surface density $\Sigma_{\rm g}$ of the host galaxy. This increase is driven by a growing fraction of all star formation that takes place in high-density regions that are capable of reaching high SFEs, thus evacuating the surrounding gas and becoming bound. At high surface densities, the CFE is inhibited by the tidal perturbation of star-forming regions due to the cruel cradle effect. This causes a peak CFE $\Gamma\sim50$--$70\%$ at a density of $\Sigma_{\rm g}\sim10^3~\msun~{\rm pc}^{-2}$, beyond which the CFE settles at $\Gamma\sim40$--$60\%$.
\item
In the region of the $\Sigma_{\rm g}$--$\Omega$ plane where most galaxies reside, increasing the angular velocity $\Omega$ very weakly affects the CFE, either yielding a modest increase (at intermediate surface densities of $10^{1.5}~\msun~{\rm pc}^{-2}<\Sigma_{\rm g}<10^{2.5}~\msun~{\rm pc}^{-2}$) or decrease (at low surface densities of $\Sigma_{\rm g}<10^{1.5}~\msun~{\rm pc}^{-2}$). The only exception arises at high surface densities ($\Sigma_{\rm g}>10^{2.5}~\msun~{\rm pc}^{-2}$) where the CFE is limited by the cruel cradle effect, of which the strong dependence on the ambient, mid-plane density $\rho_{\rm ISM}\propto \Omega^2 Q^{-2}$ implies that an increase of the angular velocity decreases the CFE.
\item
At fixed surface density and angular velocity, the variation of the \citet{toomre64} $Q$ parameter reflects the variation of the turbulent velocity dispersion $\sigma_{\rm g}$. The most important consequence of increasing $Q$ for the CFE is the corresponding decrease of the mid-plane density $\rho_{\rm ISM}\propto \Omega^2 Q^{-2}$, which leads to a smaller fraction of all star formation in high-density peaks and a weakened cruel cradle effect. The CFE thus grows with increasing disc stability. This also implies that the cruel cradle effect is most important in unstable discs.
\item
The CFE grows with time until the onset of feedback at $t=t_{\rm sn}$, after which it hardly changes. Dense clusters collapse and complete the star formation process first, after which the remaining star formation in lower-density regions contributes weakly to the global CFE.
\end{enumerate}
These characteristics persist when varying the miscellaneous parameters from Table~\ref{tab:param} that are kept fixed throughout the rest of this paper.

\section{Comparison to observed galaxies} \label{sec:obs}
Having addressed the behavior of the CFE across the parameter space, it is possible to compare modelled CFEs to observations and interpret why potential differences might arise. We first do this for the broad sample of galaxies for which CFEs have been determined, and then do the comparison on a galaxy-by-galaxy basis.

\subsection{The empirical relation between the cluster formation efficiency and the star formation rate density} \label{sec:sfrd}
The CFE has been determined observationally in several recent studies \citep{goddard10,silvavilla10,silvavilla11,adamo10,adamo11,cook12}, which together cover a galaxy sample containing dwarf, spiral, and starburst galaxies. In most cases, these studies compare the age distributions of bound clusters to the star formation rates or histories of the host galaxies. The ratio of both quantities then gives the CFE. The observational results typically cover a range of $\Gamma=1$--$50$\%, which is comparable to the outcome of our parameter survey in \S\ref{sec:param}.

Particular attention has been given to how the CFE varies with the star formation rate density of the host galaxy $\Sigma_{\rm SFR}$. The analyses by \citet{goddard10,silvavilla11,adamo11} have shown that the CFE increases with $\Sigma_{\rm SFR}$ in a way that is similar to the correlation with the gas surface density $\Sigma_{\rm gas}$ that results from our model. Both densities are related through the \citet{kennicutt98b} formulation of the \citet{schmidt59} star formation law as
\begin{equation}
\label{eq:kslaw}
\Sigma_{\rm SFR,0}=2.5\times10^{-4}\Sigma_{\rm g,0}^{1.4} ,
\end{equation}
where $\Sigma_{\rm SFR,0}\equiv\Sigma_{\rm SFR}/\msun~{\rm yr}^{-1}~{\rm kpc}^{-2}$ and $\Sigma_{\rm g,0}\equiv\Sigma_{\rm g}/\msun~{\rm pc}^{-2}$. This relation enables us to convert the model gas densities to star formation rate densities and directly contrast theory and observations. As a first comparison, the CFE can be calculated for a typical parameter set.

\begin{figure}
\center\resizebox{8cm}{!}{\includegraphics{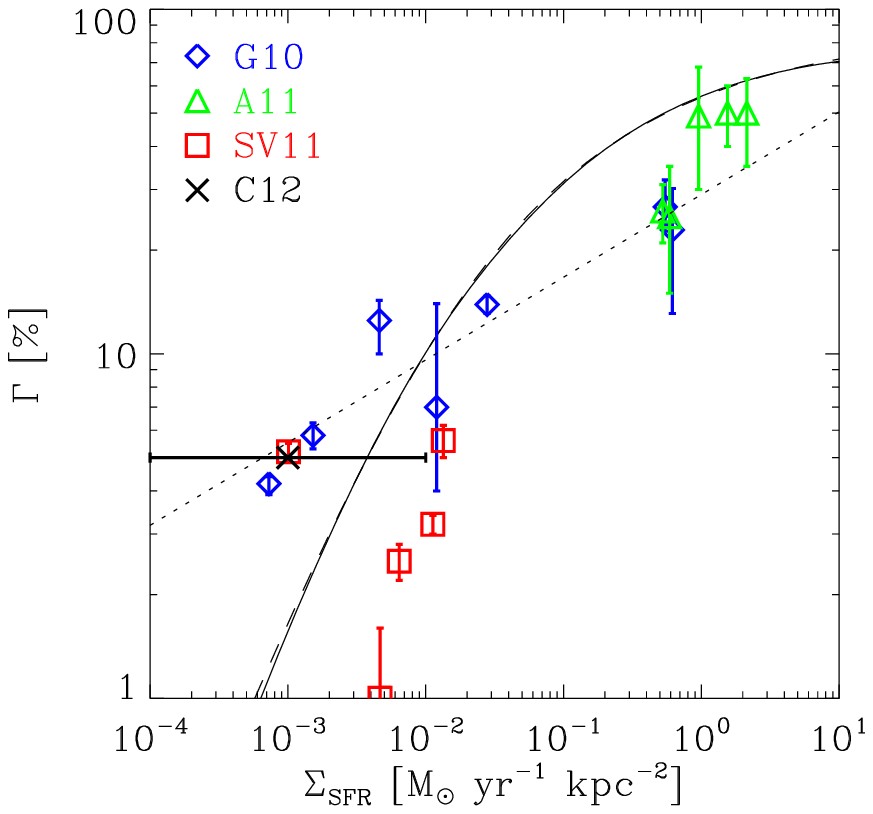}}\\
\caption[]{\label{fig:cfesfrd}
      Cluster formation efficiency (CFE) as a function of the star formation rate density. Symbols denote observed galaxies with $1\sigma$ error bars and indicate the samples from \citet[blue diamonds]{goddard10}, \citet[green triangles]{adamo11}, and \citet[red squares {indicate their $\Gamma_{\rm MDD}$}]{silvavilla11}. The black cross indicates the integrated CFE of all dwarf galaxies from the sample of \citet{cook12}, with a surface density range indicated by the horizontal error bar. The solid curve represents the modelled relation for the `typical' parameter set from Table~\ref{tab:vars}, but can vary for different galaxy and ISM properties. The dashed curve shows the fit of equation~(\ref{eq:cfefit}), and the dotted line represents the original fit by \citet{goddard10}.
                 }
\end{figure}
The observed CFEs are shown in Figure~\ref{fig:cfesfrd} for the combined galaxy sample of \citet{goddard10}, \citet{silvavilla11}, \citet{adamo11}, and \citet{cook12}, together with the modelled $\Gamma-\Sigma_{\rm SFR}$ relation for $Q=1.5$ and with $\Omega$ specified as in equation~(\ref{eq:omega}). The agreement between theory and observations is striking, and gains significance considering the modest variation of the model result with the parameters in Table~\ref{tab:param} (see \S\ref{sec:param} and \S\ref{sec:assum}), which means that the overall trend of the CFEs in Figure~\ref{fig:cfesfrd} is a solid outcome of the theory presented in this paper. We thus obtain a satisfactory explanation of the observations. As shown in \S\ref{sec:param}, the CFE increases with the surface density because at higher densities (and Mach numbers) a larger fraction of the density spectrum of star formation is pushed into the regime with short free-fall times, enabling it to reach high local SFEs and allowing it to become gravitationally bound.

For practical applications, the dashed line in Figure~\ref{fig:cfesfrd} represents a good fit to the model, and is given by
\begin{equation}
\label{eq:cfefit}
\Gamma=\left(1.15+0.6\Sigma_{\rm SFR,0}^{-0.4}+0.05\Sigma_{\rm SFR,0}^{-1}\right)^{-1}\times100\% .
\end{equation}
This fit is for one particular, `typical' parameter set and should therefore mainly be used for rough estimates. As explained below, this parameter set is not accurate for high-density galaxies ($\Sigma_{\rm SFR,0}\ga1$) and may overestimate their CFE by up to a factor of two. It also does not account for differing properties of individual galaxies, which can cause a similar variation. Recall that the star formation rate density can easily be replaced by the gas surface density using the Kennicutt-Schmidt law of equation~(\ref{eq:kslaw}).

One word of caution is in place. The observational CFEs are uncertain up to a factor of two or three due to assumptions in their derivations (see \citealt{goddard10} for an extensive discussion). For instance, they rely on the extrapolation of the observed cluster mass function down to some minimum cluster mass below the detection limit, which is affected by uncertainties on the power law slope of the cluster mass function and the lower mass limit. The estimates are also sensitive to errors in the age and mass fits with simple stellar population models, uncertainties in the metallicity, and possible selection effects. {Finally, \citet{silvavilla11} determined the CFE using two different methods. We adopt their MDD sample, but the trend only changes mildly when using their MID sample instead -- in fact, the agreement between theory and observations slightly improves when taking the mean of both methods.} The error margin resulting from all of the above is realistically shown for at least a few of the formal error margins in Figure~\ref{fig:cfesfrd}, but in some evident cases the errors are strongly underestimated. The aforementioned factor of two to three is more reasonable.

Next to the above considerations, there are also some physical grounds for the variation between the individual galaxies in Figure~\ref{fig:cfesfrd} and the model relation. Firstly, the line only shows the model for one particular parameter set, and is based on the scaling relation between $\Omega$ and $\Sigma_{\rm g}$ of equation~(\ref{eq:omega}). The scatter around that relation is about 0.5~dex, implying that it can influence the CFEs of individual galaxies to a statistically significant degree. Secondly, {high-density galaxies often undergo starbursts and are therefore typically} less stable than isolated disc galaxies, which means that there is a trend of $Q$ with surface density. This implies that at high densities the cruel cradle effect is more prominent than shown by the model here (cf. \S\ref{sec:cfeQ}). Thirdly, at low densities galaxies deviate from the simple power law Schmidt law that we assume to convert our model densities to star formation rate densities, having a larger spread at a given gas density \citep[e.g.][]{bigiel08}. Stochasticity should also play a larger role at low densities \citep{dasilva12}, and hence the increased spread of CFEs in the low-density regime is not very surprising. We address most of these considerations in \S\ref{sec:indiv} below, where we estimate the CFE of each galaxy individually.

A final remark concerns the inclusion of radiative feedback, which is explored in Appendix~\ref{sec:apprad}. A simple inclusion of radiative feedback \citep[following][]{thompson05,murray10} in the theory of this paper predicts a jump of the CFE around $\Sigma_{\rm g}\sim200~\msun~{\rm pc}^{-2}$, due to the increase of the GMC density above such densities. This gas surface density corresponds to $\Sigma_{\rm SFR,0}\sim0.5$ in Figure~\ref{fig:cfesfrd}, where a vague hint of such a jump might be present. It is clear that the uncertainties on the observations are far too large for any conclusive statements, and unfortunately the absence of galaxies around $\Sigma_{\rm SFR,0}\sim0.1$ is not very helpful either. However, this does show that better observational estimates of the CFE could aid in establishing which feedback mechanisms drive turbulence and star formation on galactic scales.

\subsection{Cluster formation efficiencies in individual galaxies} \label{sec:indiv}
\begin{table*}
 \centering
 \begin{minipage}{170mm}
  \caption{Properties and cluster formation efficiencies of galaxy discs and starbursts.}\label{tab:galaxies}
  \begin{tabular}{@{}l c c c c | c c c c c c c@{}}
  \hline
Galaxy & $\log{\Sigma_{\rm g,0}}$ & $\Omega_{-2}$ & $\sigma_{\rm g}$ & $Q$\footnote{If no observed velocity dispersion is available, a fiducial velocity dispersion of 6~km~s$^{-1}$ \citep{kennicutt89} is adopted in the calculation of equation~(\ref{eq:q}) and the value for $Q$ is shown in parentheses.} & $f_{\rm bound}$ & $f_{\rm cce}$ & $f_{\rm cce}'$ & $\Gamma_{\rm th}$ & $\Gamma_{\rm obs}$ & $\log{\Gamma_{\rm th}/\Gamma_{\rm obs}}$ & References \\
(1) & (2) & (3) & (4) & (5) & (6) & (7) & (8) & (9) & (10) & (11) & (12) \\ \hline
NGC~45 & 0.67 & 2.9 & ... & (1.9) & 0.03 & 0.71 & 0.30 & 2.3 & $5.2\pm0.3$ & $-0.4$ & 1,2 \\
NGC~628 (M74) & 0.93 & 1.3 & 6 & 0.5 & 0.06 & 0.71 & 0.30 & 4.1 & 5 & $-0.1$ & 3,4,5 \\
NGC~1313 & 0.88 & 2.1 & 14 & 2.0 & 0.06 & 0.94 & 0.61 & 5.6 & $3.2\pm0.2$ & $0.2$ & 2,6 \\
NGC~1569 & 1.33 & 6.3 & 21.3 & 3.2 & 0.15 & 0.97 & 0.73 & 14 & $13.9\pm0.8$ & $0.0$ & 3,7,8 \\
NGC~3256  & 2.1 & 7.5 & 30 & 0.9 & 0.43 & 0.97 & 0.78 & 41 & $22.9^{+7.3}_{-9.8}$ & $0.3$ & 8,9 \\
NGC~4395 & 0.59\footnote{This is a lower limit because the molecular gas mass has not been estimated.} & 1.9 & ... & (1.5) & 0.03 & 0.71 & 0.29 & 1.9 & $1.0\pm0.6$ & $0.3$ & 2,10 \\
NGC~5194 (M51) & 1.47 & 3.7 & 8 & 0.5 & 0.18 & 0.80 & 0.40 & 14 & 18 & $-0.1$ & 3,5,11 \\
NGC~5236 (M83) & 1.70 & 4.5 & 7.8 & 0.4 & 0.26 & 0.75 & 0.38 & 19 & $16.2\pm4.7$\footnote{Average of the values found by \citet[for the nuclear starburst]{goddard10} and \citet[for the spiral arms]{silvavilla11}.} & $0.1$ & 2,3,8,12 \\
NGC~5457 (M101) & 1.09 & 1.4 & ... & (0.3) & 0.08 & 0.71 & 0.30 & 5.8 & 8 & $-0.1$ & 3,5 \\
NGC~6946 (Arp~29) & 1.30 & 3.6 & 10 & 0.9 & 0.13 & 0.88 & 0.51 & 12 & $12.5^{+1.8}_{-2.5}$ & $0.0$ & 3,8,13 \\
NGC~7793 & 0.84\footnote{Because no observed surface density is available, this value is derived from the average of the star formation rate densities listed in \citet{silvavilla11}, using the \citet{kennicutt98b} star formation law.} & 4.1 & ... & (1.8) & 0.05 & 0.72 & 0.30 & 3.4 & $2.5\pm0.3$ & $0.1$ & 2,14 \\
Haro~11 & 1.98 & 4.7 & 80 & 2.0 & 0.35 & 0.99 & 0.91 & 35 & $50^{+13}_{-15}$ & $-0.2$ & 15,16,17,18 \\
SMC & 0.96 & 2.6 & 22 & 3.2 & 0.07 & 0.97 & 0.72 & 7.2 & $4.2^{+0.2}_{-0.3}$ & $0.2$ & 8,19,20 \\
LMC & 1.04 & 2.9 & 9 & 1.2 & 0.08 & 0.86 & 0.47 & 6.8 & $5.8\pm0.5$ & $0.1$ & 8,21,22 \\
Milky~Way & 0.97 & 2.6 & 7 & 2.0 & 0.07 & 0.94 & 0.62 & 6.8 & $7.0^{+7.0}_{-3.0}$ & $0.0$ & 8,23,24\\
\hline
NGC~224 (M31) & 0.50 & 2.6 & 10 & 2.9 & 0.02 & 0.77 & 0.35 & 1.7 & ... & ... & 25,26 \\
NGC~3031 (M81) & 0.85 & 4.7 & ... & (2.0) & 0.05 & 0.72 & 0.30 & 3.5 & ... & ... & 3 \\
NGC~3034 (M82)\footnote{The listed values are exclusive to the circumnuclear starburst. The gas surface density accounts for molecular hydrogen only.} & 3.52 & $1.4\times10^2$ & 40 & 1.7 & 0.88 & 0.47 & 0.41 & 41 & ... & ... & 3,27 \\
Arp~220$^{\mbox{\it e}}$ & 4.76 & $2.1\times10^2$ & 80 & 0.3 & 1.00 & 0.29 & 0.29 & 29 & ... & ... & 3,28\\
\hline
\end{tabular}\\
Col.~(1): Galaxy identifier. Col.~(2): Logarithm of the gas surface density in \msun~pc$^{-2}$. Col.~(3): Angular frequency in (100~Myr)$^{-1}$. Col.~(4): Velocity dispersion of the gas disc in km~s$^{-1}$. Col.~(5): Toomre $Q$ parameter as defined in equation~(\ref{eq:q}). Col.~(6): Theoretical bound fraction of star formation. Col.~(7): Theoretical surviving fraction of bound star formation due to the cruel cradle effect. Col.~(8): Theoretical surviving fraction of all star formation due to the cruel cradle effect. Col.~(9): Theoretical cluster formation efficiency (CFE) in \%. Col.~(10): Observed CFE in \%. Col.~(11): Logarithmic difference between theoretical and observed CFEs. Col.~(12): References as follows.\\
 (1) \citet{chemin06}, (2) \citet{silvavilla11}, (3) \citet{kennicutt98b}, (4) \citet{combes97}, (5) \citet{gieles10b}, (6) \citet{ryder95}, (7) \citet{stil02}, (8) \citet{goddard10}, (9) \citet{sakamoto06}, (10) \citet{swaters99}, (11) \citet{schuster07}, (12) \citet{lundgren04}, (13) \citet{walsh02}, (14) \citet{dicaire08}, (15) \citet{adamo11}, (16) \citet{bergvall00}, (17) \citet{ostlin99}, (18) \citet{ostlin01}, (19) \citet{wilke04}, (20) \citet{stanimirovic04}, (21) \citet{wong09}, (22) \citet{alves00}, (23) \citet{wolfire03}, (24) \citet{heiles03}, (25) \citet{chemin09}, (26) \citep{braun09}, (27) \citet{yun93}, (28) \citet{downes98}.
\end{minipage}
\end{table*}
The excellent agreement between theory and observations shown in Figure~\ref{fig:cfesfrd} warrants a more detailed comparison. Given observed values for the gas surface density, the angular velocity, and the Toomre $Q$ parameter, it is possible to calculate the theoretical CFE for a given galaxy. There are a number of studies in which the actual CFE was estimated for a small sample of galaxies \citep{goddard10,gieles10b,adamo10,silvavilla11,adamo11}. This enables a quantitative test of the theory presented in \S\ref{sec:ismpdf}--\ref{sec:cfe}.

\subsubsection{The observed properties of dwarfs, spirals and starbursts}
The sample of galaxies for which a comparison of the theoretical and observed CFEs is carried out is listed in the left-hand columns of Table~\ref{tab:galaxies}, together with the input parameters for each galaxy. The gas surface densities are obtained by averaging the total gas mass within the optical radius $R_{25}$ to ensure consistency with \citet{kennicutt98b}. In the cases in which this is not accurate because the gas disc is smaller than $R_{25}$, the listed surface density is averaged over the area covered by the gas only. In the cases of the starburst galaxies NGC~3256 and Haro~11, which should exhibit substantial internal variation of the surface density (and therefore of the CFE), the surface density is determined for the same area as for which the observed CFE has been established. For NGC~3256, most of the gas mass is confined to the inner 1.7~kpc, but the disc continues to at least 3~kpc \citep{sakamoto06} at 20\% of the central density. A total gas mass of $3.3\times10^9~\msun$ within 3~kpc is therefore adopted. For Haro~11, most of the gas mass resides within 6.5'' or 2.6~kpc, which contains about $2\times10^9~\msun$ \citep{ostlin99,bergvall00}. Whenever only angular sizes are given by the literature sources, the distance from the NASA/IPAC Extragalactic Database\footnote{{\tt http://ned.ipac.caltech.edu/}} (NED) is used to convert to physical dimensions, unless the distance is given by the original source. For the few galaxies where $R_{25}$ is not given by the corresponding literature sources, the dimensions from NED are used to calculate the surface densities.

The angular velocities are evaluated from the galaxy rotation curves at $0.5R_{25}$, which corresponds to $\Omega\equiv4\pi/t_{\rm dyn}$ (with $t_{\rm dyn}$ the dynamical time-scale) for the galaxies from \citet{kennicutt98b}. In those galaxies that are not taken from the \citet{kennicutt98b} sample and in which less than the full area of the galaxy is covered by either the gas disc or the observed CFE, the angular velocity is determined at half the maximum radius. Given a surface density and angular velocity, the Toomre $Q$ parameter is calculated from the velocity dispersion of the gas. Like the angular velocity, the velocity dispersion (obtained from H$_2$, CO or H{\sc i} observations, in that order of preference and depending on availability) is evaluated at $0.5R_{25}$ when a radial profile is provided by the literature source. Whenever the velocity dispersion is only available for certain radial intervals, an area-weighted average is listed in Table~\ref{tab:galaxies}. If no value is present in the literature at all, a fiducial velocity dispersion of 6~km~s$^{-1}$ \citep{kennicutt89} is adopted in the calculation of $Q$. It should be noted that the Milky Way values are representative of the solar neighbourhood instead of the entire disc. {Finally, we correct for the presence of spiral arms by halving the $Q$ parameters as in \citet{krumholz05} for all galaxies except the Milky Way\footnote{{The solar neighbourhood is located right between two spiral arms \citep[see e.g.][]{portegieszwart10}.}} and the circumnuclear starbursts of M82 and Arp~220.}

\subsubsection{Comparison of observed and theoretical CFEs}
\begin{figure}
\center\resizebox{8cm}{!}{\includegraphics{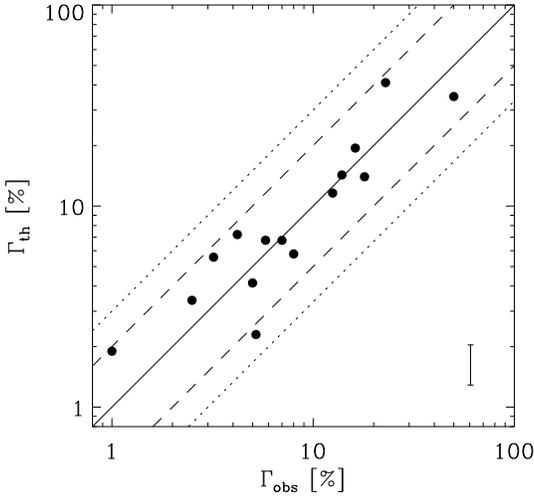}}\\
\caption[]{\label{fig:compare}
      Comparison of observed cluster formation efficiencies $\Gamma_{\rm obs}$ to the values calculated with the theoretical model $\Gamma_{\rm th}$ for the galaxies of Table~\ref{tab:galaxies}. The solid line indicates the one-to-one relation, while the dashed and dotted lines indicate deviations by factors of two and three, respectively.
                 }
\end{figure}
Using columns (2)--(5) of Table~\ref{tab:galaxies}, for each galaxy we calculate the naturally bound fraction of star formation $f_{\rm bound}$, the fraction of bound star formation surviving the cruel cradle effect $f_{\rm cce}$, the fraction of all star formation surviving the cruel cradle effect $f_{\rm cce}'$, and the resulting cluster formation efficiency $\Gamma_{\rm th}$. These quantities are listed in Table~\ref{tab:galaxies}, including the logarithmic difference between the observed and theoretical CFEs ($\log{\Gamma_{\rm th}/\Gamma_{\rm obs}}$). The comparison is also shown graphically in Figure~\ref{fig:compare}.

Given the inhomogeneity of the galaxy sample, the assumptions that were made for certain unknown observables, and the troublesome process of determining the CFE observationally, the agreement between theory and observations is remarkable. As discussed in \S\ref{sec:sfrd}, a spread of at least a factor $\sim2$ should be expected considering the uncertainties in the comparison. This is approximately consistent with the scatter of up to a factor $2.5$ around the one-to-one relation in Figure~\ref{fig:compare}. This includes the statistical errors on the observed CFE and the model input variables $\{\Sigma_{\rm g},Q,\Omega\}$ (also see \S\ref{sec:assum} and Figure~\ref{fig:cfeparam}).

Contrasting the listed values of $f_{\rm bound}$ and $f_{\rm cce}'$, it is clear that the contributions of both mechansisms to the CFE are non-negligible. However, they favour the survival of comparable overdensities (see Figure~\ref{fig:xpdf}). This overlap implies that the additional decrease of the CFE due to the cruel cradle effect typically does not exceed 30\%, as is indicated by the values of $f_{\rm cce}$. It has been mentioned throughout this paper that tidal perturbations are most important for high-density galaxies such as NGC~5236, but perhaps surprisingly the results in Table~\ref{tab:galaxies} show that it is also relevant for low-density galaxies like NGC~45, NGC~4395, and NGC~7793. The importance for low-density galaxies is caused by their relatively low Mach numbers, which imply that the influence of tidal perturbations is enhanced due to a larger influence of gravitational focusing. In galaxies with intermediate surface densities, the CFE is dominated by the naturally bound fraction of star formation.

Table~\ref{tab:galaxies} also includes a prediction for four galaxies that are obvious or interesting candidates for future observational estimates of the CFE. These include M31, which is currently undergoing an extensive survey by the {\sc phat} team \citep{dalcanton12,johnson12} for the {\it Hubble Space Telescope} treasury program. The low gas surface density of M31 leads it to have the lowest CFE ($\Gamma=1.7\%$) of the entire galaxy sample in Table~\ref{tab:galaxies}, which is a prediction that can be tested using the {\sc phat} data. Another clear candidate for a future observational determination of the CFE is M81, for which we also obtain a relatively low CFE of $\Gamma=3.5\%$. The other two listed galaxies are M82 and Arp~220. Establishing the CFE in those galaxies will be quite a challenge due to the high extinction and/or the large distance. Nevertheless, these are  extreme starburst environments, and the model predictions are correspondingly illustrative. In both galaxies, (almost) all stars are formed in naturally bound systems due to the extreme density and the resulting efficient star formation. However, the CFE ends up being much lower than 100\% as it is limited by the cruel cradle effect, which in these cases leads to $\Gamma=30$--$40\%$.

\section{The spatially resolved cluster formation efficiency} \label{sec:spatial}
The theoretical framework of \S\ref{sec:ismpdf}--\ref{sec:cfe} can be used to calculate the radial profile of the CFE within a galaxy. This requires the gas surface density profile, rotation curve and gas velocity dispersion of the galaxy in question. For several nearby galaxies such information is available, and some illustrative examples are given in this section.

\begin{figure*}
\center\resizebox{\hsize}{!}{\includegraphics{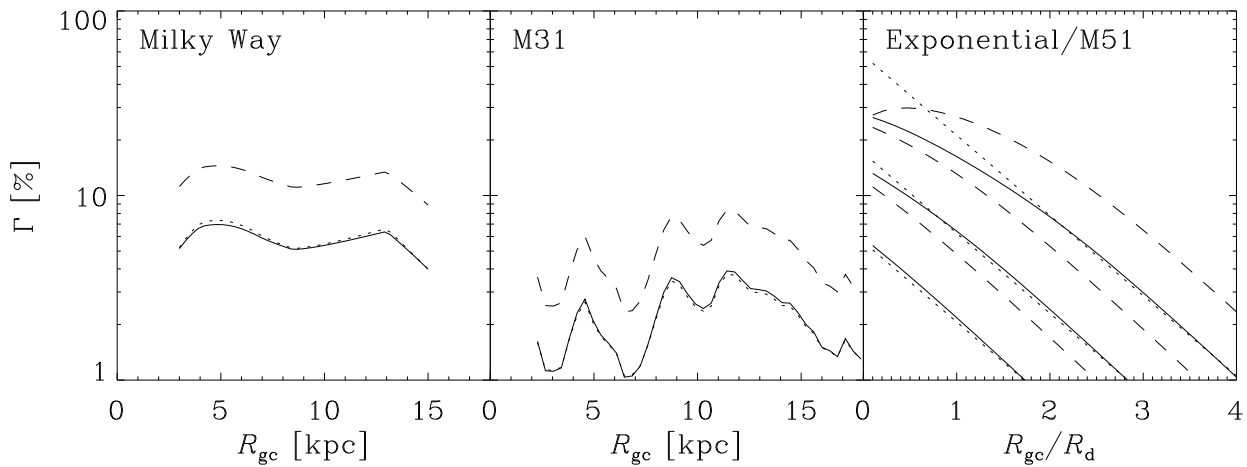}}
\caption[]{\label{fig:spatial}
      Variation of the cluster formation efficiency (CFE) with galactocentric radius $R_{\rm gc}$. Solid lines show the result for the empirically motivated star formation law of equation~(\ref{eq:ssfrff_e02}) and \citet{elmegreen02}, while dashed lines are based on the theoretical star formation law of equation~(\ref{eq:ssfrff_km05}) and \citet{krumholz05}. Dotted lines represent the gas surface density profiles, which in each case are normalized to the outermost CFE for the empirical star formation law. See text for the other adopted parameters of each galaxy. {\it Left panel}: CFE profile for the Milky Way. {\it Middle panel}: CFE profile for M31. {\it Right panel}: CFE profile for a generic exponential gas disc, with the galactocentric radius expressed in units of the scale radius $R_{\rm d}$. From bottom to top, the curves indicate central gas densities of $\Sigma_{\rm g}(0)=\{10,30,100\}~\msun~{\rm pc}^{-2}$, respectively. The two CFE profiles with $\Sigma_{\rm g}(0)=100~\msun~{\rm pc}^{-2}$ represent M51, which has $R_{\rm d}=3$~kpc.
                 }
\end{figure*}
We consider the radial variation of the CFE in the Milky Way, M31, and for a generic exponential gas surface density profile, which happens to give a good representation of M51. The input variables for the different galaxies are as follows.
\begin{enumerate}
\item
For the Milky Way, we use the \citet{wolfire03} model for the ISM, including a factor of 1.4 to account for the presence of helium, and adopt $\sigma_{\rm g}=7~{\rm km}~{\rm s}^{-1}$ \citep{heiles03} to calculate $Q$. The angular velocity $\Omega$ follows from a constant circular velocity of $v_{\rm c}=220~{\rm km}~{\rm s}^{-1}$. 
\item
For M31, \citet{chemin09} provide the H{\sc i} surface density profile, which is used here with a correction factor of 1.3 to account for neutral gas \citep{nieten06}. The adopted circular velocity is $v_{\rm c}=250~{\rm km}~{\rm s}^{-1}$ \citep{braun91,chemin09}, and the gas velocity dispersion is $\sigma_{\rm g}=8~{\rm km}~{\rm s}^{-1}$ \citep{braun09}.
\item
A generic exponential gas surface density profile is given by $\Sigma_{\rm g}(R_{\rm gc})=\Sigma_{\rm g}(0)\exp{(-R_{\rm gc}/R_{\rm d})}$. This is a reasonable description of the ISM in M51 when taking $R_{\rm d}=3$~kpc \citep{schuster07}, and therefore we set the other parameters to values that are appropriate for that galaxy. We consider central surface densities of $\Sigma_{\rm g}(0)=\{10,30,100\}~\msun~{\rm pc}^{-2}$, where the highest value is representative of M51. The circular velocity is taken to be $v_{\rm c}=200~{\rm km}~{\rm s}^{-1}$ \citep{rand93} and is converted to an angular velocity profile using the M51 scale radius of $R_{\rm d}=3$~kpc. The adopted gas velocity dispersion is $\sigma_{\rm g}=8~{\rm km}~{\rm s}^{-1}$ \citep{schuster07}.
\end{enumerate}
{As in \S\ref{sec:indiv}, we correct for the presence of spiral arms by halving the $Q$ parameters \citep[cf.][]{balbus88,krumholz05}.}

The resulting CFE profiles are shown in Figure~\ref{fig:spatial}, using the two different star formation laws used in this paper (cf. \S\ref{sec:cfesflaw}). As should be expected from our earlier analysis, the CFE profiles largely follow the surface density profiles of the galaxies, especially at low densities where $\Gamma\propto\Sigma_{\rm g}$. In the Milky Way, this gives a rather flat CFE profile, with a shallow minimum in the solar neighbourhood ($R_{\rm gc}=8.5$~kpc) and modest peaks at 5 and 13~kpc. The overall variation across a 12~kpc radius interval is about 0.15~dex. The typical CFE throughout the Milky Way disc is $\Gamma\sim6\%$ (empirical star formation law) or $\Gamma\sim12\%$ (for the \citealt{krumholz05} star formation law). M31 exhibits a pronounced variation of the CFE, which again traces fluctuations of the gas surface density with galactocentric radius. Typical CFEs range from $\Gamma=1$--$5\%$ for the empirical star formation law, with the peak values reached between 8 and 15~kpc. Note that we did not account for any radial variation of the gas velocity dispersion, which could lead to a modest ($\la0.2$~dex) increase of the CFE towards the centre of the galaxy.

The exponential surface density profiles imply a much stronger trend with galactocentric radius than in the previous two examples. As shown in the right-hand panel of Figure~\ref{fig:spatial}, the CFE covers a broad range, hitting $\Gamma=30\%$ in the central parts of M51 to $\Gamma=3\%$ at $R_{\rm gc}=10~{\rm kpc}\sim 3R_{\rm d}$. Like in the preceding cases, the CFE profiles shown here assume a flat rotation curve and no radial variation of the gas velocity dispersion. Real galaxies exhibit solid-body rotation within the central few~kpc (i.e.~$v_{\rm c}\propto R_{\rm gc}$ and $\Omega\sim{\rm constant}$) and have velocity dispersions that rise towards their centres. Of course, this can be accounted for in choosing the model input variables, but both effects induce only $\la0.2$~dex deviations of the CFE in the galaxy centres with respect to the relations shown in Figure~\ref{fig:spatial}. In general though, it is questionable to what extent the model holds in such environments. Cloud-cloud collisions might be playing an important role in determining whether star formation can proceed and structure can remain bound. Even more importantly, recent sub-millimeter observations of the central molecular zone (CMZ) of the Milky Way suggest that the tracers of dense gas are overabundant by up to a factor of a hundred compared to the star formation tracers \citep{longmore12b}. This might indicate that star formation in the Galactic centre does not obey the star formation laws that apply to galaxy discs. While this does not directly affect the CFE, it seems plausible that the cluster formation process would be affected as well. Either way, galaxy centres clearly provide important tests for current ideas about star and cluster formation.

Figure~\ref{fig:spatial} shows that the CFE directly traces the gas surface density for surface densities $\Sigma_{\rm g}\leq 20~\msun~{\rm pc}^{-2}$. This changes at higher densities, as is shown by the M51-like CFE profile. The reason is that for $\Sigma_{\rm g}\ga20~\msun~{\rm pc}^{-2}$, the increase of the CFE with the gas density becomes shallower than linear (see e.g.~Figure~\ref{fig:cfeline}). Another effect would potentially be that the central regions of the galaxies have enhanced angular velocities compared to the gas densities and therefore lie above the $\Sigma$--$\Omega$ relation from equation~(\ref{eq:omega}) that holds for galaxies as a whole. They only evolve towards that relation at increasing galactocentric radius. However, for disc galaxies this occurs in a regime where the CFE is virtually insensitive to $\Omega$ (see \S\ref{sec:cfesflaw}), and hence the influence of a varying angular velocity is negligible. More extreme spatial variations of the CFE should be present in starburst galaxies like M82 and Arp~220. The central regions of these galaxies have CFEs that are limited by the cruel cradle effect, whereas in their outskirts it should be set by the naturally bound fraction of star formation. Starburst galaxies are therefore ideal targets for exploring the boundaries of the presented theory and for understanding the CFE in more detail.

\section{Star cluster formation over the course of a Hubble time} \label{sec:hubble}
The conditions under which stellar clusters formed have changed notably with cosmic time. As shown in the preceding sections, the efficiency of cluster formation is sensitive to the properties of the host galaxy. The cosmic star formation rate density peaked at redshift $z\sim2$--$3$ \citep[e.g.][]{madau96,madau98,lilly99,perez05,hopkins06b,reddy09}, reflecting that galaxies at these times were denser and more gas-rich than in the nearby Universe. Indeed, observations of high-redshift galaxies show that the gas densities ($10^2$--$10^5~\msun~{\rm pc}^{-2}$) and star formation rates ($20$--$3000~\msun~{\rm yr}^{-1}$) commonly reached extreme values in the early Universe that are presently rare \citep[e.g.][]{chapman04,solomon05,daddi07,genzel10}. Local, extreme starburst galaxies like the merger remnants Arp~220 and NGC~6240 are the nearest examples of galaxies reaching similar conditions \citep{rieke85,scoville91}. Remarkably though, many of the high-redshift star-forming galaxies appear to be `normal' star-forming (albeit unstable) disc galaxies, of which the high star formation rates are simply caused by their substantial gas reservoir \citep{elmegreen05,erb06,genzel08,daddi10,elmegreen09,tacconi10}. With such observational input at hand, it is possible to address the CFE under the high-redshift galactic conditions.

\begin{figure*}
\center\resizebox{16.5cm}{!}{\includegraphics{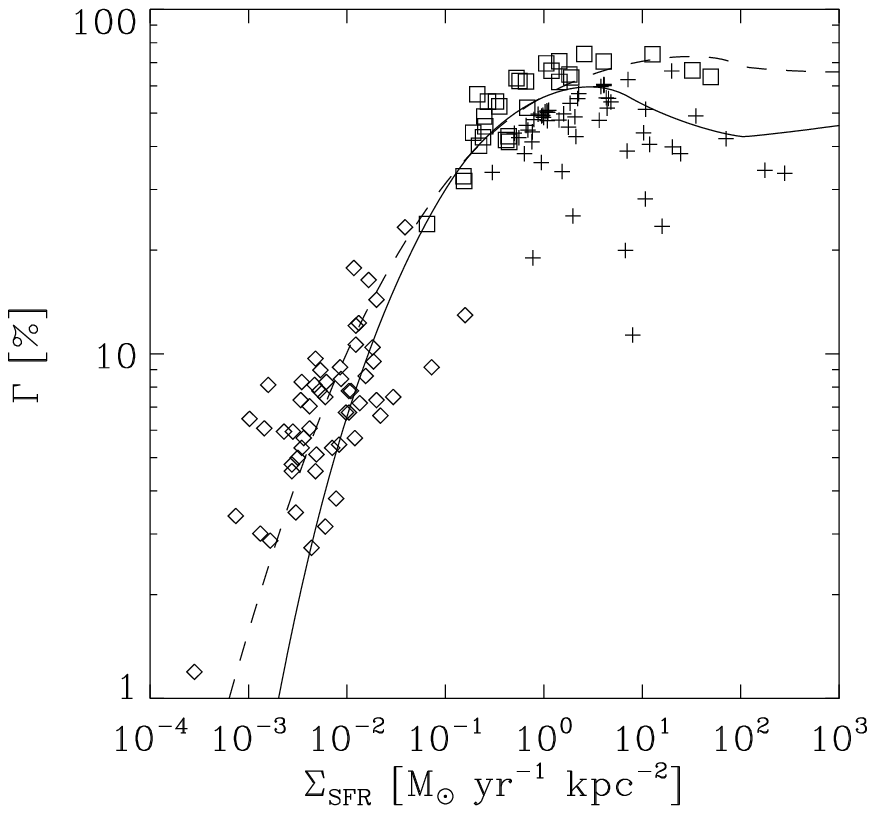}\includegraphics{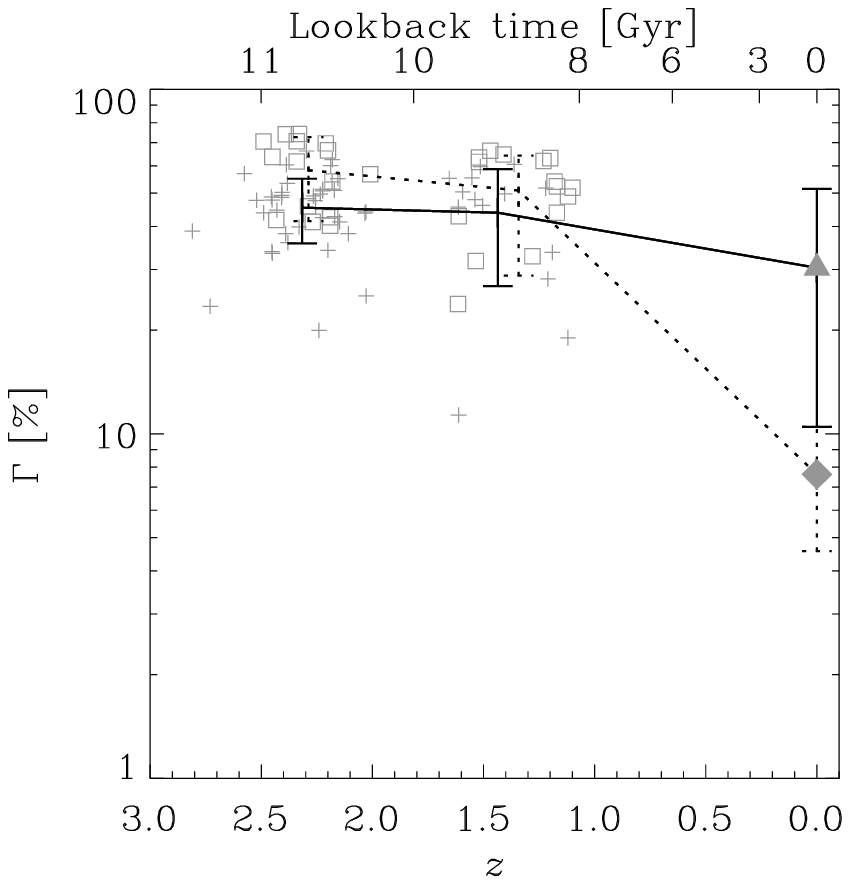}}\\
\caption[]{\label{fig:cfez}
      Variation of the cluster formation efficiency (CFE) with cosmic time. {\it Left}: Predicted CFE as a function of the observed star formation rate density for nearby disc galaxies \citep[diamonds, using input data from][]{kennicutt98b}, high-redshift ($z=1$--$2.5$) normal galaxies (squares), and high-redshift ($z=1$--$3$) starburst galaxies (plus signs). The input data ($\Sigma_{\rm g}$ and $\Omega$) for redshifted galaxies are taken from \citet{tacconi08}, \citet{forsterschreiber09}, \citet{tacconi10}, and \citet{genzel10}. To guide the eye, the curves indicate the model predictions for $Q=0.5$ (solid) and $Q=1.5$ (dashed), defining $\Omega$ and $\Sigma_{\rm SFR}$ as in equations~(\ref{eq:omega}) and~(\ref{eq:kslaw}), respectively. {\it Right}: Predicted CFE as a function of redshift and lookback time \citep[cf.][]{wright06}. Small grey symbols are the same as in the left-hand figure. Large solid symbols indicate the mean CFEs for nearby discs (diamond) and starbursts \citep[triangle, using input data from][]{kennicutt98b}. The lines indicate the mean CFEs in $\Delta z=1$ redshift bins for discs (dotted) and starbursts (solid). Error bars do not represent standard errors, but indicate the 16$^{\rm th}$ and 84$^{\rm th}$ percentiles of the underlying distributions.
                 }
\end{figure*}
We use the cluster formation theory of this paper to predict the CFEs of some $\sim150$ galaxies in the redshift range $z=0$--$3$. The sample of nearby disc galaxies is taken from \citet{kennicutt98b}, who tabulated their gas surface densities $\Sigma_{\rm g}$ and dynamical times $t_{\rm dyn}$. The angular velocities follow as $\Omega=4\pi/t_{\rm dyn}$, and we assume a constant velocity dispersion $\sigma_{\rm g}=6~{\rm km}~{\rm s}^{-1}$ to calculate $Q$. This suffices to estimate the CFE, which is a function of $\Sigma_{\rm g}$, $Q$, and $\Omega$ in the presented model (see \S\ref{sec:cfe}). For the high redshift galaxy population, the combined galaxy sample of \citet{tacconi08}, \citet{forsterschreiber09}, \citet{tacconi10}, and \citet{genzel10} is used. Surface densities are derived from the gas masses $M_{\rm gas}$ and H$\alpha$ radii $R_{\rm H\alpha}$ as $\Sigma_{\rm g}=0.5M_{\rm gas}/\pi R_{\rm H\alpha}^2$, and the angular velocity is calculated from the characteristic velocities\footnote{This velocity can either represent the circular velocity or the velocity dispersion, depending on whether the galaxy is rotation- or dispersion-dominated.} $V$ as $\Omega=2V/R_{\rm H\alpha}$. The \citet{toomre64} $Q$ parameter is chosen based on the gas depletion time-scale $t_{\rm depl}\equiv M_{\rm gas}/{\rm SFR}$. If it exceeds $300~{\rm Myr}$, the galaxy is assumed to be stable and we adopt $Q=1.5$. For $t_{\rm depl}<300~{\rm Myr}$, the galaxy is considered to undergo a starburst and we assume $Q=0.5$. While this separation is somewhat arbitrary, it does enable a rough separation between stable and unstable systems.

The calculated CFEs are shown in Figure~\ref{fig:cfez}. The left-hand panel shows the CFE in the $\Sigma_{\rm SFR}$--$\Gamma$ plane of Figure~\ref{fig:cfesfrd}, this time covering a larger range to account for the extreme star formation rate densities. For reference, two model lines for $Q=0.5$ and $Q=1.5$ are included.\footnote{Note that the symbols and lines are both theoretical predictions. As such, the `scatter' of the predicted CFEs around the theoretical lines is completely unrelated to any (dis)agreement between model and observations. It is only caused by (1) deviations of individual galaxies from the $\Sigma$--$\Omega$ relation of equation~(\ref{eq:omega}), and (2) deviations of individual galaxies from the Kennicutt-Schmidt law of equation~(\ref{eq:kslaw}), which were both used to for determining the model lines.} The predicted CFEs of nearby disc galaxies cover a range that is very similar to that of the observations in Figure~\ref{fig:cfesfrd}, increasing from $\Gamma\sim1\%$ at low star formation rate densities to $\Gamma\sim20\%$ at the high-density end. The high-redshift star-forming disc galaxies represent a natural continuation of this trend to higher densities and CFEs, increasing from $\Gamma\sim30\%$ to a saturation at $\Gamma\sim70\%$. These discs are at the point of perfect balance between a high naturally bound fraction of star formation due to their high densities, and a sufficiently low density (and adequate stability) to have star-forming regions not suffer too strongly from tidal perturbations. The high-redshift starburst galaxies are in a different regime, where in some cases the CFEs are strongly suppressed by the cruel cradle effect. This is similar to the earlier prediction for the CFE in Arp~220 (see \S\ref{sec:indiv} and Table~\ref{tab:galaxies}). For a given galaxy sample, the highest CFE should thus be attained at a star formation rate density of $\Sigma_{\rm SFR}=1$--$10~\msun~{\rm yr}^{-1}~{\rm kpc}^{-2}$ or a gas surface density of $\Sigma_{\rm g}\sim10^3~\msun~{\rm pc}^2$.

The evolution of the CFE with redshift is shown in the right-hand panel of Figure~\ref{fig:cfez}. The galaxy sample is divided into galaxy discs and starburst galaxies according to the above criterion of the gas depletion time-scale, and binned in $\Delta z=1$ redshift bins to show trends of the mean CFE. To obtain a $z=0$ data point for starburst galaxies, we add the \citet{kennicutt98b} sample of nearby circumnuclear starbursts, using the same analysis as for the local disc galaxy sample and assuming $Q=0.5$. The model predicts that the CFE in starburst galaxies has been relatively constant during the past $\sim11$~Gyr, whereas the CFE in disc galaxies experienced a strong decrease with cosmic time. Both lines cross at $z\sim1$, with starbursts being the most efficient cluster factories at lower redshifts, and disc galaxies having a slight edge at higher redshifts. Note that the error bars do not indicate the statistical significance of the plotted mean value, but instead indicate the spread of the underlying sample. The differences between the disc and starburst galaxy samples are statistically significant, particularly for the $z=0$ sample. This has been verified by running a Kolmogorov-Smirnoff test. For the hypothesis that the CFEs of both populations are drawn from the same parent sample, we obtain $p$-values of $p_{\rm KS}=\{0.004,0.148,1.73\times10^{-8}\}$ for $z=2$--$3$, $z=1$--$2$, and $z=0$--$1$, respectively.

The restricted sample of $166$ galaxies that is used for the right-hand panel of Figure~\ref{fig:cfez} is not necessarily representative of the global galaxy population. The local Universe sample relies exclusively on \citet{kennicutt98b}, of which the galaxy selection is mostly based on the availability of CO and H{\sc i} maps, yielding a galaxy sample containing a mix of field and Virgo cluster galaxies. The sample should therefore be free of strong selection biases. The high-redshift sample contains intermediate-to-high-mass ($M_{\rm star}>2\times10^9~\msun$), star-forming (${\rm SFR}>6~\msun~{\rm yr}^{-1}$) galaxies, with a large contribution from the SINS survey \citep{genzel08,forsterschreiber09}. These ranges are chosen to include a broad range of main-sequence, star-forming galaxies and hence do not impose strong restrictions on the galaxy population at $z=1$--$3$ other than being limited to the bright end of the luminosity function \citep[e.g.][]{forsterschreiber09}. For our sample of nearby main-sequence galaxies (see Table~\ref{tab:galaxies}), there is no appreciable correlation of the CFE with galaxy mass, and hence extrapolating the results for the SINS sample to fainter galaxies seems reasonable. {Nonetheless, it should be kept in mind that any low-density (i.e. $\sim$ low-SFR) galaxies are not included, and as a result the characteristic CFE at high redshifts may be lower in reality than is suggested by this sample.} We conclude that while the sample is certainly not complete, it {does highlight a change} of the star-forming population across the covered redshift range. The trend of the CFE with cosmic time of Figure~\ref{fig:cfez} is thus a tentative, but plausible prediction.

Current studies of star cluster populations can reach out to distances of $\sim100$~Mpc, corresponding to $z\sim0.024$. It is therefore not possible to confirm the predicted evolution of the CFE through direct observations. Even for an indirect verification to be possible, it would be required that some products of high-redshift cluster formation have survived until the present day. It has long been suggested that at least some fraction of the old, massive globular clusters that populate the haloes of nearby galaxies are the surviving products of the regular high-redshift star formation process, and were formed through the same mechanisms that currently lead to the formation of young stellar clusters \citep[e.g.][]{elmegreen97,kruijssen12b}. The differences between young cluster populations and these old globular clusters would then be caused by differences in the birth environment \citep[e.g.][]{kravtsov05,elmegreen10,shapiro10,kruijssen12c} and a Hubble time of dynamical evolution \citep[e.g.][]{elmegreen97,fall01,vesperini01}.

If {some part of present-day} globular clusters are indeed the relics of high-redshift, regular star cluster formation,\footnote{A common argument against this is the presence of abundance anomalies in globular clusters \citep[e.g.][]{gratton04}, which do not appear in nearby, currently forming stellar clusters. However, there are no known resolved regions locally where the formation of such high-mass ($M>10^6~\msun$) globular cluster equivalents can be seen in action. Therefore, it is unclear whether the formation of massive clusters in gas-rich galaxies could perhaps naturally lead to the observed abundance anomalies.} then we can use the framework of the present paper to address the conditions of their origin in some more detail. Figure~\ref{fig:cfez} predicts that the CFE in high-redshift disc galaxies {reached values} almost one order of magnitude larger than in present-day discs, and some factor of $2$ higher than in nearby starburst galaxies. This does not only imply that clusters forming in the early Universe were ten times more numerous at the same star formation rate. Given that young stellar clusters are formed according to a power law initial cluster mass function with an index of $-2$ \citep{portegieszwart10}, the size-of-sample effect dictates that a $1$~dex increase of the CFE would also facilitate the formation of clusters ten times more massive at the same star formation rate. Considering that the star formation rate itself attained values ${\rm SFR}>1000~\msun~{\rm yr}^{-1}$, the masses of the most massive clusters should have reached even higher. This statistical\footnote{A possibly more relevant (and currently unanswered) question is whether there are any physical mechanisms that limit the maximum cluster mass, and how these would evolve with the galactic environment \citep[e.g.][]{gieles06b,bastian08,larsen09}.} argument suggests that the `normal' cluster formation mechanisms of the nearby Universe would lead to the formation of extremely massive ($\ga10^7~\msun$) and dense stellar systems like {(metal-rich)} globular clusters in high-redshift galaxies, and that such systems were commonplace.

It is possible to make a theoretical estimate of the total fraction of all stars in the Universe that formed in bound stellar clusters by integrating the CFE over the cosmic mass assembly history. Noting that secular star formation in galaxy discs contributes a larger fraction of the cosmic mass assembly than starbursts for redshifts $z\la1$ \citep[e.g.][]{somerville01,kauffmann03}, the redshift evolution of the CFE in Figure~\ref{fig:cfez} warrants a rough, two-step approximation in which $\Gamma=50\%$ for $z>0.7$ and $\Gamma=10\%$ for $z<0.7$. {As indicated above, the former of these two values is an upper limit due to the possible incompleteness of the high-redshift sample.} Since the (comoving) stellar mass density increased by a factor of $1.6$--$2$ since $z=0.7$ \citep[e.g.][]{pozetti07,marchesini08, ilbert10}, this gives a total cosmic integrated CFE of $\Gamma_{\rm univ}\leq30$--$35\%$. In other words, the model predicts that {up to} one third of all stars in the Universe once formed in bound stellar clusters, while the remainder originated in unbound associations.

\section{Discussion} \label{sec:disc}
In this section, we discuss the possible sources of uncertainty in the presented theoretical framework, as well as future observational tests and potential applications in observations, theory, and numerical simulations.

\subsection{Influence of parameters and model assumptions} \label{sec:assum}
The first test of the uncertainties in the presented model is to assess how the adopted parameters might influence the result. While the model does not rely on a very large number of parameters (see Table~\ref{tab:param}), and even though we have shown that the variation of the CFE due to individual parameters is generally minor (see e.g.~Figure~\ref{fig:cfeline}), it is worth checking how the results change if all parameters would conspire in the same direction.

\begin{figure}
\center\resizebox{8cm}{!}{\includegraphics{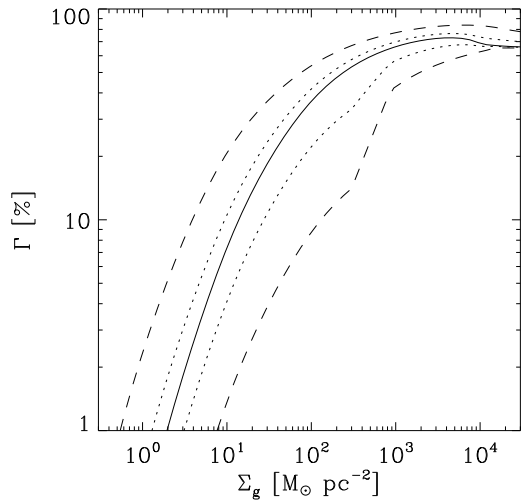}}\\
\caption[]{\label{fig:cfeparam}
      Worst-case-scenario uncertainty on the predicted cluster formation efficiency for the fiducial model (solid line). The dashed lines represent the input physics and parameters from Table~\ref{tab:param2}, which were chosen to minimize (lower dashed) or maximize (upper dashed) the CFE. Note that these parameter sets are highly improbable ($\ga2.9\sigma$), implying that the actual $1\sigma$ uncertainty is comparable to the dotted lines (see text).
                 }
\end{figure}

\begin{table}
 \centering
  \caption{Influence of the adopted parameters on the CFE.}\label{tab:param2}
  \begin{tabular}{@{}l c c c@{}}
  \hline
  Parameter & Minimum & Maximum & Impact on CFE \\
  (1) & (2) & (3) & (4) \\
 \hline
$\phi_P$ & 1 & 6 & $+$ \\
$\alpha_{\rm vir}$ & 1 & 2 & $+$ \\
$t_{\rm sn}$ & 2~Myr & 5~Myr & $+$ \\
$t$ & 5~Myr & 20~Myr & $-$ \\
$\phi_{\rm fb}$ & $0.032~{\rm cm}^2~{\rm s}^{-3}$ & $0.8~{\rm cm}^2~{\rm s}^{-3}$ & $+$ \\
$\epsilon_{\rm core}$ & 0.25 & 0.75 & $-$ \\
$f$ & 0.5 & 0.9 & $-$ \\
$g$ & 1 & 2 & $-$ \\
$\phi_{\rm sh}$ & 2 & 3 & $-$ \\
$\Sigma_{\rm GMC}^{\rm LG}$ & $30~\msun~{\rm pc}^{-2}$ & $300~\msun~{\rm pc}^{-2}$ & $-$\\
\hline
\end{tabular}
\end{table}
The typical variations or uncertainties of the model parameters is shown in Table~\ref{tab:param2}, the fourth column of which indicates whether the CFE increases ($+$) or decreases ($-$) with the parameter in question. The ranges are based on discussions in \citet{krumholz05}, \citet{gieles06}, \citet{heyer09}, \citet{portegieszwart10}, \citet{kruijssen11}, and on the appendices of this paper. Using the table, one can easily decide which extremes to choose in order to estimate the maximum variation of the CFE. Additional variation may come from the adopted star formation law (see Figures~\ref{fig:cfesflaw} and~\ref{fig:spatial}) and the inclusion of radiative feedback (see Appendix~\ref{sec:apprad}). We therefore consider two cases with the following changes to the fiducial model.
\begin{enumerate}
\item
The low-CFE case, including the prescription for radiative feedback from Appendix~\ref{sec:apprad} and adopting the minimum and maximum parameters of Table~\ref{tab:param2} leading to the lowest CFEs.
\item
The high-CFE case, using the \citet{krumholz05} star formation law of equation~(\ref{eq:ssfrff_km05}) and adopting the minimum and maximum parameters of Table~\ref{tab:param2} leading to the highest CFEs.
\end{enumerate}
The resulting extremes of the CFE are shown as a function of the gas surface density in Figure~\ref{fig:cfeparam}, together with the predicted CFE for the fiducial parameter set. The `uncertainty' around the fiducial $\Gamma$--$\Sigma_{\rm g}$ relation is of the same order as the scatter of the observations around the `typical' model in Figure~\ref{fig:cfesfrd}. While in principle this implies that differences between theory and observations could be accounted for by model uncertainties, we remind the reader that the discrepancies are smaller when choosing appropriate input variables for each individual galaxy (Figure~\ref{fig:compare}). This should be expected, because the two extreme cases of Figure~\ref{fig:cfeparam} are highly improbable. For illustration, if one interprets the ranges of Table~\ref{tab:param2} conservatively as $1\sigma$ limits, the chance that all ten parameters conspire to give the extremes of Figure~\ref{fig:cfeparam} is $0.16^{10}\sim10^{-8}$. Of course, this assumes that all parameters are independent and influence the CFE to an equal extent. In practice, the uncertainty on the CFE is mainly caused by three independent variables, being $\phi_P$, $t_{\rm sn}$ and $\epsilon_{\rm core}$. But even in this case, the probability of the extremes is $0.16^3\sim0.004$, which is a $2.9\sigma$ limit. Assuming a log-normal distribution of errors, the actual $1\sigma$ logarithmic uncertainty on the model predictions (dotted lines in Figure~\ref{fig:cfeparam}) should therefore be about $0.35$ times the logarithm of the extreme case. This uncertainty decreases with the gas surface density, from $\sigma_\Gamma\sim0.3$~dex at $\Sigma_{\rm g}\sim2~\msun~{\rm pc}^{-2}$ to $\sigma_\Gamma<0.1$~dex at $\Sigma_{\rm g}\sim10^3~\msun~{\rm pc}^{-2}$.

In the discussion so far, we solely focused on the uncertainty of the adopted model parameters. However, other assumptions that are physical in nature might also influence the results of this paper.
\begin{enumerate}
\item
Magnetic fields are not included explicitly, i.e.~there are no physical terms for magnetic fields in any of the equations. This means that the influence of magnetic fields on cluster formation cannot be studied using the theoretical framework of this paper. However, their influence is implicitly present to some degree in the assumption of an ISM overdensity PDF that is consistent with weak-field magnetohydrodynamics simulations, and likewise in the adopted specific star formation rate per free-fall time. The model predictions should thus not be strongly affected by any (future) explicit formulation. A possible, first-order modification of the theory to allow for strong magnetic fields would be to add a factor $\beta_0/(\beta_0+1)$ to the second term of equation~(\ref{eq:sigma}), where $\beta_0\equiv P_{\rm th}/P_{\rm mag}$ is the ratio of the thermal pressure to the magnetic pressure \citep{padoan11,molina12}. However, this does not account for the influence of a strong magnetic field on the physics of gravitational collapse.
\item
When considering individual galaxies, the overdensity PDF of the ISM might not follow the log-normal shape of equation~(\ref{eq:pdf}). It is known from numerical work that deviations arise mainly at low overdensities \citep[e.g.][]{wada07,tasker09}, which is in the range where the formation of stars and stellar clusters proceeds slowly. As a result, the impact on the absolute star formation rate and CFE are generally well within the other uncertainties.
\item
The model assumes that {star-forming regions reside} in gas discs that are in hydrostatic equilibrium. This could pose a problem when modeling star formation in galaxy mergers or in dynamic, high-redshift environments that might be out of equilibrium. Acknowledging this, it is important to note that when gas cools to form stars, the energy dissipation drives the formation of a disc \citep[e.g.][]{hopkins09b}. This means that while the presence of a disc might not be obvious on a global scale, the main loci of star formation {should follow a disc-like morphology}. This means that our assumption is satisfied at least locally.
\item
In our model, star formation {in intermediate-density regions} is halted by the large-scale obstruction of further gas inflow by supernova feedback. The role of different feedback mechanisms in truncating the star formation process is traditionally a very active topic in the literature, and hence alternatives such as stellar winds and radiative feedback have been discussed extensively \citep[e.g.][]{murray10}. It is important to note that the total energy inputs from these feedback mechanisms are in the same ballpark, which implies that the influence of this approach on the CFE is not major (see Appendix~\ref{sec:appfb} and~\ref{sec:apprad}). The description of feedback mainly serves the purpose of truncating star formation on a time-scale that is broadly consistent with observations. With $t_{\rm fb}\sim3$~Myr, this is indeed the case \citep[e.g.][]{portegieszwart10}.
\item
The cruel cradle effect is modelled by calculating a critical overdensity below which tidal perturbations are able to inhibit the collapse of a star-forming region. In doing so, we neglect the density increase of such regions during collapse and the correspondingly increased resistance to tidal disruption. This is especially relevant in view of the stochasticity of tidal perturbations, which are unlikely to start precisely at $t=0$ and hence do not act on the initial ISM density of a star-forming region, but on a (somewhat) higher density. The quoted impact of the cruel cradle effect is therefore an upper limit. If we consider the extreme case of only a single perturbation per free-fall time,\footnote{Even lower encounter rates would imply the cruel cradle effect is negligible due to the adiabatic dampening of equation~(\ref{eq:phiad}).} then on average regions have the potential to collapse halfway before experiencing their first tidal perturbation, causing at most a factor-of-eight density increase. This is similar in magnitude to the variation of the product $fg\phi_{\rm sh}$ in Table~\ref{tab:param2}, which contributes only a small fraction of the total uncertainty in Figure~\ref{fig:cfeparam}. We conclude that the treatment of tidal perturbations is adequate for the problem at hand.
\end{enumerate}

\subsection{Observational tests and applications} \label{sec:applobs}
Because the presented theory makes a number of clear assumptions and predictions, there are several possible ways in which (components of) it can be tested. The model can also be applied as a tool in interpreting observational data. Note that Fortran and IDL routines for calculating the CFE with our model are publicly available (see Appendix~\ref{sec:appsupp}).

\subsubsection{Observational verification with Gaia using kinematics}
On small spatial scales, the model postulates that the formation of bound stellar clusters proceeds hierarchically, reaching high SFEs in high-density regions. In this picture, the gas and stars decouple dynamically as the stellar component accretes and shrinks, allowing it to attain virial equilibrium while embedded in a cocoon of gas \citep{kruijssen12}. The direct implication is that the classical picture of early cluster disruption by infant mortality does not work as such. Gas expulsion does not unbind the dense, spherically symmetric systems that went through hierarchical merging and violent relaxation. Such systems should largely survive gas expulsion. Instead, gas removal only affects star-forming regions on the scales where they are still gas-rich and substructure is still present. This prevents the further merging of individual sub- or protoclusters. With kinematic data from {\it Gaia}, it is possible to derive from dispersed structure whether gas expulsion acted as classical infant mortality on a spherically symmetric cluster, or whether the stars still followed a hierarchical structure at the onset of the dispersal. In the former case, the velocity vectors would be radial and have a common origin, whereas in the latter case, the velocity vectors should be traceable to multiple centres and filamentary structure.

{\it Gaia} will also be able to measure the relative occurrence of the dispersion of stellar structure due to being naturally unbound, and dispersion due to the cruel cradle effect. Tidal perturbations induce structure in velocity space, in that the stellar velocity vectors tend towards the plane of the interaction \citep{kruijssen11b}. Such kinematic evidence would be visible until the stars reach the tidal boundary of the region, and allows one to quantify the fraction of early cluster disruption due to the cruel cradle effect. In the Milky Way, we estimate that while about half of all stellar structure is affected by tidal perturbations, a large fraction of this is already naturally unbound. Therefore, the cruel cradle effect is critical only for $\la1\%$ of all young stellar structure, and unbinds some $10\%$ of all naturally bound structure (see Table~\ref{tab:galaxies}).

Our model can also be considered using the relation between the local SFE $\epsilon$ (on any scale) and the fraction of stellar mass that will remain bound upon gas removal $\gamma$ (see Appendix~\ref{sec:appbound}). The model predicts a certain distribution of SFEs, which through the adopted $\gamma(\epsilon)$ relation yields a distribution of bound mass fractions. With {\it Gaia}, it will be possible to identify expanding haloes of escaping stars around young, unembedded clusters and associations. Correlating the masses of these haloes with the cluster masses will provide a direct measure of $\gamma$. The measured distribution of bound fractions should be similar to the predicted distribution of SFEs that follows from the properties of the ISM. Another prediction of the model is that expanding stellar haloes should be sparse or absent around massive and dense stellar clusters.\footnote{Low-mass, dense clusters could still have a substantial halo because of rapid evaporation due to two-body relaxation \citep{moeckel12}. Also note that due to the hierarchical nature of star and cluster formation, it is possible that a non-negligible mass fraction may form dispersed in the periphery of a compact cluster \citep[e.g.][]{bressert12}}

\subsubsection{Observing gas-embedded regions with ALMA}
The theory of this paper causally connects the properties of the ISM to the eventual population of bound stellar clusters. This is well-suited for verification with the sub-millimeter capacity of ALMA. Within the Milky Way, ALMA observations can trace the dynamical decoupling of gas and stars in protoclusters. It should also enable a statistical analysis of the protocluster population, and establish how the density PDF of bound stellar structure follows from the density PDF of the ISM. Specifically, the model predicts that the high-density tail of the density spectrum should promote bound cluster formation (see Figure~\ref{fig:xpdf}). With ALMA observations it is possible to witness and quantify this process as it takes place.

In extragalactic systems, the study of populations of protoclusters and their dynamical states will provide a direct measure of the dependence of cluster formation on the galactic environment. Our theory predicts that the key factor that sets the CFE is the gas surface density, with angular velocity and disc stability being the secondary variables. Are gravitationally bound protoclusters more numerous in high-density galaxies than in gas-poor galaxies? An obvious nearby example is M31, where we predict that only 2\% of all star formation eventually produces bound stellar clusters. Contrasting a sub-millimeter census of the protocluster population in this galaxy with a similar survey of systems like M83, M82 or NGC~3256 would be very valuable for our understanding of how the galactic environment influences the conversion of gas into bound stellar clusters.

\subsubsection{Application to observed cluster populations}
Another evident avenue for the application of the presented theory is in the study of observed cluster populations. Firstly, our framework provides a theoretical benchmark to test observational estimates of the CFE against. The model requires input variables that are easily obtained for large samples of galaxies\footnote{As mentioned throughout the paper, only the gas surface density is essential for calculating the CFE. The angular frequency and Toomre $Q$ parameter can be approximated if no observational input is available. The default $\Omega$ follows from equation~(\ref{eq:omega}), and $Q\sim1.5$ for regular disc galaxies.} and hence any new observed value of the CFE can be compared to the model prediction (see Table~\ref{tab:galaxies}). Secondly, this approach can also be reversed. When the star formation history and cluster age distribution are known, having an estimate for the CFE will enable one to constrain the cluster disruption law in that particular galaxy \citep[see e.g.][]{silvavilla11}. Thirdly, it is straightforward to convert observed cluster age distributions \citep[e.g.][]{gieles05,smith07} to star formation rates using the modelled CFE. A case in point is the recent study by \citet{ford12}, who use the cluster age distribution to derive extremely low star formation rate densities $\Sigma_{\rm SFR}\sim10^{-5}~\msun~{\rm yr}^{-1}~{\rm kpc}^{-2}$ in nearby elliptical galaxies, but do so assuming that all stars form in clusters. The model relation of Figure~\ref{fig:cfesfrd} shows that at such low surface densities only a minor fraction of all stars forms in clusters. A self-consistent solution is obtained by taking $\Gamma\sim0.01$, which increases the estimated star formation rate density to $\Sigma_{\rm SFR}\sim10^{-3}~\msun~{\rm yr}^{-1}~{\rm kpc}^{-2}$. This example illustrates that only when used together, the CFE and the cluster age distribution provide an accurate, independent tracer of star formation activity.

The CFE is also a potentially useful quantity for globular cluster studies. We have shown in \S\ref{sec:hubble} and Figure~\ref{fig:cfez} that the CFE was much higher ($\Gamma\sim50\%$) during the formation epoch of metal-rich globular clusters than it is in the local Universe ($\Gamma\sim10\%$). When studying globular cluster populations it is relevant to know that these objects represent the survivors of a cluster population that initially constituted half of all star formation, rather than just a minority. Combining this with prescriptions for cluster disruption, it is possible to derive the coeval stellar mass of present-day globular cluster populations, which refines their use as tracers of galaxy assembly.

\subsection{Future applications in theoretical and numerical work} \label{sec:applth}
The model is applicable to a range of theoretical and numerical problems. Fortran and IDL routines for calculating the CFE with our model are publicly available (see Appendix~\ref{sec:appsupp}).

\subsubsection{Models of star and planet formation}
There is some steadily accumulating evidence that the stellar initial mass function \citep[IMF, e.g.][]{kroupa01,chabrier03} is universal in the local Universe, and does not vary between clustered and dispersed star-forming regions \citep{dewit05,parker07,bressert12}. However, indirect IMF measurements in giant elliptical galaxies have suggested that these systems could have an excess of low-mass stars \citep[e.g.][]{vandokkum10,cappellari12}. Since giant ellipticals assembled at very high redshift \citep[e.g.][]{cimatti04,daddi05,bundy06,naab07} and at correspondingly high surface densities, the CFE must have been high as well. It is unclear whether a possible variation of the IMF is caused by the variation of the large-scale galactic environment, or by an influence of high-density clustered environments on the star formation process. The absence of IMF variations in Galactic stellar clusters favours the former \citep{bastian10}, but the possibility that the high-redshift conditions of cluster formation may have affected star formation cannot be ruled out. In such a case, the CFE would determine to what extent the IMF of an entire galaxy changes.

On the planet formation side, the influence of clustered star formation is more evident. It has been known for some time that circumstellar discs and planet formation are affected by external photoevaporation \citep{scally01,adams04} and dynamical encounters with other stars \citep{olczak06,dukes12,parker12,dejuanovelar12}. Because both effects occur on an encounter time-scale $t_{\rm enc}\propto\rho^{-1}$, external perturbations of planetary systems or circumstellar discs are more important in bound clusters, and hence in galaxies with high CFEs. Recent observational work suggests that environmental effects can truncate circumstellar discs above stellar number densities $\Sigma>10^{3.5}~{\rm pc}^{-2}$ on a $1$-Myr time-scale \citep{dejuanovelar12}. A natural extension of this argument would be that in galaxies with gas surface densities $\Sigma_{\rm g}\sim10^3~\msun~{\rm pc}^{-2}$ only few planets may form in relative isolation because $\Gamma\sim70\%$. When modeling the properties of the integrated planet population of such a galaxy or galactic region, it may be necessary to account for the fraction of stars (and planets) that was born in bound stellar clusters.

\subsubsection{The chemical composition of stellar clusters}
A range of recent papers focus on the chemical (in)homogeneity of globular and open clusters in the Milky Way \citep[e.g.][]{gratton04,randich06,desilva06,blandhawthorn10}. While homogeneity may be desirable to identify cluster remains in stellar streams (\citealt{desilva07,wyliedeboer10,blandhawthorn10}, although see \citealt{majewski12} for a counterexample), there appears to be a natural transition to chemically inhomogeneous clusters at higher masses, as evidenced by old globular clusters \citep{gratton04}. This trend seems to be qualitatively consistent with the picture presented in our model. For a given density and free-fall time shorter than the star formation time-scale, the spatial extent from which mass is hierarchically accumulated increases with GMC mass, thus allowing for a larger degree of inhomogeneity in the eventual cluster. Alternatively, it has been argued that the collapse time-scale increases weakly with the GMC mass \citep{blandhawthorn10}. This would enable self-enrichment and increasing chemical inhomogeneity at higher cluster masses, which is the prevailing explanation for the chemical inhomogeneities in globular clusters \citep[e.g.][]{gratton04}. However, we should remind the reader that this relies on the identification of mass quanta \citep[cf.][]{tan06}, whereas there is no intrinsic mass-scale present in the continuous formulation of this paper. Quantized clusters only appear after the star formation process has ended, the time of which depends principally on the density. As a result, a mass-dependence of the collapse time-scale can only be obtained from our model using additional assumptions about the relation between mass and density or radius. The above considerations imply that high-mass, low-density collapse should lead to the largest chemical inhomogeneities, provided that such inhomogeneity increases with both the spatial extent of the region and the star formation time-scale.

\subsubsection{Numerical simulations of star cluster populations} \label{sec:num}
Analytic models for the evolution of star cluster populations have recently been implemented in numerical simulations of galaxy formation and evolution \citep{prieto08,muratov10,kruijssen11,kruijssen12c}. This is a promising new avenue for investigating the origin of old globular cluster populations, and for relating the properties of these populations to the high-redshift galactic environments they formed in. So far, these simulations have only included a physical treatment of the dynamical evolution and disruption of stellar clusters. For cluster formation, these studies have needed to assume that all stars (or a certain arbitrary fraction thereof) formed in bound stellar clusters. This assumption is especially inconvenient because the properties of the surviving cluster population are strongly influenced by how the cluster disruption rate varies with space and time \citep{kruijssen11,bastian12}, which could be offset (or enhanced) by similar variations of the CFE. Numerical simulations that follow the star cluster population with a subgrid, semi-analytic model should thus benefit from the theoretical framework that is presented in this paper.

Contrary to observed galaxies, the state of the ISM in numerical simulations is ill-described by quantities such as the gas surface density, angular velocity and Toomre $Q$ parameter, which all rely on a suitable projection or choice of coordinate system. Instead, for implementation in numerical simulations it is more useful to consider strictly local quantities. The best way to do this is to discard the assumption of an equilibrium disc (see \S\ref{sec:disk}), and to determine a subgrid overdensity PDF of the ISM from the local gas density $\rho_{\rm loc}$, gas velocity dispersion $\sigma_{\rm loc}$, and sound speed of the cold ISM $c_{\rm s,loc}$. The local Mach number then follows as ${\cal M}_{\rm loc}=\sigma_{\rm loc}/c_{\rm s,loc}$, which means that the PDF of the overdensity $x_{\rm loc}\equiv\rho_{\rm g}/\rho_{\rm loc}$ is uniquely determined (see \S\ref{sec:ismpdf}). The conversion to a PDF of the absolute density $\rho_{\rm g}$ then proceeds simply through multiplication with $\rho_{\rm loc}$, which is determined by integrating over the smoothing kernel (for {\it Smoothed Particle Hydrodynamics} or SPH codes) or averaging over neighbouring cells (for grid codes). Most other steps in our model are formulated locally, except for the feedback-limited SFE $\epsilon_{\rm fb}$ and the critical overdensity for surviving the cruel cradle effect $x_{\rm cce}$. Both can be rewritten to be a function of the local variables introduced here. For the SFE we write
\begin{equation}
\label{eq:sfeloc}
\epsilon_{\rm fb,loc}=\frac{{\rm sSFR}_{\rm ff}}{t_{\rm ff}}\frac{t_{\rm sn}}{2}\left(1+\sqrt{1+\frac{4t_{\rm ff}\sigma_{\rm loc}^2}{\phi_{\rm fb}{\rm sSFR}_{\rm ff}t_{\rm sn}^2x_{\rm loc}}}\right) .
\end{equation}
The critical overdensity becomes
\begin{equation}
\label{eq:xcceloc}
x_{\rm cce,loc}=\frac{87.5\sqrt{\pi}Gfg\phi_{\rm sh}\Sigma_{\rm GMC}t}{\sigma_{\rm loc}}\phi_{\rm ad}(x_{\rm cce,loc}) ,
\end{equation}
where the GMC surface density can either be taken to be constant or be written as $\Sigma_{\rm GMC}=\max{\{\Sigma_{\rm GMC}^{\rm LG},\Sigma_{\rm g}\}}$ as in the rest of this paper. In that case, $\Sigma_{\rm g}$ can be estimated very roughly by again assuming an equilibrium disc and writing
\begin{equation}
\label{eq:sigmag}
\Sigma_{\rm g}\sim\sqrt{\frac{2\rho_{\rm loc}\sigma_{\rm loc}^2}{\pi G\phi_P}} .
\end{equation}

Note that the aforementioned models of \citet{prieto08} and \citet{kruijssen11} account for the tidal disruption of clusters from the moment they are born. In such a context, the cruel cradle effect implies that clusters have high chances of being disrupted at young ages, due to still residing in their dense natal environment. This is complementary to the manifestation of the cruel cradle effect that is treated in this paper, in which it prevents the global collapse of star-forming regions to form bound clusters. The best ways of connecting both forms in numerical work are either (1) to delay the tidal disruption of clusters until the time $t$ at which the CFE is evaluated, or (2) to exclude the cruel cradle effect in the calculation of the CFE (i.e.~$\Gamma/100\%=f_{\rm bound}$) without delaying the tidal disruption of clusters. Because lower-density regions are more easily disrupted, the first option underestimates the CFE by neglecting the density increase of collapsing regions (although this effect is minor, see \S\ref{sec:assum}). Conversely, the second option may overestimate the CFE if the modelled evolution of the cluster radius does not include the collapse of clusters during their formation, and their correspondingly lower densities at $t=0$.

The local and global prescriptions for the CFE are both included in publicly available Fortran and IDL routines (see Appendix~\ref{sec:appsupp}). This enables a straightforward implementation into existing numerical models.

\subsubsection{The feedback efficiency in cosmological simulations}
In numerical simulations of hierarchical galaxy formation and evolution, the injection of feedback energy into the ISM generally includes an efficiency factor $\chi_{\rm fb}<1$ that indicates the fraction of the energy that successfully couples to the gas. This feedback efficiency is often taken to be $\chi_{\rm fb}\sim0.05$ \citep[see the discussion in Appendix~\ref{sec:appfb} and, e.g.,][]{efstathiou00,abadi03,robertson05,dubois08}, but unfortunately, a detailed understanding of this parameter is still lacking \citep{silk97,maclow99,navarro00,dib06}. It is not hard to imagine that a systematic variation of the feedback efficiency during certain episodes of galaxy evolution could affect the properties of the resulting galaxy population \citep[e.g.][]{sales10}, and therefore it is relevant to investigate how small-scale physics may influence large-scale galactic outflows.

If feedback sources are clustered, then some fraction of the feedback energy is deposited within existing bubbles, which can allow the energy to accumulate at a rate higher than the cooling rate, and hence increases the feedback efficiency \citep[e.g.][]{strickland99,krause12}. Quantitatively, the relative change of the feedback efficiency for clustered ($\chi_{\rm fb,cl}$) as opposed to dispersed ($\chi_{\rm fb,0}$) feedback can be parameterized as $\phi_{\rm cl}\equiv\chi_{\rm fb,cl}/\chi_{\rm fb,0}$. According to the recent numerical work of \citet{krause12} $\phi_{\rm cl}\sim2$, indicating that feedback from clustered star formation is more effective than if it were dispersed. Combining this with the model of the present paper, it is possible to formulate a composite feedback efficiency from the clustered and dispersed parts of star formation. When simulating the assembly of galaxies and their evolution, the CFE can easily be calculated for each star-forming particle (in SPH) or cell (in a grid code) as detailed in \S\ref{sec:num}. The composite feedback efficiency $\chi_{\rm fb}$ then follows to first order by noting that a fraction $\Gamma$ of the particle (or cell) mass is clustered and a fraction $(1-\Gamma)$ is dispersed:
\begin{eqnarray}
\label{eq:chifb}
\nonumber \chi_{\rm fb}&=&(1-\Gamma)\chi_{\rm fb,0}+\Gamma\phi_{\rm cl}\chi_{\rm fb,0} \\
&=&\left[1+(\phi_{\rm cl}-1)\Gamma\right]\chi_{\rm fb,0} .
\end{eqnarray}
The maximum variation of $\chi_{\rm fb}$ is a factor of $\phi_{\rm cl}$ per definition. While for $\phi_{\rm cl}\sim2$ this is not substantial, the systematic rise of the CFE at high densities (and star formation rates) could imply that the galaxy population is affected over the course of a Hubble time. With equation~(\ref{eq:chifb}) it will be straightforward to investigate this.\footnote{Note that this prescription only applies to numerical resolutions at which the clustering of star formation is not resolved. At higher resolution, the influence of clustered star formation may begin to emerge naturally.}

\subsubsection{Semi-analytic galaxy modeling}
Galaxy populations are often modelled semi-analytically to alleviate the computational effort required for the numerical simulation of their formation \citep[e.g.][]{white91,kauffmann93,somerville99,cole00,bower06,guo11}. These analyses are covering the characteristics of the modelled galaxies in increasing detail \citep[see e.g.][]{guo11}. It seems only natural that the origin and evolution of the star cluster populations hosted by the modelled galaxies will be included in future work. This will enable an understanding of how the co-formation and co-evolution of (old) star cluster populations and galaxies may eventually produce the present-day cluster systems across the entire galaxy mass range \citep[for typical observables of interest, see e.g.][]{peng06,jordan07,peng08,burkert10}.

A self-consistent analytic model for cluster populations that can be implemented in semi-analytic models of galaxy formation requires (1) a prescription for the formation of cluster populations and (2) a prescription for their dynamical evolution. Both components of the problem are still unsolved. While the theory of the CFE presented in this paper provides the fraction of star formation occurring in bound stellar clusters, the initial characteristics of the individual clusters within the population are yet to be understood in more detail. Their mass spectrum can be taken from observations \citep{portegieszwart10} and has been shown to arise from the hierarchical nature of the ISM \citep{elmegreen96}, but the understanding of the corresponding upper \citep[e.g.][]{gieles06b,bastian08,larsen09} and lower \citep[e.g.][]{lada03,moeckel12} cluster mass limits requires more work, as does the distribution of initial cluster radii \citep[e.g.][]{larsen04b}. These quantities all need to be described with reasonable accuracy to enable the inclusion of cluster formation and disruption in a single semi-analytic framework. The evolution of cluster populations due to dynamics should be addressed by building on more detailed numerical modeling, both accounting for the internal cluster dynamics \citep[e.g.][]{baumgardt03,kruijssen09c,gieles11b} and the disruptive influence of the galactic environment \citep{gieles06,kruijssen11}. While some of the above ingredients can only be roughly estimated at present, the recent progress of the field is encouraging. By addressing the above questions, a first inclusion of cluster populations in semi-analytic galaxy population modeling should become possible in the next couple of years.

\section{Conclusions} \label{sec:concl}
We have described and applied an analytic theory of the cluster formation efficiency (CFE), i.e.~the fraction of star formation occurring in bound stellar clusters. The conclusions of the paper are as follows.
\begin{enumerate}
\item
The presented theory of cluster formation provides a framework in which the galaxy-scale environment is related to the properties of star-forming regions. Bound stellar clusters naturally arise from the high-density end of the density spectrum of the interstellar medium (ISM). Due to short free-fall times, these high-density regions can achieve high star formation efficiencies (SFEs). This makes them insensitive to gas expulsion and enables them to form bound clusters. In regions of lower density, the SFE is lower and the collapse into spherically symmetric structure is not completed. As a result, such regions (or associations) remain gas-rich and are unbound upon gas expulsion. (\S\ref{sec:ismpdf}--\ref{sec:fbound})
\item
In the picture described here, gas expulsion does not affect centrally concentrated stellar clusters like in the classical scenario of `infant mortality', but inhibits the merging of hierarchically structured stellar groups and associations. (\S\ref{sec:fbound} and Appendix~\ref{sec:appbound})
\item
Additionally, we have included the `cruel cradle effect', which represents the tidal perturbation of star-forming regions by surrounding GMCs. This is shown to give a second-order decrease of the CFE in normal disc galaxies, and is most important in high-density galaxies. (\S\ref{sec:cce})
\item
The CFE is obtained by integrating the probability distribution function (PDF) of the densities at which star formation proceeds, and the corresponding part thereof that is both bound locally and sufficiently dense to survive the cruel cradle effect. (\S\ref{sec:cfe} and Figure~\ref{fig:xpdf})
\item
Our theoretical framework allows the prediction of the CFE as a function of galaxy-scale observables: the gas surface density $\Sigma_{\rm g}$, the angular velocity $\Omega$, and the Toomre $Q$ parameter. For use in numerical simulations, a local formulation is also given, which uses the gas volume density $\rho_{\rm loc}$, the gas velocity dispersion $\sigma_{\rm loc}$, and the sounds speed $c_{\rm s,loc}$. (\S\ref{sec:cfe} and \S\ref{sec:num})
\item
The model predicts that only some fraction (1--70\%) of all star formation results in bound stellar clusters. This CFE increases with the gas surface density of the galaxy, because in denser galaxies a larger fraction of the ISM is pushed into the high-density regime where bound cluster formation occurs. However, at surface densities $\Sigma_{\rm g}\ga10^{3}~\msun~{\rm pc}^{-2}$ the CFE is limited by the cruel cradle effect. For $\Sigma_{\rm g}\ga10^{2.5}~\msun~{\rm pc}^{-2}$, the CFE decreases with increasing angular velocity $\Omega$, while at lower densities it is generally insensitive to $\Omega$. The CFE increases with Toomre $Q$, and hence with the stability of the galaxy disc. (\S\ref{sec:cfesflaw}, \S\ref{sec:cfeQ}, and \S\ref{sec:summparam})
\item
We find remarkable agreement between our theoretical model and the observed relation between the CFE and the star formation rate density $\Sigma_{\rm SFR}$ in dwarf, spiral and starburst galaxies. Also when modeling individual galaxies, the model predictions are in excellent agreement with the observed CFEs. (\S\ref{sec:obs})
\item
The model is applied to investigate the spatial variation of the CFE within galaxies, using parameter sets that resemble the Milky Way, M31, and M51. At the low densities of the Milky Way and M31, the CFE simply follows the relatively flat surface density profile of the gas. However, for the exponential gas density profile of M51, the CFE increases towards the centre of the galaxy, but (at high densities) does so less than the gas density profile. (\S\ref{sec:spatial})
\item
We use samples of nearby and high-redshift galaxy discs and starbursts to predict the evolution of the CFE across a Hubble time of star formation. For galaxy discs, the CFE is found to decrease from $\Gamma=30$--$70$\% to $\Gamma=5$--$10$\% between redshifts $z\sim2$ and $z=0$, whereas starburst galaxies exhibit a moderate decrease of $\Gamma=20$--$60$\% to $\Gamma=10$--$50$\% over the same redshift range. High-redshift disc galaxies may thus have been the most efficient cluster factories in the history of the Universe, although presently the CFE is generally higher in starburst galaxies than in discs. We estimate that {up to} $\Gamma_{\rm univ}=30$--$35$\% of all stars in the Universe was formed in bound stellar clusters. (\S\ref{sec:hubble})
\item
The predictions of our theoretical framework are well-suited for verification with {\it Gaia} or ALMA, and also for application to observational studies of cluster populations. We provide specific recipes for use of the model in theoretical and numerical work, such as numerical simulations of galaxies and their star cluster populations, the feedback efficiency in cosmological simulations, and semi-analytic galaxy models. (\S\ref{sec:applobs} and \S\ref{sec:applth})
\end{enumerate}
We have presented a new model that describes how populations of bound stellar clusters form, and that predicts which fraction of all star formation is constituted by these clusters (the cluster formation efficiency or CFE). The model enables the calculation of the CFE over a range of galactic environments, and shows that star clusters are no fundamental unit of star formation, but instead are a possible outcome.

\section*{Acknowledgments}
The anonymous referee is thanked for a thoughtful report. I am indebted to Nate Bastian, Eli Bressert, Andreas Burkert, Cathie Clarke, James Dale, Bruce Elmegreen, Guinevere Kauffmann, Martin Krause, Mark Krumholz, Henny Lamers, Steve Longmore, Nora L\"{u}tzgendorf, Thorsten Naab, Jeremiah Ostriker, Laura Sales, Esteban Silva-Villa, Todd Thompson, Stefanie Walch, and Simon White for stimulating discussions and/or helpful comments on the paper. The SAO/NASA Astrophysics Data System greatly accelerated the development of this paper. I am grateful to the Aspen Center for Physics for their hospitality and to the National Science Foundation for support, Grant No.~1066293.

\appendix
\section{Routines for calculating the cluster formation efficiency in Fortran or IDL} \label{sec:appsupp}
Fortran and IDL routines for calculating the CFE with the presented model are publicly available at http://www.mpa-garching.mpg.de/cfe. For both languages, two versions of the model are provided. The first version of the routines adopts the global formulation that is followed throughout this paper, with as required input variables the gas surface density of the galaxy $\Sigma_{\rm g}$, the \citep{toomre64} $Q$ parameter, and the angular velocity $\Omega$. Note that $Q$ and $\Omega$ are optional. If they are left unspecified, the routines set $Q=1.5$ and define $\Omega$ according to equation~(\ref{eq:omega}). Additional optional input parameters allow the user to adjust the star formation law or several of the parameters from Table~\ref{tab:param}. The second version of the routines adopts the local formulation from \S\ref{sec:num}, with as required input variables the local density $\rho_{\rm loc}$, the local gas velocity dispersion $\sigma_{\rm loc}$, and the sound speed of the cold ISM $c_{\rm s,loc}$.

\section{The feedback time-scale and a gradual truncation of feedback} \label{sec:appfb}
This appendix considers the scenario in which the inflow of gas into a star-forming region is halted by supernova feedback. The energy injection rate into the cold ISM per unit stellar mass $\phi_{\rm fb}$ due to supernovae can be estimated by writing it as
\begin{equation}
\label{eq:phifbapp}
\phi_{\rm fb}=\frac{\chi_{\rm fb}\phi_{\rm sn}E_{\rm sn}}{\Delta t} ,
\end{equation}
where $\chi_{\rm fb}$ is the fraction of the supernova energy that couples to the cold ISM, $\phi_{\rm sn}$ is the number of supernovae per unit stellar mass, $E_{\rm sn}=10^{51}$~erg is the typical energy per supernova, and $\Delta t\equiv t_{\rm sn}^{\rm last}-t_{\rm sn}$ represents the time interval during which supernovae occur, with $t_{\rm sn}^{\rm last}$ the time at which the last supernova explodes, and $t_{\rm sn}$ the time of the first supernova as in \S\ref{sec:sfe}. This assumes that the supernova rate is approximately constant during $\Delta t$. The efficiency of the kinetic coupling of the supernova energy to the ISM $\chi_{\rm fb}$ has been the subject of a wide range of studies \citep[e.g.][]{silk97,efstathiou00,navarro00,abadi03,robertson05,dib06}. It is smaller than unity due to (1) rapid cooling in the cold ISM and (2) the porosity of the hierarchically structured ISM. In the literature, values in the range $\chi_{\rm fb}=0.01$--$0.25$ have been reported, which illustrates that this quantity is the most uncertain of all variables in equation~(\ref{eq:phifbapp}). The remainder of this appendix assumes $\chi_{\rm fb}=0.05$, which is the logarithmic mean of the quoted range and is a commonly used value in numerical simulations. Physically, this corresponds to a scenario in which about 10\% of the supernova energy goes into the kinetic energy of an expanding shell \citep{thornton98} and the total surface angle of the porous, cold ISM takes up about 50\% of the spherical surface around the star-forming region.

\begin{figure*}
\center\resizebox{14cm}{!}{\includegraphics{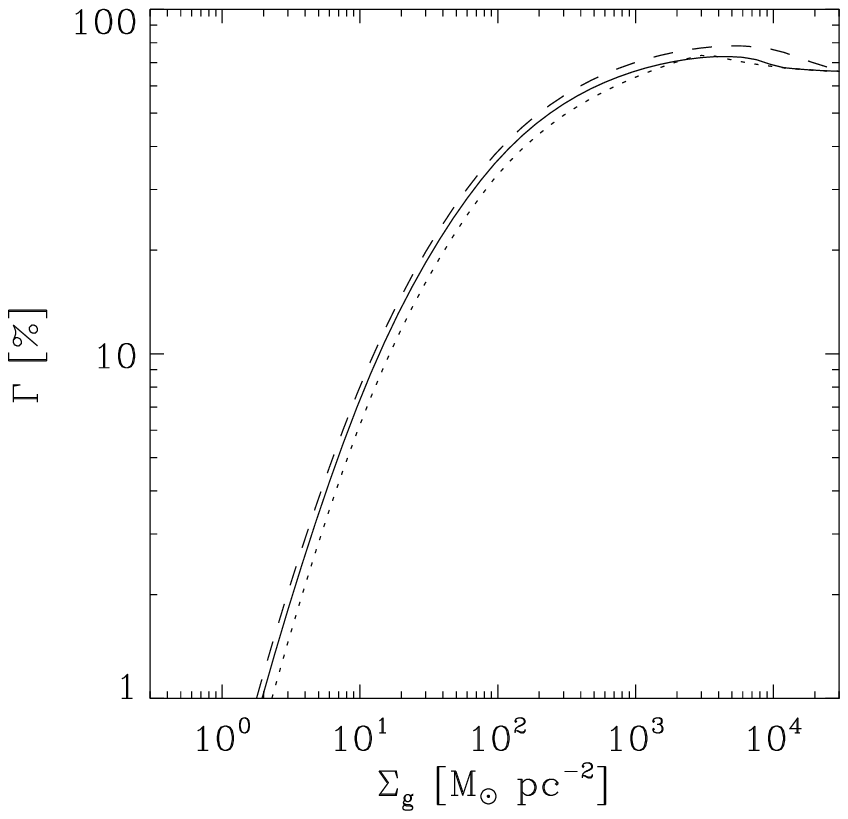}\includegraphics{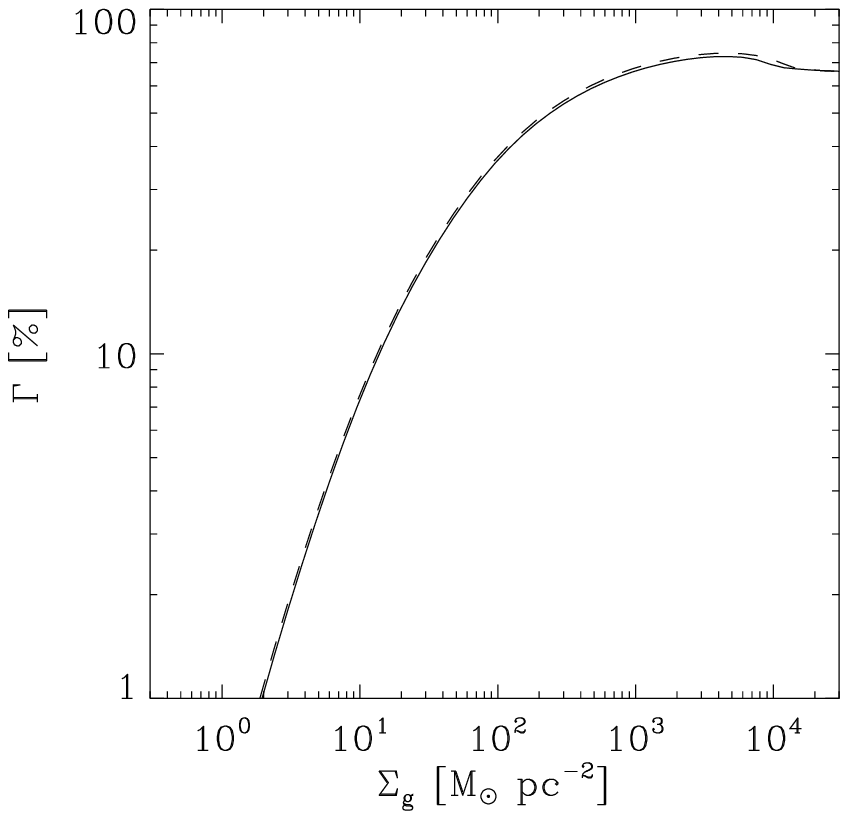}}\\
\caption[]{\label{fig:cfefb}
      {\it Left}: Influence of $\phi_{\rm fb}$ on the global cluster formation efficiency (CFE) as a function of the gas surface density of the disc $\Sigma_{\rm g}$, assuming $Q=1.5$ and writing $\Omega$ as a function of $\Sigma_{\rm g}$ according to equation~(\ref{eq:omega}). The dotted, solid, and dashed lines show the relation for $\phi_{\rm fb}=\{0.016,0.16,1.6\}~{\rm cm}^2~{\rm s}^{-3}$, respectively, hardly affecting the CFE despite varying by two orders of magnitude. {\it Right}: Same as the left-hand panel, but showing the influence on the CFE of a gradual truncation of star formation by feedback. The solid line shows the abrupt truncation as in \S\ref{sec:sfe}, while the dashed line represents a linear decrease of the specific star formation rate per free-fall time, obtained by substituting $t_{\rm eq}\rightarrow t_{\rm eq}/2$ in equation~(\ref{eq:sfefb}). The CFEs are almost indistinguishable.
                 }
\end{figure*}
The number of supernovae per unit stellar mass is obtained by assuming a stellar initial mass function (IMF) and integrating over the appropriate mass interval. For a \citet{salpeter55} IMF ${\rm d} n/{\rm d} m\propto m^{-2.35}$ in the mass range 0.1--$100~\msun$ and a minimum mass of $8~\msun$ for a star to result in a supernova, this implies
\begin{equation}
\label{eq:phisnapp}
\phi_{\rm sn}=\frac{\int_{8}^{100} m^{-2.35}{\rm d} m}{\int_{0.1}^{100} m^{-1.35}{\rm d} m}=7.4\times10^{-3}~\msun^{-1} .
\end{equation}
The time interval $\Delta t$ during which the supernovae occur is set by $t_{\rm sn}=3$~Myr (see \S\ref{sec:sfe}) and $t_{\rm sn}^{\rm last}=40$~Myr \citep[for the Padova stellar evolutionary isochrones, see][]{marigo08}. This range is very insensitive to the metallicity, as $t_{\rm sn}^{\rm last}$ only varies by 0.05~dex in the metallicity range $z=0.0004$--$0.03$.

Substituting all variables into equation~(\ref{eq:phifbapp}) now provides $\phi_{\rm fb}=0.16~{\rm cm}^2~{\rm s}^{-3}=3.2\times 10^{32}~{\rm erg}~{\rm s}^{-1}~\msun^{-1}$, with a total uncertainty of $\sim1.5$~dex. While this might sound substantial, its effect on the CFE is almost negligible, as is shown in the left-hand panel of Figure~\ref{fig:cfefb}. The curves represent the CFE as a function of gas surface density for different $\phi_{\rm fb}$ spanning two orders of magnitude, yet the variation of the CFE is always smaller than 0.1~dex. The reason is that $\phi_{\rm fb}$ influences the star formation efficiency at low overdensities. It mainly controls the critical overdensity where the feedback-induced truncation of star formation becomes so inefficient that it is still ongoing at the time of observation. This critical point happens to lie in the overdensity range where the PDF is quite flat and where no bound stellar clusters are formed due to the cruel cradle effect. The overdensities where bound clusters do form is characterized by such powerful feedback that typcally $t_{\rm eq}\ll t_{\rm sn}$ regardless of the value of $\phi_{\rm fb}$, or by such short free-fall times that star formation is near-optimal, with $\epsilon\approx\epsilon_{\rm core}$. In other words, the uncertainty on $\phi_{\rm fb}$ may translate into a substantial shift of the critical overdensity that separates $\epsilon_{\rm inc}$ from $\epsilon_{\rm fb}$ in equation~(\ref{eq:sfefinal}), but this hardly influences the integration of equation~(\ref{eq:fbound}) and thus leaves the CFE unaffected.

So far, the truncation of star formation by supernova feedback has been considered to be instantaneous once pressure equilibrium between the feedback energy and the cold ISM has been reached. As a first-order correction for this assumption, it is possible to adopt a linear decrease of the specific star formation rate per free-fall time between $t_{\rm sn}$ and $t_{\rm fb}$. This is mathematically equivalent to writing $t_{\rm eq}\rightarrow t_{\rm eq}/2$ in equation~(\ref{eq:sfefb}). Clearly, this reflects a much smaller change than the above variation of $\phi_{\rm fb}$, and indeed the right-hand panel of Figure~\ref{fig:cfefb} shows that the CFE for a gradually decreasing star formation rate is nearly identical to the case of instantaneous truncation.

\section{The truncation of star formation by radiation pressure} \label{sec:apprad}
In \S\ref{sec:sfe}, the local SFE is determined by allowing star formation to proceed until a certain point at which it is halted by feedback.\footnote{If this takes longer than the age at which the cluster formation efficiency is determined, star formation is considered to be still ongoing. If star formation is completed before feedback can halt star formation, the local SFE is optimal (see \S\ref{sec:sfe}).} In the main body of this paper, this is assumed to happen due to supernova feedback. However, recent work has shown that actually radiative feedback may be the responsible mechanism, especially on small scales \citep{thompson05,murray10,fall10,dale12}. To verify to what extent this can influence the CFE, we focus on the truncation of star formation by radiation pressure in this appendix.

\begin{figure}
\center\resizebox{6.8cm}{!}{\includegraphics{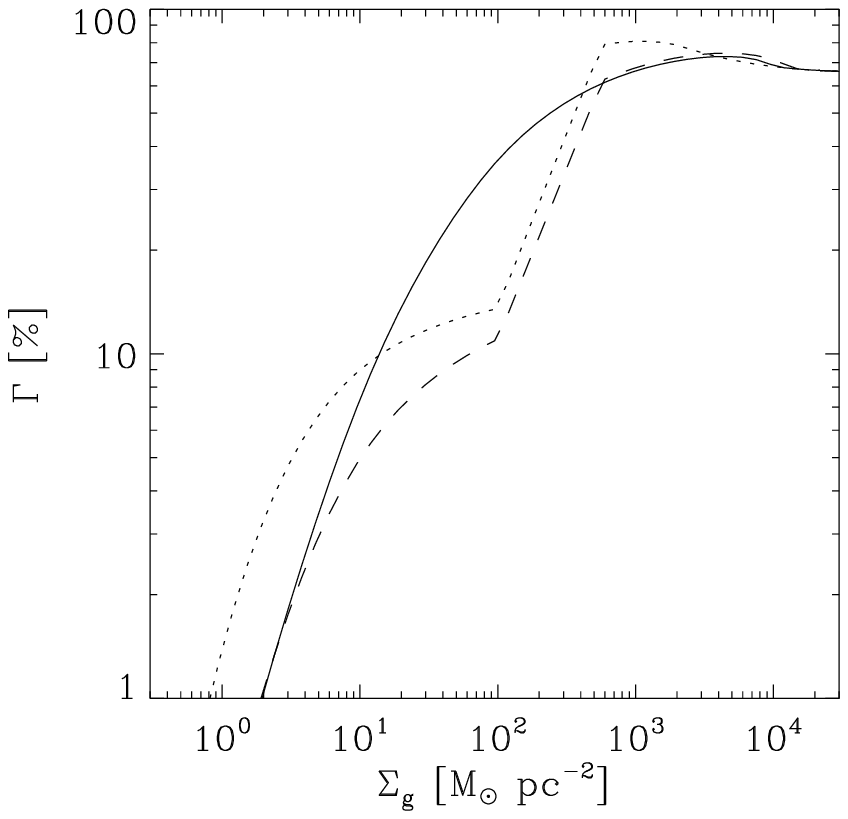}}\\
\caption[]{\label{fig:cferad}
      Influence of radiative feedback on the global cluster formation efficiency (CFE) as a function of the gas surface density of the disc $\Sigma_{\rm g}$, assuming $Q=1.5$ and writing $\Omega$ as a function of $\Sigma_{\rm g}$ according to equation~(\ref{eq:omega}). The solid line shows the fiducial model with supernova feedback, while the result for radiative feedback only is given by the dotted line. The dashed line represents the case in which star formation is halted by whichever mechanism generates the required feedback energy first, and thus combines both feedback processes.
                 }
\end{figure}

We require pressure equilibrium between radiation pressure $P_{\rm rad}$ and turbulent pressure within a GMC $P_{\rm GMC}$. The radiation pressure reads
\begin{eqnarray}
\label{eq:prad}
\nonumber P_{\rm rad}&=&(1+\phi_{\rm tr}\tau)\frac{{\cal F}}{c}\sim(1+\phi_{\rm tr}\kappa_{\rm R}\Sigma_{\rm GMC})\frac{\Psi M_\star}{4\pi R_{\rm GMC}^2c}\\
&=&(1+\phi_{\rm tr}\kappa_{\rm R}\Sigma_{\rm GMC})\frac{\epsilon_{\rm rad}\Psi\Sigma_{\rm GMC}}{4c} ,
\end{eqnarray}
where $\tau\sim\kappa_R\Sigma_{\rm GMC}$ is the optical depth, $\phi_{\rm tr}\equiv f_{\rm tr}/\tau\sim0.2$ \citep{krumholz12b} is a constant that indicates the fraction of infrared radiation that is trapped at an optical depth $\tau=1$, ${\cal F}$ is the radiative flux, $c$ is the speed of light, $\Psi\sim0.3~{\rm m}^2~{\rm s}^{-3}$ is the light-to-mass ratio \citep{thompson05}, $M_\star$ is the stellar mass, $R_{\rm GMC}$ is the GMC radius, $\epsilon_{\rm rad}$ is the SFE when star formation is limited by radiative feedback, and $\kappa_{\rm R}$ is the Rosseland-mean dust opacity \citep[cf.][]{thompson05,murray10}, which for $T<200~{\rm K}$ is expressed as
\begin{eqnarray}
\label{eq:kappaR}
\kappa_{\rm R}&\sim&\kappa_0\left(\frac{T}{{\rm K}}\right)^2=\kappa_0\sqrt{\frac{\tau{\cal F}}{\sigma_{\rm SB}}}\\
\nonumber &=&\kappa_0\sqrt{\frac{\kappa_R\epsilon_{\rm rad}\Psi\Sigma_{\rm GMC}^2}{4\sigma_{\rm SB}}}
\rightarrow\kappa_R\sim\frac{\kappa_0^2\epsilon_{\rm rad}\Psi\Sigma_{\rm GMC}^2}{4\sigma_{\rm SB}} ,
\end{eqnarray}
where $\kappa_0\sim2.4\times10^{-5}~{\rm m}^2~{\rm kg}^{-1}~{\rm K}^{-2}$ is a proportionality constant and $\sigma_{\rm SB}$ is the Stefan-Boltzmann constant. The turbulent pressure in a GMC is taken to be
\begin{equation}
\label{eq:pgmc}
P_{\rm GMC}=\frac{\pi}{2}G\Sigma_{\rm GMC}^2 .
\end{equation}
By demanding pressure equilibrium $P_{\rm rad}=P_{\rm GMC}$, equations~(\ref{eq:prad})--(\ref{eq:pgmc}) can be solved for the SFE $\epsilon_{\rm rad}$. This gives a critical SFE above which radiation pressure inhibits further star formation:
\begin{equation}
\label{eq:sferad}
\epsilon_{\rm rad}=\frac{2\sigma_{\rm SB}}{\phi_{\rm tr}\kappa_0^2\Psi\Sigma_{\rm GMC}^3}\left(\sqrt{1+\frac{2\pi c G\phi_{\rm tr}\kappa_0^2\Sigma_{\rm GMC}^4}{\sigma_{\rm SB}}}-1\right) .
\end{equation}
In the context of our model, $\epsilon_{\rm rad}$ sets the maximum SFE if it is smaller than the core efficiency $\epsilon_{\rm rad}<\epsilon_{\rm core}$ (see \S\ref{sec:sfe}). For $\Sigma_{\rm GMC}=\Sigma_{\rm GMC}^{\rm LG}=100~\msun~{\rm pc}^{-2}$, we indeed obtain $\epsilon_{\rm rad}\sim0.1<\epsilon_{\rm core}$. The resulting CFE is shown for the entire density range in Figure~\ref{fig:cferad} and is compared to the fiducial model of \S\ref{sec:sfe}. Radiative feedback strongly limits the SFE and CFE until a density $\Sigma_{\rm g}=\Sigma_{\rm GMC}^{\rm LG}$, above which the GMC density changes from being approximately constant to $\Sigma_{\rm GMC}=\Sigma_{\rm g}$. As shown by equation~(\ref{eq:sferad}), the SFE then scales as $\epsilon_{\rm rad}\propto\Sigma_{\rm GMC}\propto\Sigma_{\rm g}$ if the interstellar medium (ISM) is optically thin, and as $\epsilon_{\rm rad}\propto\Sigma_{\rm GMC}^{-1}\propto\Sigma_{\rm g}^{-1}$ if the ISM is optically thick. This makes the SFE (and hence the CFE) peak at $\Sigma_{\rm g}\sim10^{3.5}~\msun~{\rm pc}^{-2}$. At higher densities, the cruel cradle effect and radiative feedback in an optically thick ISM both decrease the CFE.

While the details of the CFE--$\Sigma_{\rm g}$ relation are affected by assuming that radiative feedback halts star formation, the overall behavior is comparable to the fiducial model of this paper. We conclude that while the effects should be kept in mind (see \S\ref{sec:sfrd} and \S\ref{sec:assum}), there is no immediate need to adjust the fiducial model.

\section{The bound stellar fraction and the star formation efficiency} \label{sec:appbound}
In \citet{kruijssen12}, we used a minimum spanning tree \citep[MST, see e.g.][]{zahn71,maschberger10,kirk11} to identify subclusters in the \citet{bonnell08} simulation of turbulent fragmentation. This technique is commonly used in the analysis of structure in star-forming regions due to its advantage of not imposing any symmetry on the data set. The MST is the unique connection of all points such that there are no closed loops and the total length of all connecting lines is minimized. Subclusters are then identified by adopting a break distance $d_{\rm break}$, which is the maximum distance between connected points. Any connecting lines longer than $d_{\rm break}$ are removed, which breaks the MST of an entire star-forming region up into individual subclusters. The fiducial break distance of $d_{\rm break}=0.035$~pc in \citet{kruijssen12} was chosen to result in subclusters that are comparable to those identified by eye, but a range of $d_{\rm break}=0.020$--$0.100$~pc was used to validate the results. Additionally, a minimum of 12 stars per subcluster was postulated.

\begin{figure*}
\center\resizebox{14cm}{!}{\includegraphics{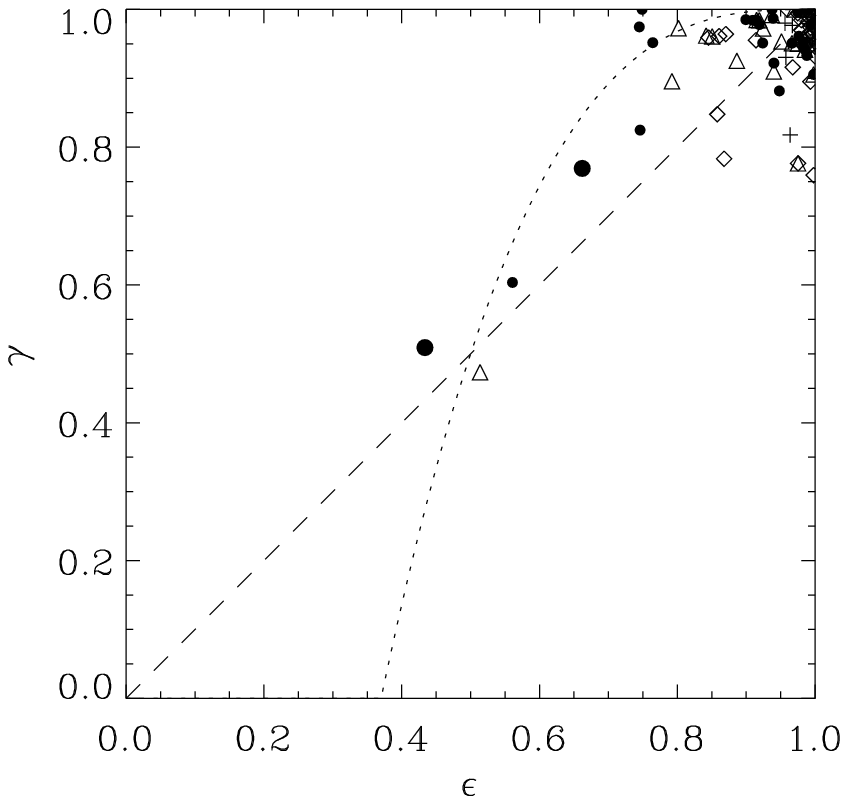}\includegraphics{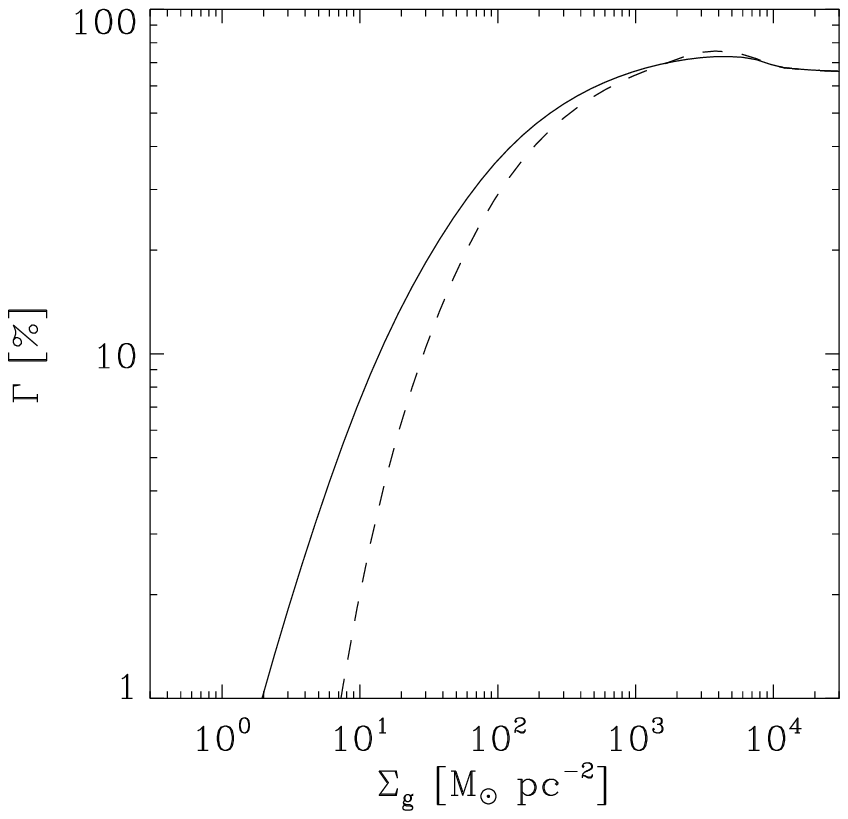}}\\
\caption[]{\label{fig:cfelocal}
      {\it Left}: Local cluster formation efficiency $\gamma$ as a function of the local star formation efficiency $\epsilon$ in the \citet{bonnell08} simulation. The symbols indicate the values for subclusters that were identified in the simulation with different minimum spanning tree break distances -- plus signs, diamonds, triangles and dots correspond to $d_{\rm break}=\{0.020,0.040,0.075,0.100\}$~pc, respectively, with big dots indicating the entire simulation (see text). The dashed line shows the 1:1 relation, and the dotted curve gives a parameterized $\gamma(\epsilon)$ relation that is typically found in equilibrium-type, spherically symmetric simulations of gas expulsion \citep[e.g.][]{baumgardt07}. {\it Right}: Influence of the choice of $\gamma(\epsilon)$ relation on the global cluster formation efficiency $\Gamma$ as a function of the gas surface density of the disc $\Sigma_{\rm g}$, assuming $Q=1.5$ and writing $\Omega$ as a function of $\Sigma_{\rm g}$ according to equation~(\ref{eq:omega}).
                 }
\end{figure*}

In this appendix, the subcluster data from \citet{kruijssen12} are revisited to determine how the local CFE is related to the local SFE in a hierarchically structured star-forming region. To this end, the stellar-to-total mass ratios within the half-mass radii of the subclusters represent the SFE $\epsilon$, while the fraction of the stars that is bound to the subcluster when ignoring the gravitational potential of the gas is considered to reflect the local CFE $\gamma(\epsilon)$. The result is shown for break distances of $d_{\rm break}=\{0.020,0.040,0.075,0.100\}$~pc in the left panel of Figure~\ref{fig:cfelocal}, and includes the subclusters at times $t_1=0.442$~Myr (briefly after the onset of star formation) and $t_2=0.641$~Myr (after one global free-fall time). The figure also shows data points for which the calculation was applied to the whole box of the simulation -- the largest (and thus most gas-rich) scale on which the CFE can be obtained.

Figure~\ref{fig:cfelocal} shows that the subclusters follow a single trend of
\begin{equation}
\label{eq:cfelocalapp}
\gamma(\epsilon)=\epsilon ,
\end{equation}
independently of the time, break distance or location. The scatter around the relation is modest, about 30\%. A continuous relation such as the figure seems to suggest, implies that the fraction of the stellar mass that ends up being bound is given by the local stellar mass fraction itself. There is no obvious reason as to why the $\gamma(\epsilon)$ relation would be linear, let alone why $\gamma$ and $\epsilon$ are simply approximately equal. Qualitatively speaking, a high local stellar mass fraction translates into a higher gravitational attraction of stars towards the stellar loci, where the resulting increase of the stellar density may lead to the runaway accretion and gas depletion that was found by \citet{kruijssen12} and \citet{girichidis12}. But again, the details are not clear {\it a priori}, and it is evident that additional study is needed to further elucidate the $\gamma(\epsilon)$ relation in a hierarchical interstellar medium.

The $\gamma(\epsilon)$ relation that is found in simulations of the influence of gas expulsion on star cluster survival with spherically symmetric initial conditions, in which the gas is modelled with an analytic background potential \citep{geyer01,boily03,goodwin06,baumgardt07}, differs from the relation that is obtained here. In such simulations, the central concentration of the gas implies a threshold SFE below which a stellar cluster is unable to remain bound after gas expulsion. Because these are equilibrium models, such a gas-rich star-forming region is then also unable to produce locally bound agglomerates of stars, implying that below the threshold SFE they predict $\gamma=0$. This implies that the transition from a zero CFE to almost 100\% is more sudden, and is generally found to occur in the 30-50\% SFE range \citep[e.g.][]{baumgardt07}. It should be noted that this is exactly the range that is not well-resolved by the above analysis of the \citet{bonnell08} simulation. However, the behavior of $\gamma(\epsilon)$ at SFEs of $\epsilon>0.5$ differs as well. Due to the sudden transition from $\gamma=0$ to $\gamma=1$ in spherically symmetric simulations, they show an elevated $\gamma(\epsilon>0.5)$ with respect to the relation of equation~(\ref{eq:cfelocalapp}).

From a physical perspective it is preferable to adopt the hierarchical $\gamma(\epsilon)$ relation over the spherically symmetric one, but to assess the impact of this choice, the influence of the two different $\gamma(\epsilon)$ relations on the CFE is shown in the right-hand panel of Figure~\ref{fig:cfelocal}. For this purpose, the $\gamma(\epsilon)$ relation for spherically symmetric clusters is parameterized with a simple function that reflects the typical form found in \citet{baumgardt07}:
\begin{equation}
\label{eq:cfelocalapp2}
\gamma(\epsilon)=\max{[0,1-4(1-\epsilon)^3]} ,
\end{equation}
where in the model calculation $\epsilon$ is replaced by $\epsilon/\epsilon_{\rm core}$ as in equation~(\ref{eq:cfelocal}). The right-hand panel of Figure~\ref{fig:cfelocal} shows $\Gamma$ as a function of the gas surface density $\Sigma_{\rm g}$. The difference between CFEs for the hierarchical and spherically symmetric $\gamma(\epsilon)$ relations is typically $\la0.6$~dex. However, the steep transition from $\gamma=0$ to $\gamma=1$ at intermediate SFEs causes $\Gamma$ in the spherically symmetric case to be two times more sensitive to the feedback time-scale than when using the hierarchical $\gamma(\epsilon)$ relation. It is therefore not only physically preferable to adopt a gradual $\gamma(\epsilon)$ relation, but also for the purpose of minimizing the model uncertainties.

\section{The efficiency and stochasticity of the tidal disruption of star-forming regions} \label{sec:appcce}
To determine the contribution of the cruel cradle effect to the cluster formation efficiency, it is important to determine the fraction $f$ of the energy injected by tidal shocks into star-forming regions that is used to unbind the regions. As explained in \S\ref{sec:cce}, this fraction is smaller than unity due to the energy dissipation by the cold gas. If this turbulent energy decay is assumed to be exponential, it follows that the disruption is only possible if a sequence of tidal perturbations has the appropriate strengths and time spacing to cause a runaway effect. Since this is a highly stochastic process, $f$ should be determined by modeling a sequence of tidal perturbations in a Monte-Carlo simulation.

Each Monte-Carlo experiment carried out here uses three randomly drawn variables: the time interval between successive encounters $t_{\rm diff}$, the impact parameter $b$, and the relative velocity $v$. For a stochastic process that occurs on a characteristic time-scale $t_{\rm enc}$, the probability distribution function (PDF) of the time interval $t$ between successive events is
\begin{equation}
\label{eq:tpdf}
\frac{{\rm d} p}{{\rm d} t_{\rm diff}}=t_{\rm diff}^{-1}{\rm e}^{-t_{\rm diff}/t_{\rm enc}} ,
\end{equation}
where $t_{\rm enc}$ is taken from equation~(\ref{eq:tenc}). The PDF for the impact parameter follows from geometric considerations as
\begin{equation}
\label{eq:bpdf}
\frac{{\rm d} p}{{\rm d} b}=\frac{2b}{b_{\rm max}^2} ,
\end{equation}
with $b_{\rm max}$ defined as in \S\ref{sec:cce}. The relative velocity of each encounter is drawn from a Maxwellian distribution of relative velocities with a dispersion of $\sqrt{2}\sigma_{\rm g}$:
\begin{equation}
\label{eq:vpdf}
\frac{{\rm d} p}{{\rm d} v}=\frac{v^2}{2\sqrt{\pi}\sigma_{\rm g}^3}{\rm e}^{-v^2/4\sigma_{\rm g}^2} .
\end{equation}

The example for which the results are shown below was set up for solar neighbourhood-type conditions, i.e.~$\Sigma_{\rm g}=12~\msun~{\rm pc}^2$, $Q=1.5$ and $\Omega=2.6\times10^{-2}~{\rm Myr}^{-1}$, but a broad parameter range (cf. Table~\ref{tab:vars}) was considered in other experiments to verify the universality of the results. For a given sequence of encounters, the energy after each encounter is calculated by writing
\begin{equation}
\label{eq:energy}
E=E_{\rm prev}+(\Delta E)_{\rm tid}+(\Delta E)_{\rm diss} ,
\end{equation}
where $E_{\rm prev}$ is the energy after the previous perturbation, $(\Delta E)_{\rm tid}$ is the energy change due to the tidal shock, and $(\Delta E)_{\rm diss}$ is the energy loss since the last shock due to dissipation. For a given overdensity $x$, the tidal disruption equations from \S\ref{sec:cce} are used to calculate the injected energy per encounter as
\begin{equation}
\label{eq:deltaetid}
(\Delta E)_{\rm tid}=\frac{16G^2g_{\rm close}\phi_{\rm sh}MM_{\rm GMC}R_{\rm h}^2\phi_{\rm ad}}{3b^4v^2} ,
\end{equation}
in which $M$ is the mass of the perturbed region, $R_{\rm h}$ its half-mass radius, and all other variables are defined as previously. For the example below, $M=M_{\rm GMC}$ and $R_{\rm h}=(3M_{\rm GMC}/8\pi\rho_{\rm g})^{1/3}$. The density $\rho_{\rm g}$ is set by $x$ and the density of the ambient interstellar medium $\rho_{\rm ISM}$ as given by equation~(\ref{eq:rhoism}). The initial energy of the region $E_0$ is set by a \citet{plummer11} profile as in equation~(\ref{eq:e}), and the energy that is dissipated between successive encounters is written as an exponential decay:
\begin{equation}
\label{eq:deltaediss}
(\Delta E)_{\rm diss}=-|E_{\rm prev}-E_0|\left(1-{\rm e}^{-t_{\rm diff}/t_{\rm diss}}\right) ,
\end{equation}
with $t_{\rm diss}$ given by equation~(\ref{eq:tdiss}). Note that many of the details of these equations (such as the mass and radius of the region) are actually irrelevant to the question at hand, which strictly involves the fractional energy loss due to dissipation and the stochasticity of the disruption process. Multiplication of equation~(\ref{eq:deltaetid}) by a random factor may change the lifetimes of the modelled star-forming regions, but does not necessarily alter the fraction of the injected energy that is used to unbind the region.

\begin{figure}
\center\resizebox{6.8cm}{!}{\includegraphics{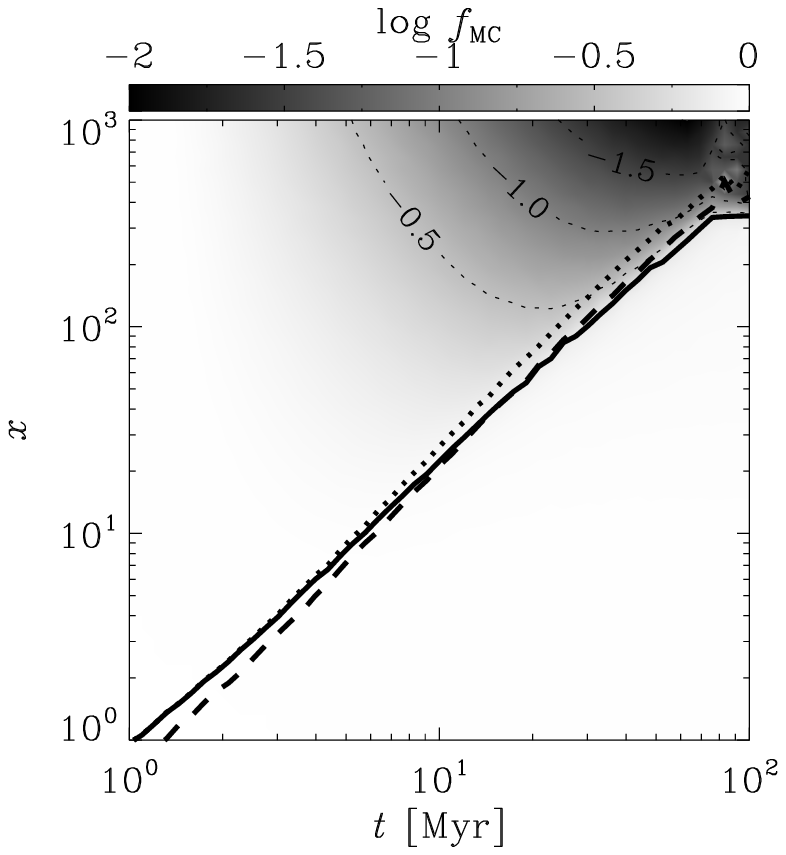}}\\
\caption[]{\label{fig:fitf}
      Fraction $f_{\rm MC}$ of tidally injected energy that is used in unbinding the perturbed region instead of being dissipated away as a function of the time of observation $t$ and the overdensity with respect to the ambient interstellar medium $x$. The grayscale indicates the median value of $f_{\rm MC}$ out of a sample of 30,000 Monte-Carlo experiments per grid point. The thick solid line indicates where the median total disruption time due to the cruel cradle effect $t_{\rm cce,MC}$ is equal to $t$, whereas the thick dotted line represents the loci where the median $t_{\rm cce,MC}'$ (i.e.~without energy dissipation) equals $t$, and the thick dashed line shows the same for $t_{\rm cce,MC}^{\rm fit}$ and $f_{\rm fit}=0.7$. As the figure shows, the agreement between $t_{\rm cce,MC}$ and $t_{\rm cce,MC}^{\rm fit}$ is reasonable, and the difference between them never exceeds 0.3~dex in $x$ across the full parameter space probed in the performed Monte-Carlo experiments.
                 }
\end{figure}
For each Monte-Carlo experiment, the region is subjected to tidal perturbations until either it is disrupted (i.e.~$E>0$) or a time $t$ is reached. The fraction of the energy that was used to unbind the region then follows as
\begin{equation}
\label{eq:fapp}
f_{\rm MC}\equiv\left|\frac{E-E_0}{(\Delta E)_{\rm tid,tot}}\right| ,
\end{equation}
where the subscript `MC' indicates the value is obtained from the Monte-Carlo experiment, $(\Delta E)_{\rm tid,tot}=\sum{(\Delta E)_{\rm tid}}$, and the summation is over all perturbations. For each experiment, the total disruption time $t_{\rm cce,MC}$ is recorded:
\begin{equation}
\label{eq:tcceapp}
t_{\rm cce,MC}\equiv t_{\rm obs}\left|\frac{E_0}{E-E_0}\right| ,
\end{equation}
where $t_{\rm obs}=\min{\{t_{\rm dis},t\}}$ and $t_{\rm dis}$ is the time of the tidal shock that causes $E>0$. The definition of equation~(\ref{eq:tcceapp}) is appropriate for regions that are disrupted before time $t$ (for which $t_{\rm obs}=t_{\rm dis}$) and corrects the disruption time-scale for any excess energy, but it also gives an estimate for regions with disruption times $t_{\rm dis}>t$ (for which $t_{\rm obs}=t$). Without energy dissipation, the total disruption time would read
\begin{equation}
\label{eq:tcceapp2}
t_{\rm cce,MC}'\equiv t_{\rm obs}'\left|\frac{E_0}{(\Delta E)_{\rm tid,tot}'}\right| ,
\end{equation}
where $t_{\rm obs}'$ and $(\Delta E)_{\rm tid,tot}'$ are set at the time of the shock that causes $|(\Delta E)_{\rm tid,tot}'|>|E_0|$. The dissipational energy loss can then be parameterized by decreasing the total injected energy by a fixed factor $f_{\rm fit}$, i.e.~writing $t_{\rm cce,MC}^{\rm fit}=t_{\rm cce,MC}'/f_{\rm fit}$ and choosing $f_{\rm fit}$ such that $t_{\rm cce,MC}^{\rm fit}\approx t_{\rm cce,MC}$. This formulation is useful when dealing with ensembles of star-forming regions as in the context of this paper. We determine the constant $f_{\rm fit}$ from the results of the Monte-Carlo experiments below.

The described Monte-Carlo experiment is executed 30,000 times for each combination of the times $1<t/{\rm Myr}<100$ and overdensities $10^0<x<10^4$. The median value of $f$ according to equation~(\ref{eq:fapp}) is shown in Figure~\ref{fig:fitf} as a function of $t$ and $x$. The figure shows that the energy loss is typically largest for high overdensities and late times. This is not surprising, because high-density regions have short dissipation time-scales and take the longest to be tidally disrupted. Additionally, the energy loss should be low at young ages, because the energy simply has not had the time to be dissipated.

Because the contribution of the cruel cradle effect to the CFE is formulated in equation~(\ref{eq:xcce}) as a threshold value $x_{\rm cce}$ above which star-forming regions survive until time $t$, Figure~\ref{fig:fitf} includes the line that satisfies ${\rm median}(t_{\rm cce,MC})=t$. For comparison, the line where the median of $t_{\rm cce,MC}^{\rm fit}$ equals $t$ is also shown, assuming $f_{\rm fit}=0.7$. For this choice of $f_{\rm fit}$, both lines never deviate by more than 0.3~dex, indicating that it provides a good representation for the fraction of the tidally injected energy that is used to unbind the region. As the discrepancy between the solid and dashed lines illustrates, this slightly underestimates the impact of the cruel cradle effect at low $x$ and $t$, while slightly overestimating it at high $x$ and $t$. This reflects the increase of the energy dissipation with $x$ and $t$. However, the differences are so minor that $f=0.7$ is adopted throughout this paper.

The reason that the tidally injected energy is mainly used for unbinding a region (i.e.~$f>0.5$) if the disruption time $t_{\rm cce,MC}<t$ is partially caused by a selection effect. Due to the energy dissipation, a sequence of tightly-spaced tidal perturbations is required to unbind a gas-rich region. Because the energy from such a sequence of perturbations is not dissipated, any region that is disrupted before time $t$ will have a high efficiency factor $f$. Conversely, regions that survive have not had a runaway sequence of tidal shocks and radiated away most of the energy, thus giving a low value of $f$. In a statistical ensemble of tidal shocks, the probability of having a sequence of encounters that successfully disrupts a region is set by its disruption time-scale. This probability is close to unity if $t_{\rm cce,MC}\leq t$, and hence $f$ will be close to unity as well.

\bibliographystyle{mn2e2}
\bibliography{mybib}

\bsp

\label{lastpage}

\end{document}